\documentclass[12pt]{article}
\usepackage{graphicx}
\usepackage{subfig}
\usepackage{float}
\usepackage{multirow}
\usepackage{amsmath}

\def\beq{\begin{equation}}
\def\eeq{\end{equation}}
\def\beq{\begin{eqnarray}}
\def\bea{\end{eqnarray}}
\begin{document}
\baselineskip18pt
%\paraskip10pt
\topmargin=-8pt

\begin{center}
{\huge Evolution of the vibrational spectra of single-component solids with
pressure: some universalities}
\end{center}

\vskip0.9in

\begin{center}
{\bf Divya Srivastava and Subir K. Sarkar}

{\it School of Physical Sciences, Jawaharlal Nehru University, New Delhi - 110 067, INDIA}
\end{center}

\vskip0.9in

\begin{center}
\section*{ABSTRACT}
\end{center}

We have studied numerically the evolution of the zero temperature
vibrational spectra of single-component solids with pressure using various
model potentials with power law (type A) or exponential (type B) repulsive part.
Based on these data and some semi-analytical calculations our principal
results may be summarized as follows. For type A potentials: (i) The average 
frequency has a power law dependence on the pressure;(ii) The normalized vibrational 
density of states (NVDOS), with the average frequency as the unit of frequency, will
saturate as the pressure keeps increasing. This asymptotic NVDOS is
independent of the attractive component of the potential and hence define a
universality class; and (iii) At higher pressures the Debye frequency and the
average frequency have the same pressure dependence and this dependence is
identical for the amorphous form and the two crystalline forms studied (FCC
and HCP). For type B potentials, the above phenomenology will hold
good to a good approximation over a wide range of intermediate pressures. We suggest a scaling
form of the dispersion relations that would explain these observations. Various
aspects of the evolution of sound speed with pressure are also studied. In particular
we show that  the Birch's law prescribing linear relationship between density and
sound speed will hold good at very high pressures only in exceptional cases.
We have also analyzed the data available in the literature from
laboratory experiments and {\it ab initio} calculations and find that there is  
agreement with our conclusions which are derived from the study of model 
potentials. We offer explanation for this agreement.

\vskip0.5in

PACS numbers: 05.70.Jk, 63.50.-x, 63.20.-e, 62.50.-p, 62.65.+k

\newpage
\begin{center}
{\bf I. Introduction}
\end{center}
\vskip0.2in
\noindent Study of the vibrational spectra of solids provides a rich variety of 
information.
It helps us probe the interaction amongst the constituent particles of the
solid. It also leads to an understanding of many properties measured at the
macroscopic level such as sound propagation, specific heat etc.. In this sense
vibrational spectra provide a bridge between the microscopic and
the macroscopic aspects of solids. Unfortunately, understanding these
spectra theoretically  poses a very difficult challenge. As far as {\it computation} of
the vibrational spectrum of any real material under a specified set of
external conditions is concerned {\it ab initio} procedures based on electronic structure
calculations are eminently possible
nowadays for crystals if the unit cell size is not too big. But from the point of
developing an understanding of the physics involved model-based studies offer some
advantages. However, even if the
interaction amongst the constituents is given, analytical calculation
is made difficult first by the need to generate the geometry of the solid and
secondly, in the case of amorphous systems, by having to compute the vibrational
eigenvalue spectrum of a system with no spatial periodicity. It is not
surprising therefore that a  majority of even the model-based calculations are done on
computers -- especially when the amorphous state is involved[1-14]. Many details 
of the spectrum naturally depend on the potential
of interaction applicable to the particular solid. Our interest in this paper
is  a study of the evolution
of the vibrational spectra (and any other information derivable from
it) of solids with externally applied pressure -- with special emphasis on
those aspects, qualitative or quantitative, that do not depend on some of
the details of the potential. This study is done  in the context of
model single-component solids. The model potentials do not necessarily
describe any particular material realistically. But we will argue
that some predictions derived from studies based on these model potentials are
applicable to real materials and we will make comparisons of these predictions
with experimental data and {\it ab initio} calculations whenever permissible.

Pressure dependence of the crystalline vibrational spectrum has been studied
in the literature both because of its fundamental physical interest and its
relevance to problems of planetary geophysics [15-28]. Iron has been studied
the most since it is the primary ingredient of the core of the earth.
Interpretation of the data from experiments on geomagnetic or seismological
properties of earth's interior depend crucially on the availability of
reliable data on the relevant properties of iron at high pressure. Analysis
of data on the evolution of the vibrational spectrum has to be done keeping
in mind that there can be a change of crystal structure as pressure is
increased. For example, iron changes from body-centered-cubic to
hexagonal-close-packing at around 13 GPa. 

For amorphous systems these studies have focussed largely on the boson
peak [29-40] which
stands for an excess of vibrational modes somewhere in the very low frequency
part of the spectrum and the excess is in comparison to what would be expected
from the Debye model. In particular we draw attention to [37] . In this
laboratory experiment the vibrational spectrum of a disordered polyisobutylene
sample is measured at various pressures ranging from ambient to $1.4$ GPa. The
highest pressure applied causes a twenty percent increase of density and about
a factor of two increase in the characteristic vibrational frequency. One
result presented in this paper (see also [38-39]), analogues of which 
will be of central importance in our work,
is that the plot of the reduced density of vibrational states against
frequency  can be made independent of applied pressure in the region of the
boson peak with proper choice of units for the two axes. Also
the variation of the Debye frequency with pressure was studied via measurements
of longitudinal and transverse sound velocities (along with density). Our work
is a computational version of experiments such as this -- with the qualification 
that we study both crystalline and amorphous states and boson peaks are not
studied at all due to the lack of adequate accuracy in the computation of the density
of states at the lowest frequencies.

The other aspect that we investigate here is the speed of propagation of longitudinal and
transverse acoustic waves. Study of sound velocities is of great interest in many physical 
situations 
and they may or may not include measurements of vibrational 
spectrum. In many cases the experimentally measured quantities are the elastic
constants and this is then transcribed into statements on sound velocities. In
yet another set of experiments inelastic scattering from phonons is used to
determine sound speeds. A rather well established law regarding sound speed at
high pressures is the Birch's law which specifies linear relationship between density
and sound velocity at high pressures. In the present work we address this aspect also
and demonstrate that at extreme pressures this law holds only in exceptional situations.

Some of the key features of our work are as follows:
(i) The applied pressure is always hydrostatic;
(ii) We deal with a single-component system for which the interactions are
given and have a relatively simple functional form;
(iii) We study both amorphous and crystalline forms of matter for a given
model potential. In the crystalline phase both FCC and HCP forms are
supported by all the potentials used by us;
(iv) The choice of potentials is a continuation of a previous set of studies
where we demonstrated some universal properties of the statistical aspects
of the vibrational spectra of amorphous clusters [41-42];
(v) Our calculations of vibrational modes are at zero
temperature and hence we use the harmonic approximation. However, this should not be too much
of a compromise since in the
limit of very high pressure effect of a finite temperature is expected to
be rather weak unless of course the temperature is very high, and  finally,
(vi) Considering the changes in density or characteristic vibrational
frequency, our highest pressures are far higher than in any previous
laboratory study or in any presently conceivable laboratory study. 

The organization of this paper is as follows: In section II we present
the methodologies of
computation and analysis. Section III presents raw and scaled data on the
evolution (with pressure) of vibrational spectra for representative
potentials. We provide some representative data 
that make all the essential points. Also presented are data on the power law scaling
of the average vibrational frequency with pressure for various potentials.
Section IV describes the results of a
study of a nearest neighbor model of the crystalline vibrational spectrum in the case
of FCC and HCP lattices.
Section V contains a scaling ansatz and a discussion of its applicability
in various situations. Section VI reports on the study of
sound velocity with particular reference to the Birch's law - especially the
domain and limits of its validity.
Section VII contains some additional results relating to our study of
amorphous systems. The main result  that is reported relate to the
existence of a pair of isosbestic points in the normalized vibrational spectrum 
when the exponent of the power law repulsive potential
is varied ( in the infinite pressure limit). Section VIII contains an
analysis of data that is available in the literature and comparison with our
predictions. Finally, section IX contains some concluding remarks.
%\newpage
\vskip0.3in

\begin{center}
{\bf II.  Methodology}
\end{center}
\vskip0.3in

\noindent In order to prepare solid 'samples' in the
model studies first we have to choose the interaction among the
particles. All the potential energy expressions that we have used have the
structure of an attractive part plus a repulsive part. The latter
is always a sum over pairs i.e. is of the form $\sum_{All\, pairs}u(r_{ij})$
where u is the central pair potential; $i$ and $j$ label the particles in a pair
for which $r_{ij}$ denotes the pair separation.
The attractive component is sometimes of the
sum-over-pairs type. However, we also use non-sum-
over-pairs potential in some cases for this part. The expression 
for $u$, the repulsive pair potential,
will be of great importance in subsequent discussions. This function will
always be of a power law form (i.e. $u(r)$ proportional to $r^{-m}$) or
an exponentially  decaying one (i.e. $u(r)$ proportional to $\exp(-r/r_{0})$). We call
them type A and type B potentials, respectively. The explicit forms of the potential
energy expressions that we have used are: (i) Generalized Lennard-Jones potential:
$\sum_{All\, pairs}(1/r_{ij}^{m_1} - 1/r_{ij}^{m_2})$. Here $m_1$ and $m_2$ are positive
integers with $m_1$ greater than $m_2$ ($m_1$ is typically 12 or more). We will refer to 
this potential as
GLJ$(m_1,m_2)$. (ii) Morse potential: $\sum_{All\, pairs}(\exp(-2\alpha(r_{ij} - 1)) - \exp(-\alpha
(r_{ij} - 1)))$. In this study the value of $\alpha$ will always be taken to be 9.0. (iii) Gupta
potential: $A\sum_{All\, pairs} \exp[-p(r_{ij}-1)]-\sum_{i=1}^{N}\sqrt{\sum_{j\neq i}\exp[-2q(r_{ij}-1)]} \,\mathrm{with}\, A=0.0376,\, p=16.999, \, \mathrm{and}\, q=1.189 $    , and
(iv) Sutton-Chen potential: $\sum_{All\,pairs}(1/r_{ij}^9) -\beta \sum_{i=1}^{N}\sqrt{\sum_{j\neq i}(1/r^6_{ij})}\,\mathrm{with}\, \beta = 39.432$. 
The attractive parts of the expressions for the Gupta and Sutton-Chen potentials, as written here, are
for a finite cluster with $N$ particles. We, of course, use them in the context of the infinitely
extended solid.
The numerical values used for the parameters in the Sutton-Chen and Gupta
potentials are those applicable to nickel. For all the potentials  a suitable pair separation cutoff is
employed.  Also used is an appropriate interpolating function at this point. The latter ensures that
derivatives upto second order are continuous at the cutoff distance. It can be seen that GLJ and Sutton-Chen
potentials are of type A whereas Morse and Gupta potentials are of type B. The choice of these 
particular potentials has been made basically as a continuation of their usage in
some previous investigations into universal aspects of vibrational spectra [41-43]. Their
analytical structures are more or less of the simplest types that incorporate the
minimal requirements of physical plausibility and also lend themselves to efficient computation.

Face centered cubic (FCC) or hexagonal close packed (HCP)
crystalline configurations at a particular pressure $P_0$ are
produced by minimizing $(U + P_{0} {\cal V})$ with respect to the adjustable lattice
parameters. Here $U$ denotes the potential energy per unit cell of volume $\cal V$.
For HCP structures the lattice parameters in and perpendicular to the plane of
hexagonal symmetry are allowed to vary independently. But empirically it is found that
at higher pressures the geometry of the HCP lattice is essentially indistinguishable from the ideal HCP
lattice.  This is a
reflection of the isotropic non-bonding nature of the interactions that we
are dealing with. In some parts of our analysis we will make use of this observation of ideality
of the HCP lattice.

Following standard practice we approximate the amorphous state by a periodic
system with as large a unit cell as possible. For us the unit cell contains
$343$ particles. By comparison with spectra with only 125 particles and
spectra with as high as 6980 [44] particles in the unit cell we know that
the spectra with 343 particles is very close to the infinite system limit
except in the lowest eight percent or so of the spectral range (containing
about two percent of all the modes) where the difference
is somewhat noticeable. However, as we shall see, most of our results do not
require a knowledge of this small part of the spectrum with high accuracy.

To prepare the amorphous solid 'samples' at various pressures we adopt the
procedure outlined next. First choose the potential energy function. The
unit cell containing 343 particles is
defined by the edges ${\bf a}_1 $, ${\bf a}_2$ and ${\bf a}_3$. We
start with the stable FCC lattice at some relatively low pressure $Q$.
This is then melted via a $NPT$-type configurational Monte-Carlo simulation
with $P = Q$. In the simulation the variables
are the three unit cell vectors and the 343 position vectors. From the liquid
state we select a certain number of configurations  spaced equally in time
by a timescale substantially longer than the correlation time. Finally, using these
configurations as initial guesses, conjugate gradient
minimization is performed for the function $(U + Q {\cal V})$ with respect
to the variables of the $NPT$ simulation. Every state obtained via this 
minimization is characterized by the
volume ($\Omega$) and potential energy ($\epsilon$) per particle. The states
in the $\epsilon$-$\Omega$ plane with the highest energies and the lowest
densities correspond to the amorphous 'samples'(figure 1(a)). This is also
crosschecked
by examining the pair correlation function (figure (1b)). \\
\begin{figure}[H]
  \centering
  \subfloat[]{\label{fig:(a)}\includegraphics[width=0.4\textwidth]{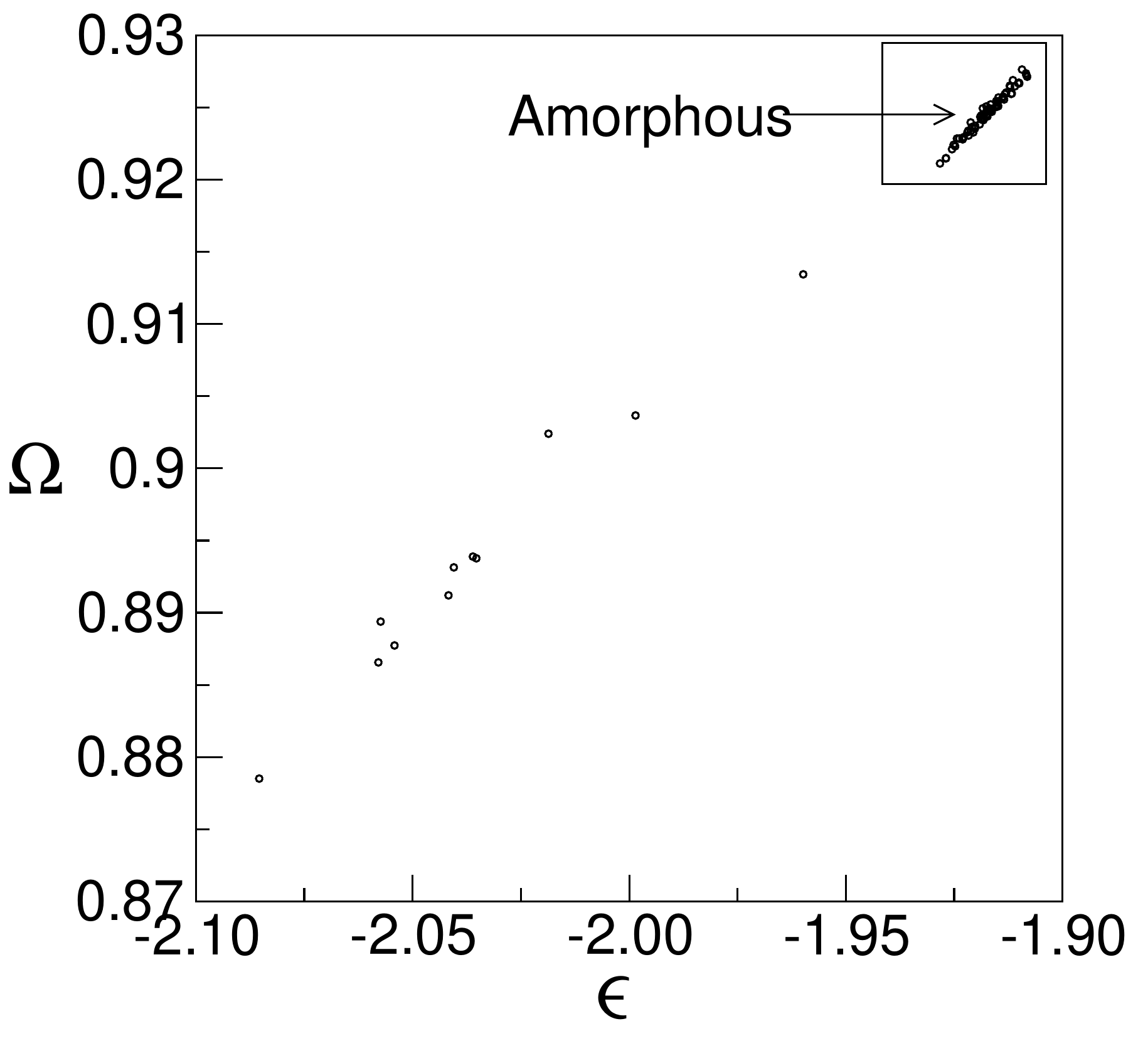}}
  \subfloat[]{\label{fig:(b)}\includegraphics[width=0.4\textwidth]{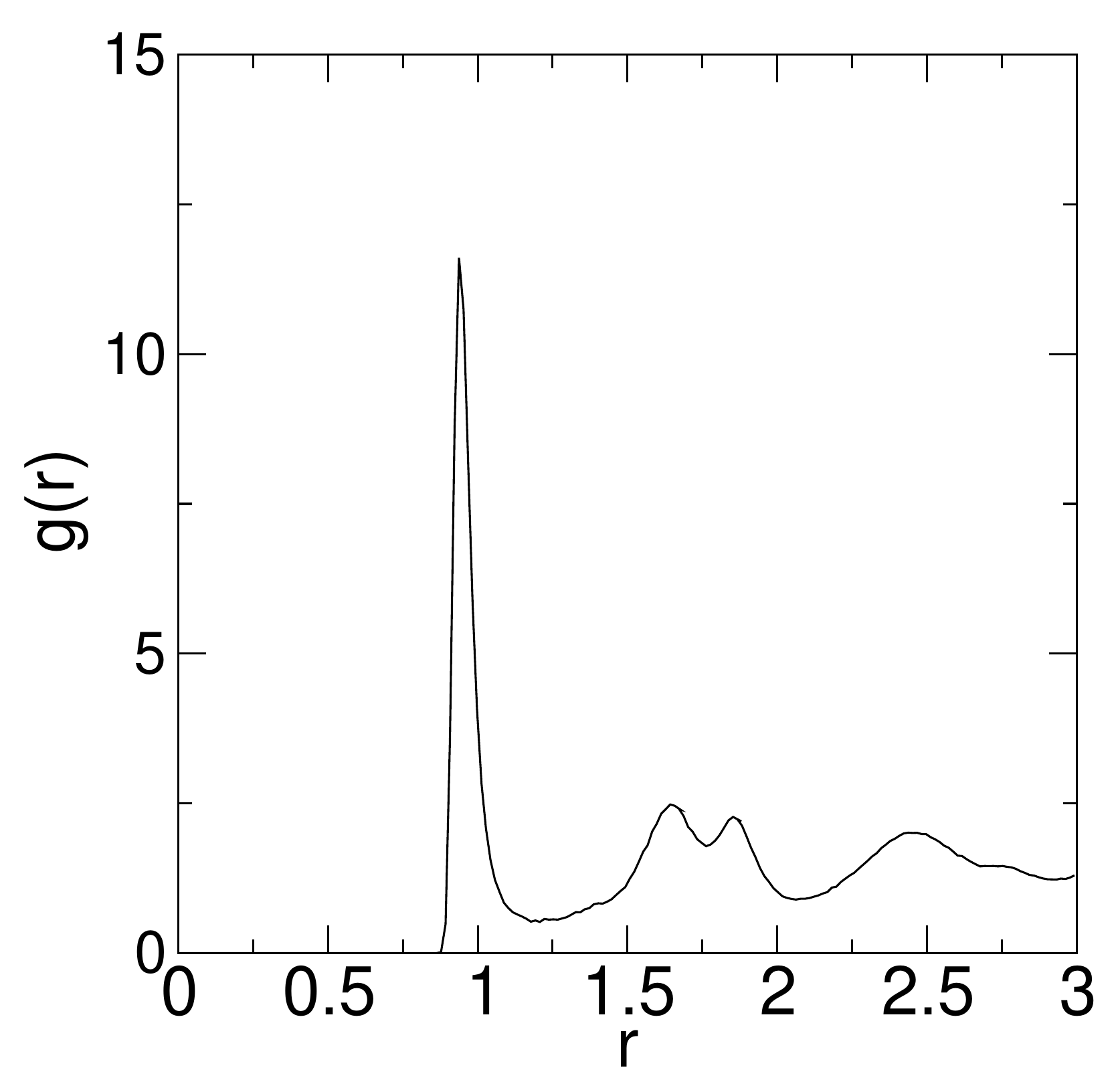}}
  \caption{(a)Volume per particle($\Omega$) versus energy per particle($\epsilon$) for
GLJ(12,6) at $P = 1$. The region inside the box corresponds to amorphous 
states. (b) Pair correlation function averaged over all the states in the
amorphous region of (a)}
  \label{fig:fig1}
\end{figure}

Once the $T = 0$ amorphous solid geometries are obtained at the pressure $Q$, amorphous states at other
pressures are generated by  slow variation of pressure i.e. configurations for the
minima at a
particular pressure are used as the initial guesses for minimizations at
a slightly different pressure. For each new pressure only the states with
proper amorphous character are accepted -- through an examination of the pair
correlation function. We have checked that there is no statistically significant hysteresis
in the generation of the amorphous states during pressurization
and depressurization. Thus these states are reasonably uniquely defined as a
function of pressure. 

Once the stable solid geometries are generated computation of the
vibrational density of states (VDOS) is done by diagonalizing the dynamical
matrix and by integrating the resulting spectrum over the first Brillouin
zone . Number of $\bf k$ -points taken in the irreducible part of the first Brillouin
zone for this integration varies from 108 in the amorphous case to not
less than 11760 for the crystalline cases. For the amorphous case the DOS is
averaged over all the local minima (between fifty and hundred in number).

To calculate the transverse ($c_{T}$) and longitudinal ($c_{L}$) speeds of
sound propagation we use a method that avoids having to numerically evaluate
the ratio of (frequency/length of wavevector) for a progressively shrinking sequence
of wavevectors in a fixed direction -- thus saving computational time and eliminating 
a significant source of
numerical error. For the sake of completeness we briefly summarize
the method [45] for a situation where there are $N$ 
particles in the
unit cell and they are not necessarily all of the same type -- although in our applications
the particles are identical in all respects. Since we approximate the amorphous state 
by a periodic geometry
with a sufficiently large unit cell the method described in the following is
applicable to both crystalline and amorphous states.\\
%******************************************************************************
For the unit cell associated 
with the Bravais lattice vector $\bf R$ let the equilibrium position and the 
displacement from equilibrium position of the $j^{th}$ atom $(j=0,1,2,......,(N-1))$
of mass $m_{j}$ be ${\bf P}(^{\bf R}_{j})$  and ${\bf u} (^{\bf R}_{j})$, 
respectively.\\
If V is the overall potential energy of the system, define
\begin{equation}
\Phi_{\alpha \beta}(^{\,\bf S}_{jj'}) =  \frac{\partial^{2}V}{\partial 
u_{\alpha}(^{\bf R_{2}}_{\,j})\,\partial u_{\beta}(^{\bf R_{1}}_{\,j'})} \end{equation}
and
\begin{equation}\qquad {\bf x}(^{\,\bf S}_{jj'}) =  {\bf P}(^{\bf R_{2}}_{\,j})-
{\bf P}(^{\bf R_{1}}_{\,j'}) \end{equation}
where $\bf S \equiv \bf R_{2}-\bf R_{1}$ and the derivatives are evaluated at the  
equilibrium configuration. Now define 
\begin{equation}C_{\alpha \beta}^{(0)}(jj')=\frac{1}{\sqrt{m_{j}m_{j'}}}\sum_{\bf R\in BL}
\Phi_{\alpha \beta}(^{\,\bf R}_{jj'})\end{equation}
\begin{equation}C_{\alpha \beta, \gamma}^{(1)}(jj')=\frac{-2\pi}{\sqrt{m_{j}m_{j'}}}\sum_{\bf R\in BL}
\Phi_{\alpha \beta}(^{\,\bf R}_{jj'})\,x_{\gamma}(^{\,\bf R}_{jj'})\end{equation} and
\begin{equation} C_{\alpha \beta, \gamma\lambda}^{(2)}(jj')=\frac{-4\pi^{2}}{\sqrt{m_{j}m_{j'}}}
\sum_{\bf R\in BL}\Phi_{\alpha \beta}(^{\,\bf R}_{jj'})\,x_{\gamma}(^{\,\bf R}_{jj'})
\,x_{\lambda}(^{\,\bf R}_{jj'})\end{equation}
where {\bf BL} stands for the Bravais lattice.

Now consider the $(3N-3)\times(3N-3)$ matrix $C_{\alpha \beta}^{(0)}(jj')$ with $j$ and 
$j'=1,2,..........,(N-1)$. Denote the inverse of this matrix by $\Pi$ and then 
define the $3N\times3N$ matrix $\Gamma$ as follows
\begin{equation}\begin{array}{c c c}\Gamma_{\alpha \beta}(jj')&
=\Pi_{\alpha \beta}(jj') &\qquad if  j,j'\neq 0\\
&=0\qquad\quad &\qquad otherwise
\end{array} \end{equation}
With these definitions the squares of the three sound velocities in the direction of 
the unit vector $\hat{s}$ are given by the eigenvalues of a $3\times3$ matrix $G$ where
\begin{equation}G_{\alpha \beta}=\frac{1}{(\sum m_{j})}\sum_{\gamma\lambda}s_{\gamma}\,s_{\lambda}\,
([\alpha \beta,\gamma\lambda]+\{\alpha\gamma, \beta\lambda\})\end{equation} with
\begin{equation}[\alpha \beta,\gamma\lambda]=\frac{1}{8\pi^{2}}\sum_{jj'}(m_{j}m_{j'})^\frac{1}{2}
\,C_{\alpha \beta, \gamma\lambda}^{(2)}(jj')\end{equation} and
\begin{equation} \{\alpha\gamma, \beta\lambda\}=-\frac{1}{4\pi^{2}}\sum_{jj'}\sum_{\mu \nu}
\Gamma_{\mu \nu}(jj')\, \Big(\sum_{j''}C^{(1)}_{\mu \alpha,\gamma}(jj'')\sqrt{m_{j''}}\Big)
\,\Big(\sum_{j'''}C^{(1)}_{\nu \beta,\lambda}(j'j''')\sqrt{m_{j'''}}\Big)\end{equation}

%******************************************************************************
\noindent In general these
speeds will not be isotropic -- although for amorphous states the direction
dependence, caused by the finite size of the unit cell, will be very weak.
Hence, unless otherwise stated, average speed of sound (transverse or
longitudinal)  will always mean an average over ten thousand randomly (and
isotropically) choosen directions. For amorphous states a further averaging
is done over the configurations.

Finally, the Debye frequency $\omega_{D}$ is calculated through the formula
${{\omega}_{D}}^{-3} = B/(18 n {{\pi}^{2}})$ where $B$ denotes the direction
averaged value of ($1/{{{c}_{1}}^3}$ +  $1/{{{c}_{2}}^3}$ +
$1/{{{c}_{3}}^3}$) --  ${c}_{1}$, ${c}_{2}$ and ${c}_{3}$ being the three
speeds of sound in a particular direction. $n$ denotes the number density of
particles.
\vskip0.3in
%\newpage
\begin{center}
\bf III.  Results for the vibrational spectrum
\end{center}
\vskip0.3in
\begin{center}
{\bf A. Scaling law for the average frequency}
\end{center}
\vskip0.2in

\noindent Given the VDOS for a given potential and state of aggregation
calculation of the average vibrational frequency is straightforward.
Average frequency is defined as
$<\omega> = \int \omega G(\omega)d\omega / \int G(\omega)d\omega$  where
$G(\omega)$ is the vibrational density of states. Please note that $G(\omega)$
is proportional to the size of the sample but $<\omega>$ is not. This
average is an indicator of the characteristic vibrational frequency of the
system -- although one can define other measures. Studies of variation
of boson peak frequency or spectrum averaged frequency (or square of frequency
, in case of studies in superconductivity) with pressure do exist.
However, we are not aware of
any investigations in the ultrahigh pressure regime that we are
studying here. We find that in all the cases we have studied there is very
strong support for a power law scaling of the average frequency with pressure
, if we exclude the relatively low values of pressure. Thus $<\omega>$ is
proportional to $P^\delta$ where $P$ is the applied hydrostatic pressure
and $ \delta$ is the scaling exponent. And this power law
regime extends over a range where the applied pressure varies
by a factor of the order of 100 or so and the frequency varies by a factor
of about
10 (typically the range of our numerics is somewhat less for amorphous states since
computation has to be done for a large number of local minima). Here we
present the data only for one potential each of types A and B.
Figures 2(a),
2(b) and 2(c) show the data for the GLJ(12,6) potential for the FCC, HCP and the
amorphous state, respectively. In each case data for the full range of
pressures is shown in the main display and that for the regime of high
quality power law scaling is shown in the inset (in all cases the data points
in the inset correspond to the high pressure end of the main display with no
alteration in the number of data points). Figure 3 shows the
corresponding data for the Morse potential. The quality of the power law
fit should be evident from these figures. 
%*****************************************************
\begin{figure}[H]
  \centering
  \subfloat[]{\label{fig:(a)}\includegraphics[width=0.32\textwidth]{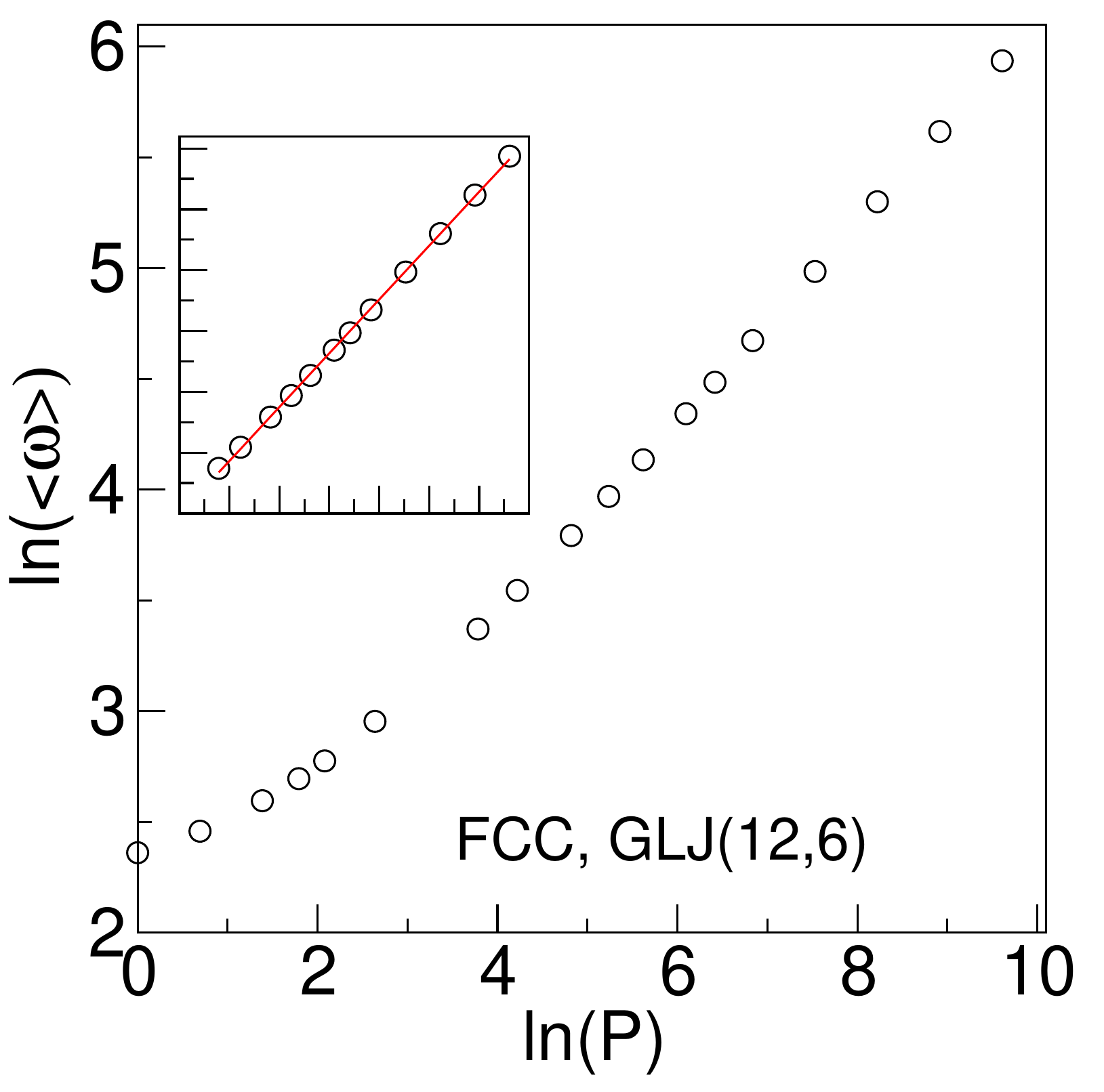}}
  \subfloat[]{\label{fig:(b)}\includegraphics[width=0.32\textwidth]{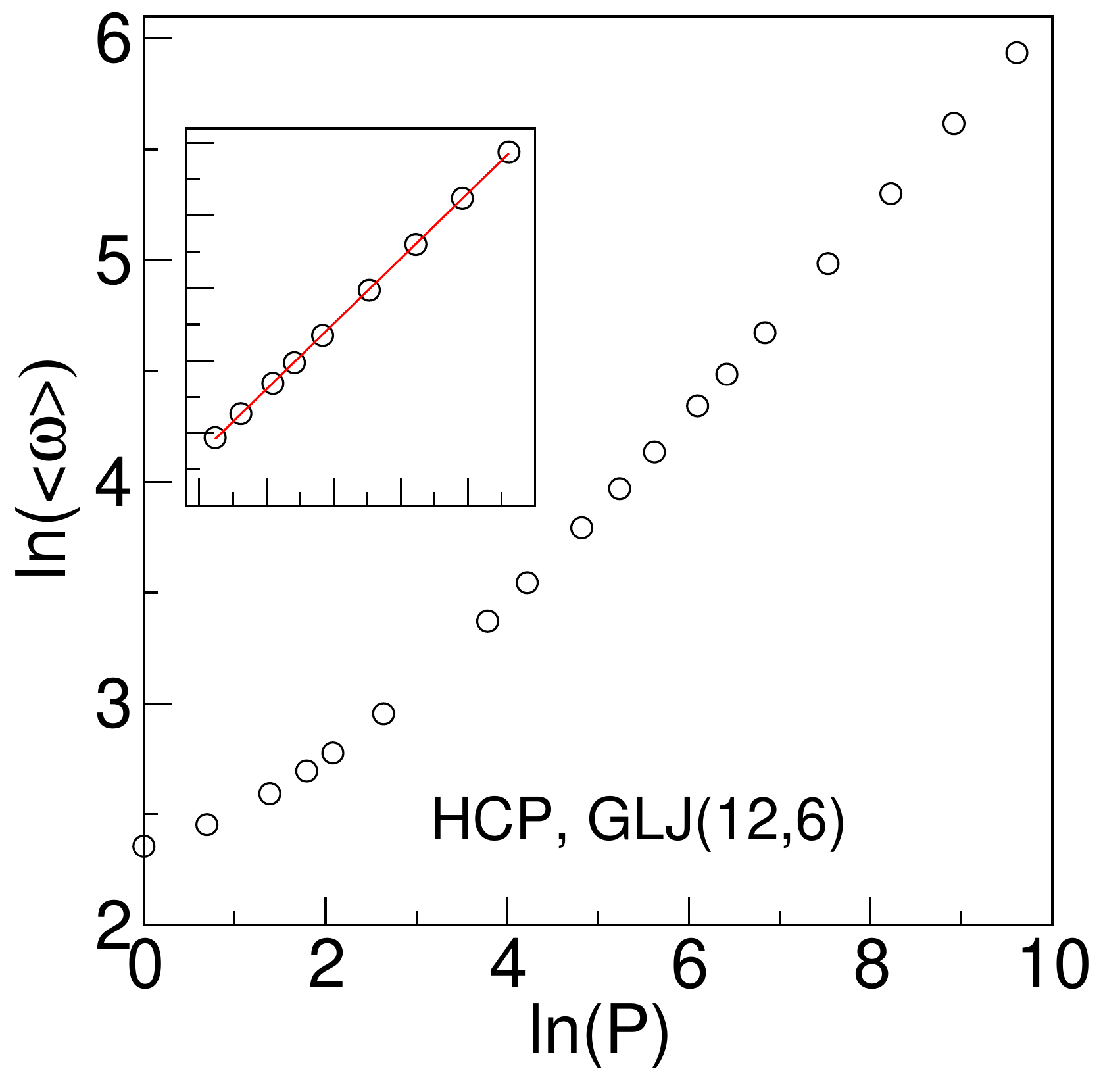}}
  \subfloat[]{\label{fig:(c)}\includegraphics[width=0.32\textwidth]{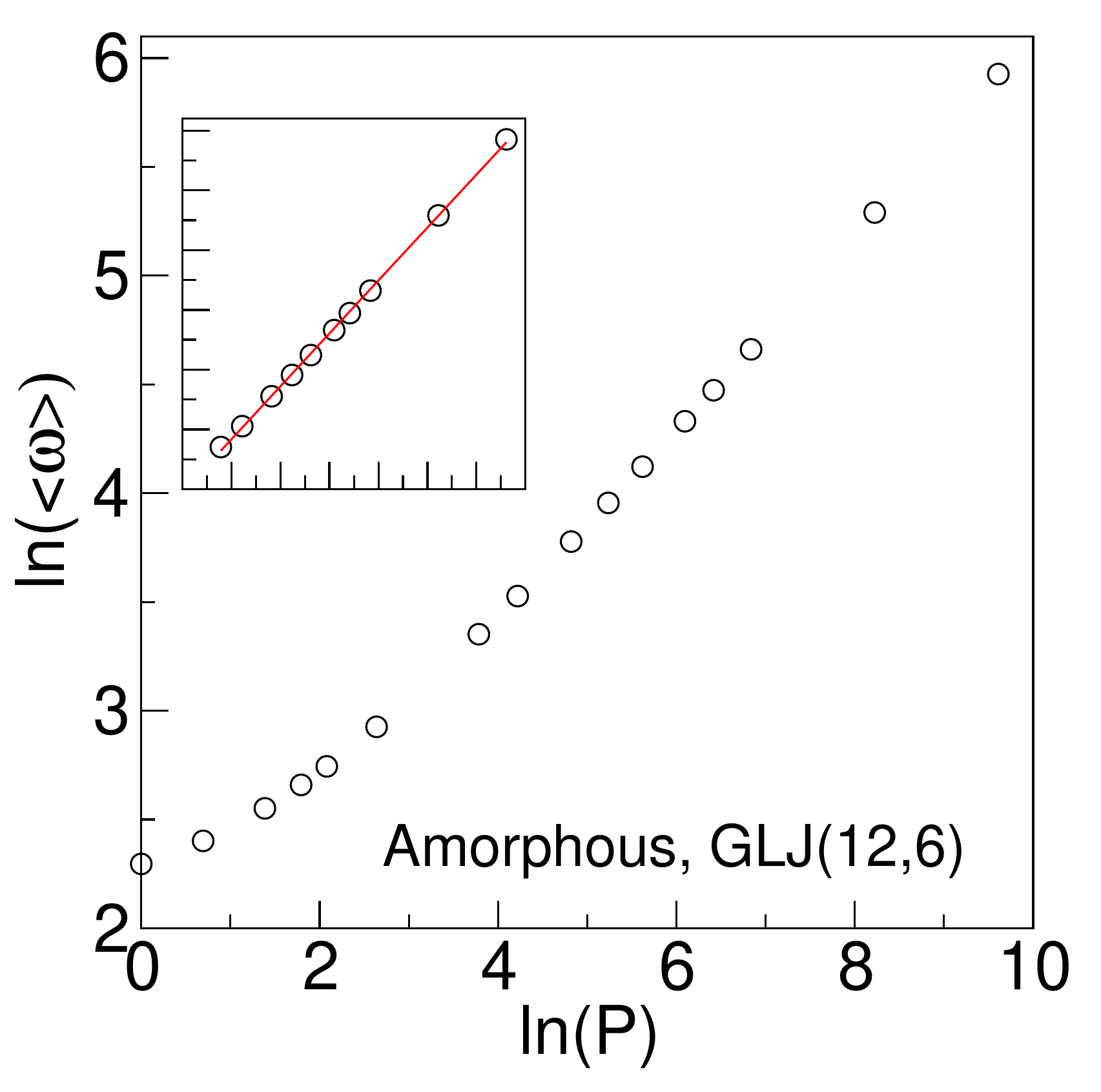}}
  \caption{\small{(a) Log of average frequency $(<\omega>)$ plotted as a function of 
log of pressure $(P)$ for the FCC state of GLJ(12,6). The inset shows the precise linear
dependence  in the highest pressure region. (b) Same as in (a) but for the
HCP state. (c) Same as in (a) but for the amorphous state.}}
  \label{fig:fig2}
\end{figure}

\begin{figure}[H]
  \centering
  \subfloat[]{\label{fig:(a)}\includegraphics[width=0.32\textwidth]{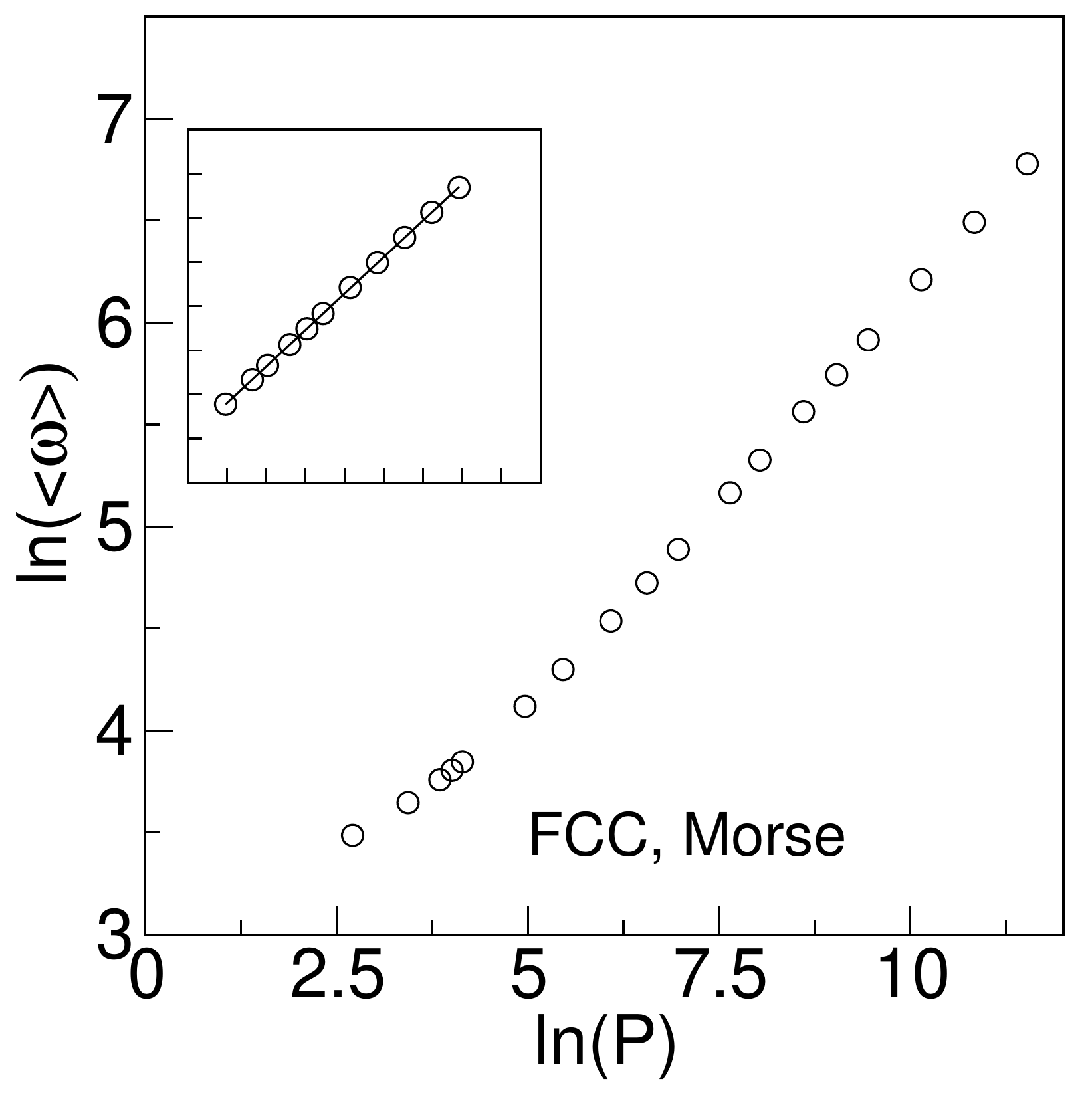}}
  \subfloat[]{\label{fig:(b)}\includegraphics[width=0.32\textwidth]{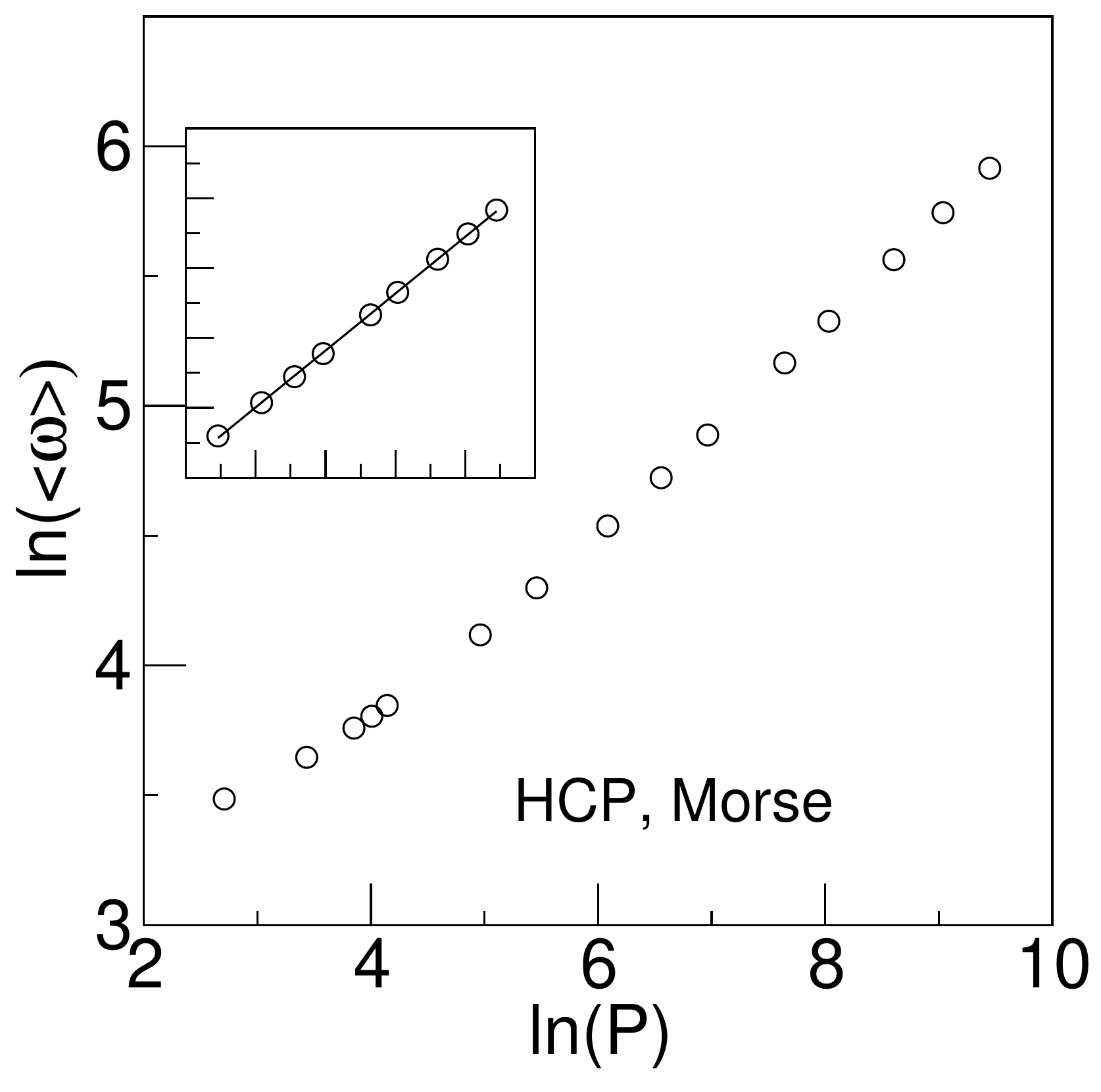}}
  \subfloat[]{\label{fig:(c)}\includegraphics[width=0.32\textwidth]{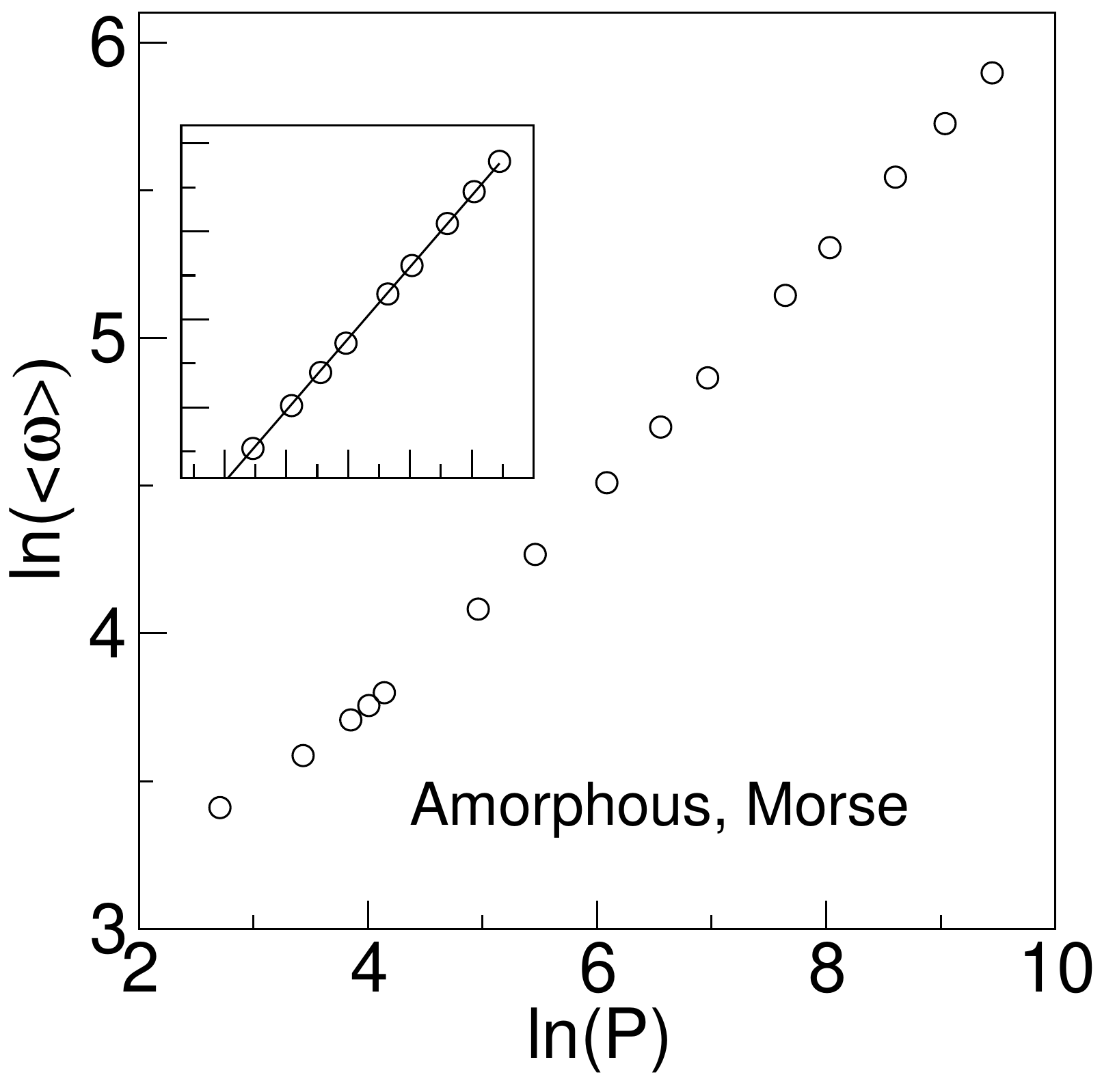}}
  \caption{\small{(a) Log of average frequency $(<\omega>)$ plotted as a function of 
log of pressure $(P)$ for the FCC state of Morse potential. The inset shows the precise linear
dependence  in the highest pressure region. (b) Same as in (a) but for the
HCP state. (c) Same as in (a) but for the amorphous state.}}
  \label{fig:fig3}
\end{figure}

%*****************************************************

Table I provides the best fit
values of the power law exponent ($ \delta$) for five different potentials. Two
observations that can be made on the basis of the data in this table are: (1) The
exponent seems to not depend on the state of aggregation, and (2) In all the cases
the exponent is in a relatively narrow range below 1/2.
\begin{table}[H]
\begin{center}
\caption{\small{Scaling exponent for the variation of average frequency with pressure}}
\vskip0.3in

\begin{tabular}{|cc|c|c|c|l}
\cline{1-5}
\multirow{2}{*}{Potential} && \multicolumn{3}{|c|}{State of aggregation} \\ \cline{3-5}
 & & AMORPHOUS &\hspace{0.026cm} FCC \hspace{0.026cm} &\hspace{0.026cm}  HCP\hspace{0.026cm}  \\ \cline{1-5}

GLJ(12,6)&& 0.44 & 0.44 & 0.45\\ \hline
GLJ(12,10)&& 0.45& 0.45 & 0.45 \\ \hline
Sutton-Chen    & &0.44 &0.45 &0.44\\ \hline
Morse && 0.41 & 0.41 & 0.41 \\ \hline
Gupta && 0.41 & 0.39 & 0.40 \\ \hline

\end{tabular}
\end{center}
\end{table}

%\newpage
\vskip0.3in
\begin{center}
{\bf B. Shape convergence of the density of states function}
\end{center}
\vskip0.3in
\noindent Absolute vibrational density of states is proportional to the size of the
sample and the scale of frequency depends on the particular potential at hand.
Thus to compare VDOS at different pressures or perhaps for different
potentials it is necessary to introduce a description that uses only
dimensionless variables. This can be done in several ways but the one we use
is to choose the unit of frequency to be $<\omega>$ , the average frequency.
The dimensionless normalized frequency thus defined is denoted by
$\nu ( \equiv \omega/<\omega>$). The next step is to define a normalized
density of states for normalized frequency (NDOSNF) $g(\nu)$ =
${\beta}G({\nu}<\omega>)$ where $\beta$ is choosen so that the integral
$\int g(\nu)d{\nu}$ has a fixed value in all cases. It is clear that 
$g(\nu)$ describes the shape of the VDOS function. When the applied pressure goes up
there is a steady increase in density as well as the characteristic frequency.
In fact density increases by a factor which is often higher than four
(and typically a little less for amorphous states) and the characteristic
frequency goes up by a factor of around 15 ( Density and characteristic
frequency serve as  physical measures of how large an applied pressure is).
In fig.4(a) we show the VDOS function, denoted by $G(\omega)$, at five
different pressures for the FCC state of the GLJ(12,6) potential.
For each pressure the unit of VDOS is such that every curve has the same
area under it (i.e. size of the 'sample' is fixed). Average frequency for
the highest pressure is approximately twelve times the value for $P = 0$.
This factor of increase in raw frequency is far higher than for any previous
study (experimental or numerical) we are aware of. The raw VDOS plots of
fig. 4(a) look quite disparate. However, if we now rescale the
frequency variable in each curve by $<\omega>$ for that curve and again plot
the NDOSNF curves (area under each curve being unity) we get fig.4(b). It is
seen that the VDOS function with {\it normalized} frequency as the argument
changes rather little even with huge variations in the external pressure --
even though the absolute scale of vibrational frequency does change by a large
factor. In fact it seems that the NDOSNF saturates asymptotically as pressure 
keeps increasing. This is
suggested by fig.4(c) which reproduces the data in fig.4(b) but only for the
three highest pressures. As a justification of the inference of saturation it
may be noted that the absolute scale of frequency changes by a factor of two
between the highest and the lowest pressure cases in fig.4(c). Figures 5 and
6 are the analogues of figure 4 for the HCP crystalline state and the
amorphous state, respectively. It can be seen that the inference drawn from
fig.4 for the FCC state is applicable also to the HCP and amorphous states i.e.
there is saturation in the {\it shape} of the vibrational spectrum as pressure
keeps growing. 
%*******************************************************************************
\begin{figure}[H]
  \centering
  \subfloat[]{\label{fig:(a)}\includegraphics[width=0.32\textwidth]{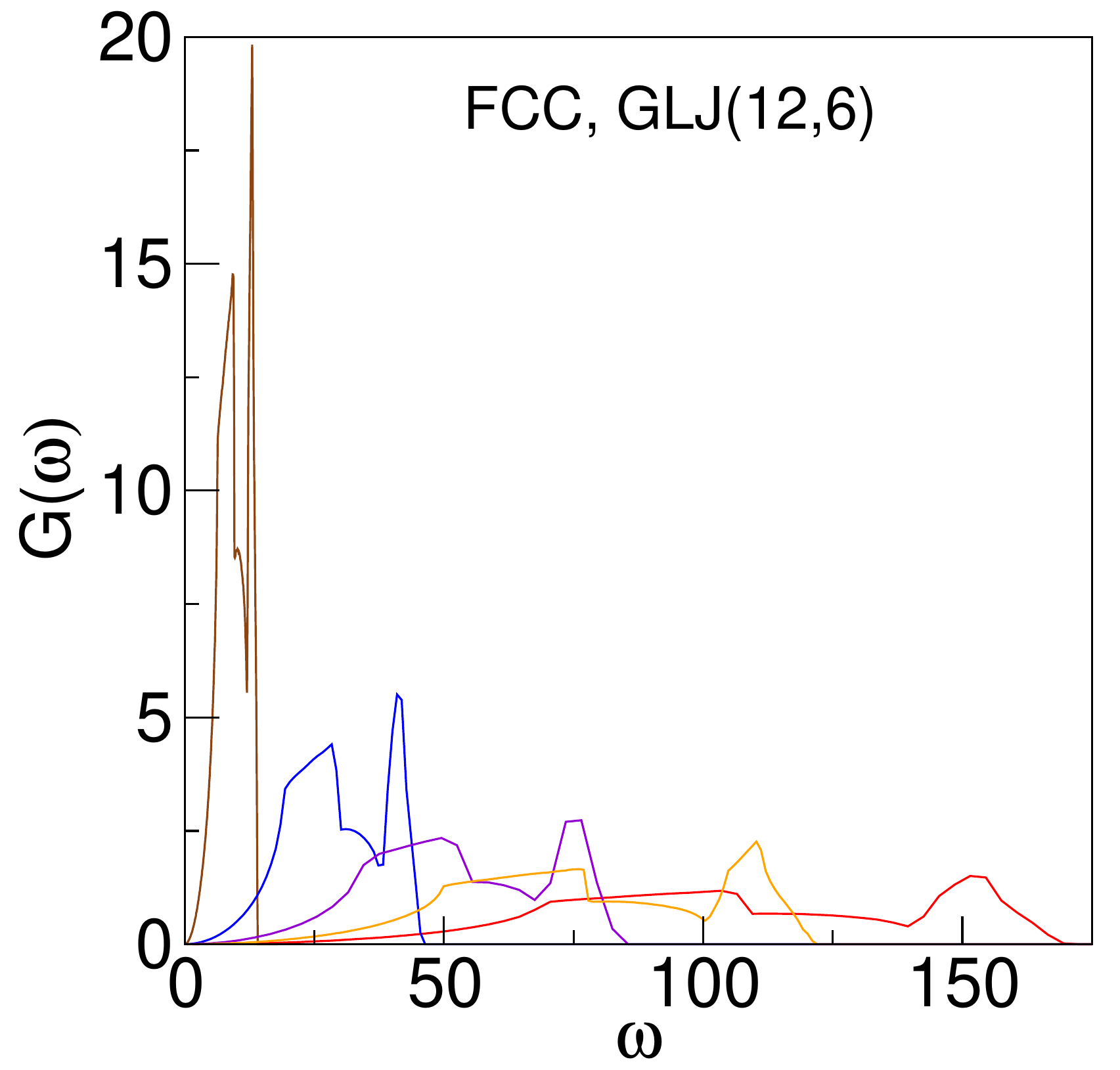}}
  \subfloat[]{\label{fig:(b)}\includegraphics[width=0.32\textwidth]{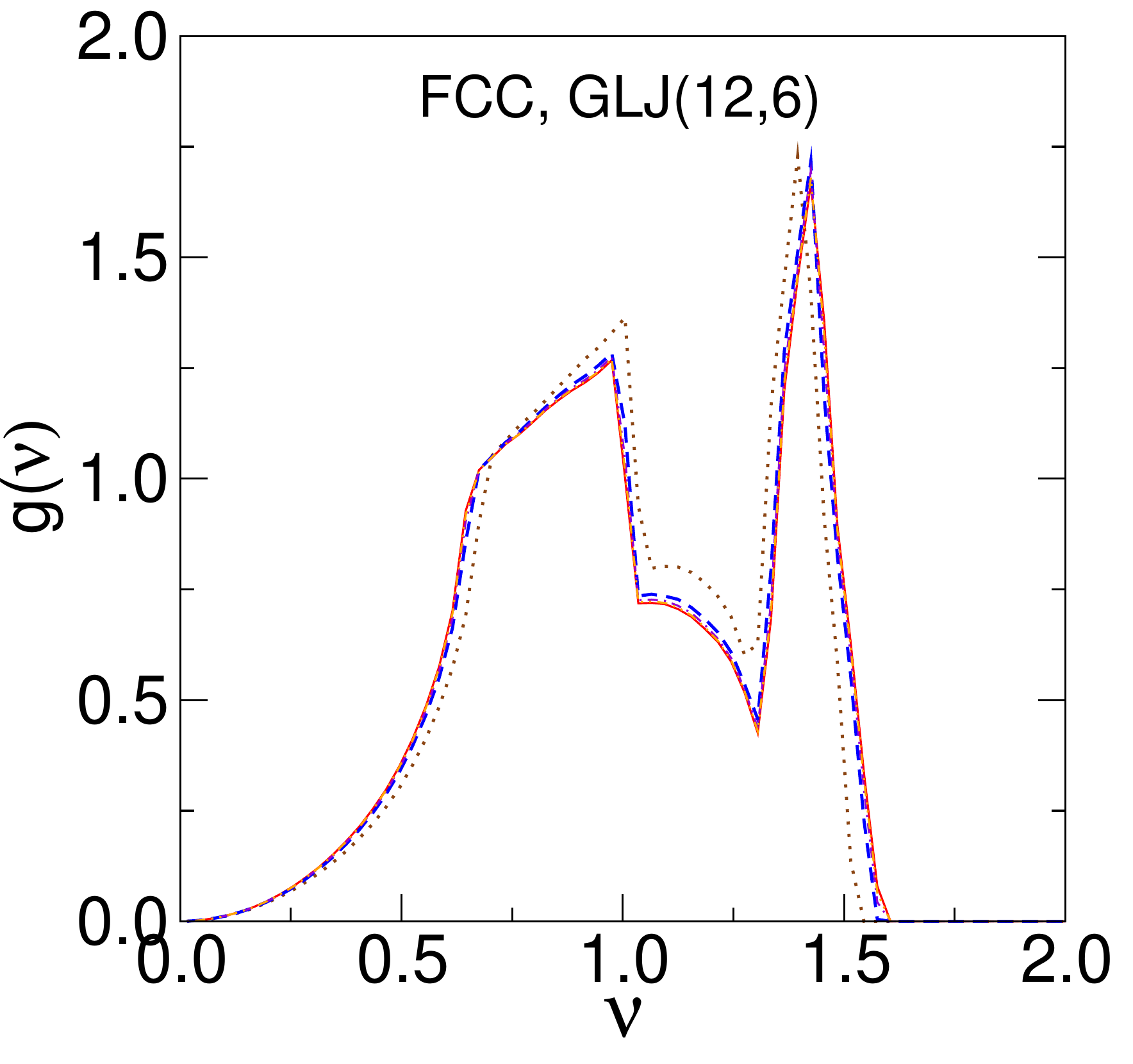}}
  \subfloat[]{\label{fig:(c)}\includegraphics[width=0.32\textwidth]{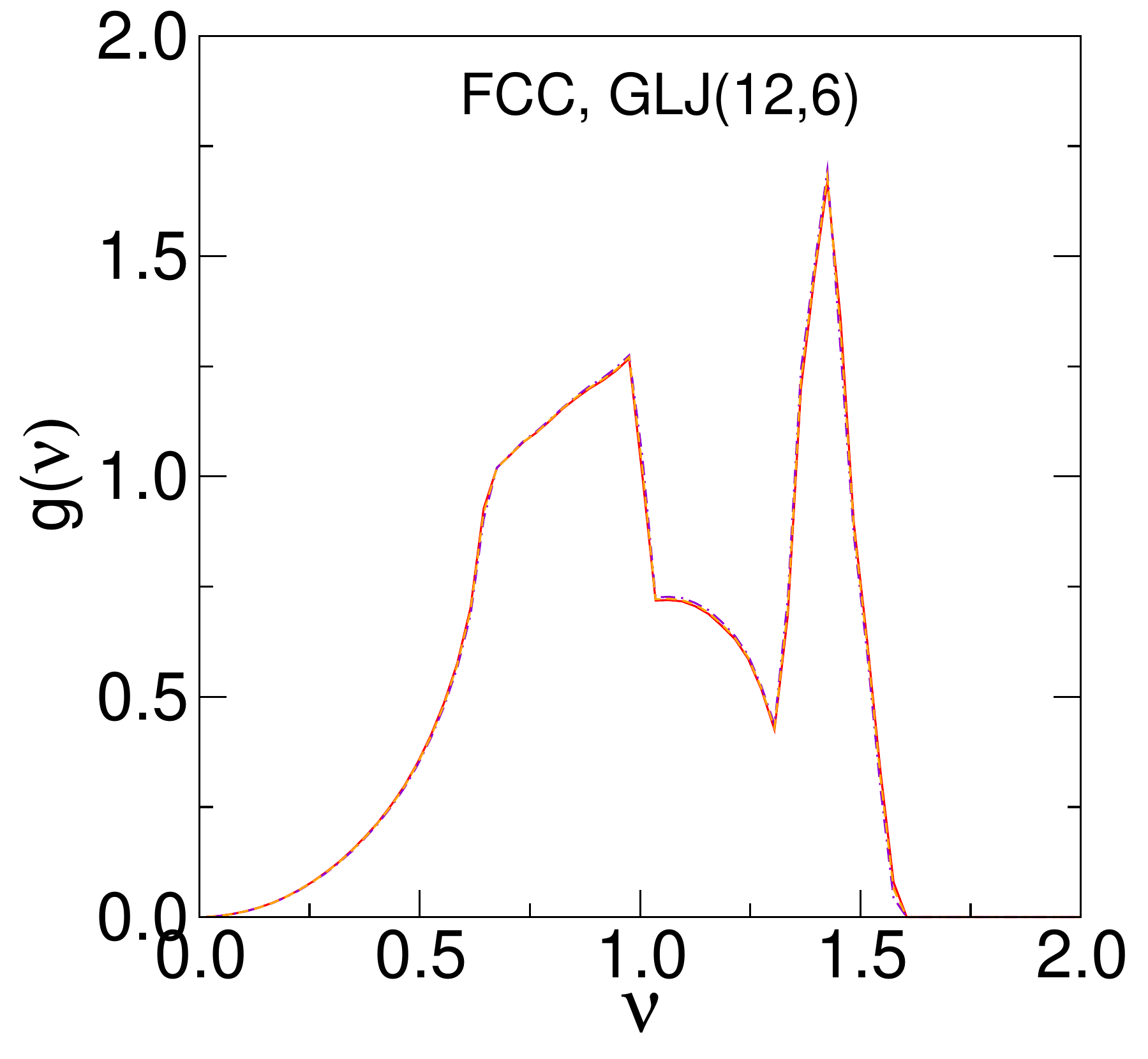}}
  \caption{\small{(a) Density of states (DOS) $G(\omega)$ for the FCC
state of GLJ(12,6) is plotted against frequency ($\omega$) for (from left 
to right) pressure $P$ = 0, 44, 188, 444 and 932. Area under every curve is 
the same. (b) For the FCC state normalized DOS $g(\nu)$ is
plotted against normalized frequency ($\nu$) for the same pressures as in
(a). Area under every curve is unity. (c) The three highest pressure cases
from (b) are reproduced to demonstrate apparent asymptotic convergence in
the normalized DOS function as pressure keeps increasing.}}
  \label{fig:fig4}
\end{figure}
\begin{figure}[H]
  \centering
  \subfloat[]{\label{fig:(a)}\includegraphics[width=0.32\textwidth]{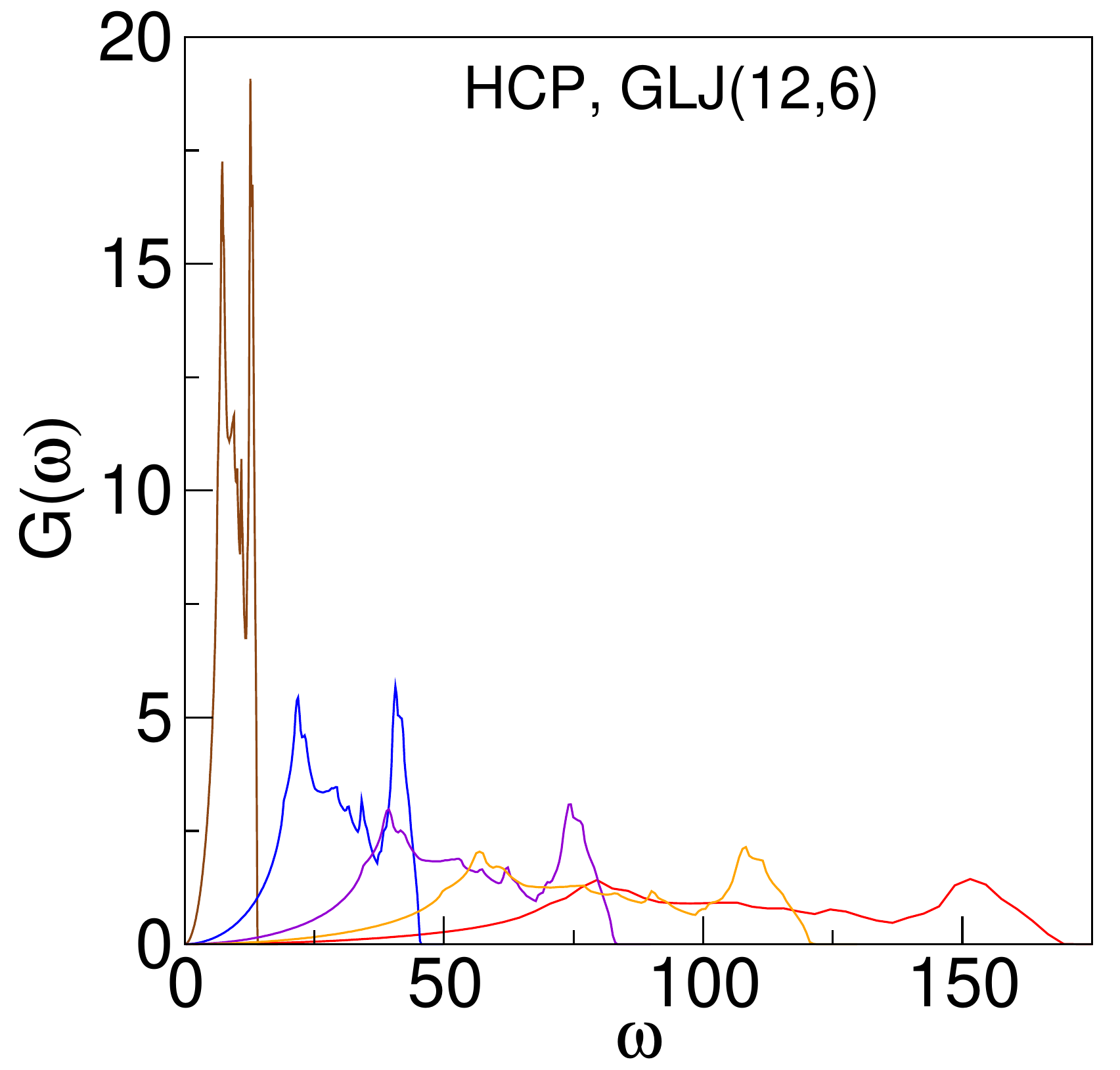}}
  \subfloat[]{\label{fig:(b)}\includegraphics[width=0.32\textwidth]{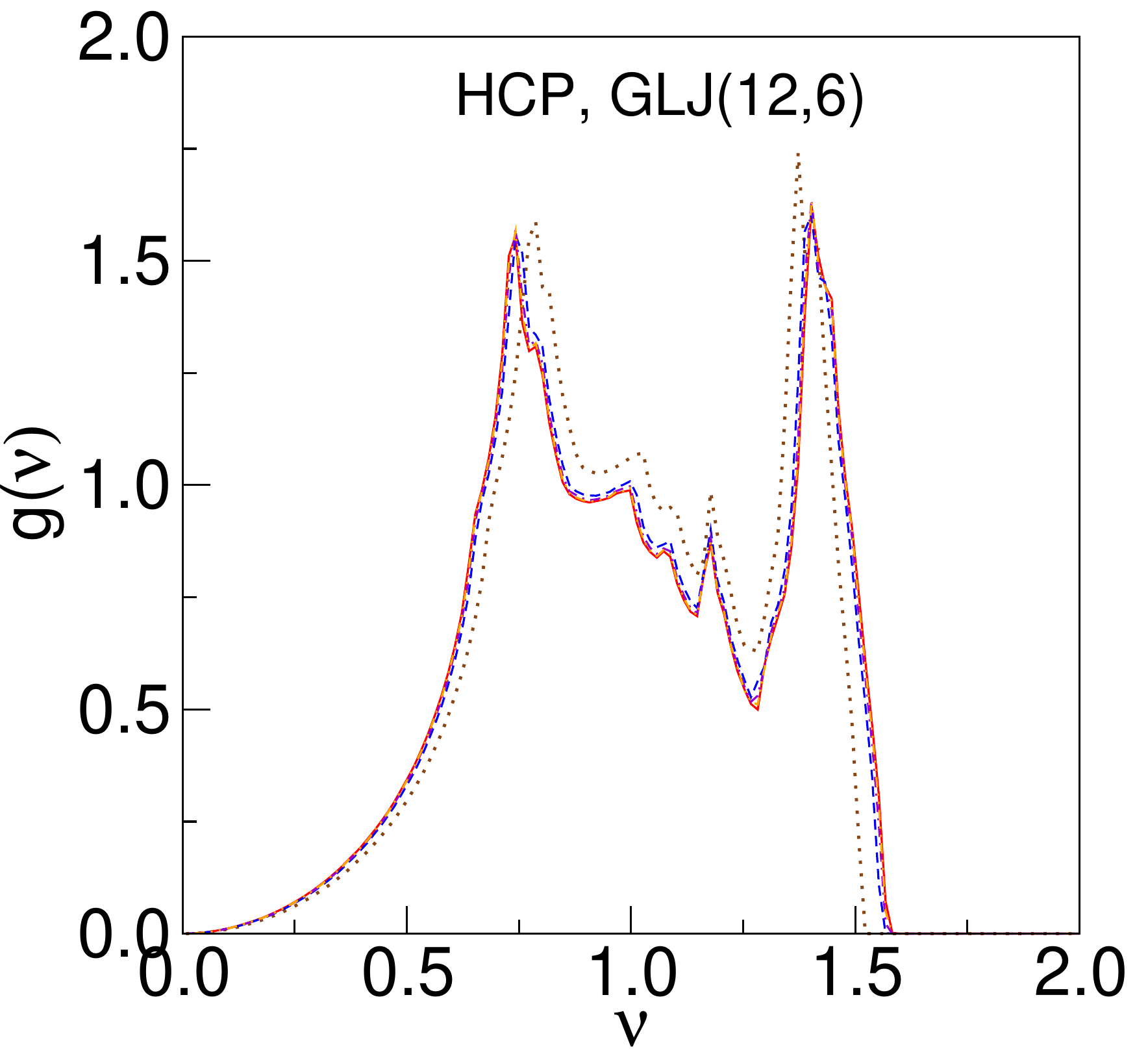}}
  \subfloat[]{\label{fig:(c)}\includegraphics[width=0.32\textwidth]{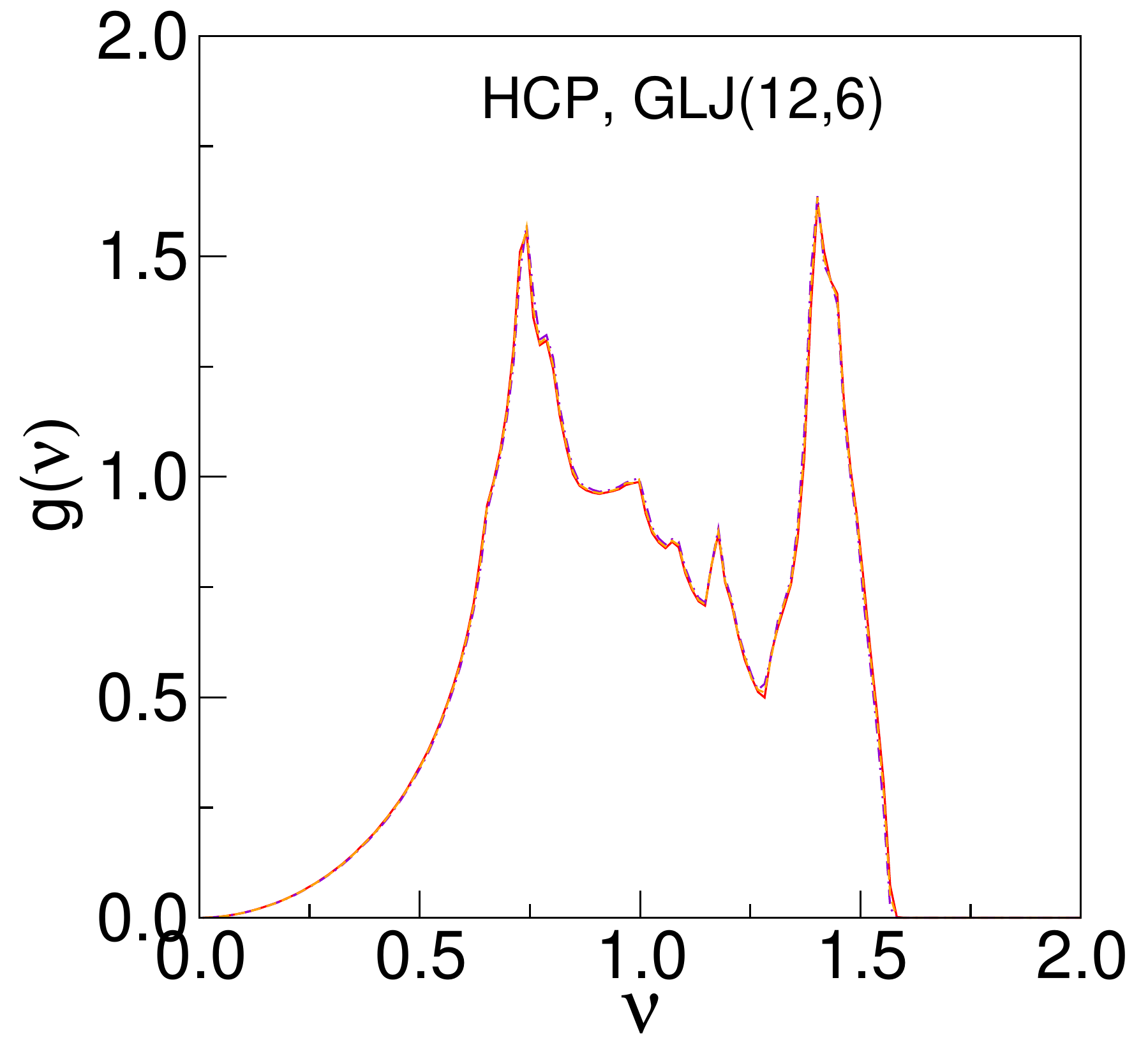}}
  \caption{\small{(a) Density of states (DOS) $G(\omega)$ for the HCP
state of GLJ(12,6) is plotted against frequency ($\omega$) for (from left 
to right) pressure $P$ = 0, 44, 188, 444 and 932. Area under every curve is 
the same. (b) For the HCP state normalized DOS $g(\nu)$ is
plotted against normalized frequency ($\nu$) for the same pressures as in
(a). Area under every curve is unity. (c) The three highest pressure cases
from (b) are reproduced to demonstrate apparent asymptotic convergence in
the normalized DOS function as pressure keeps increasing.}
}
  \label{fig:fig5}
\end{figure}
\begin{figure}[H]
  \centering
  \subfloat[]{\label{fig:(a)}\includegraphics[width=0.32\textwidth]{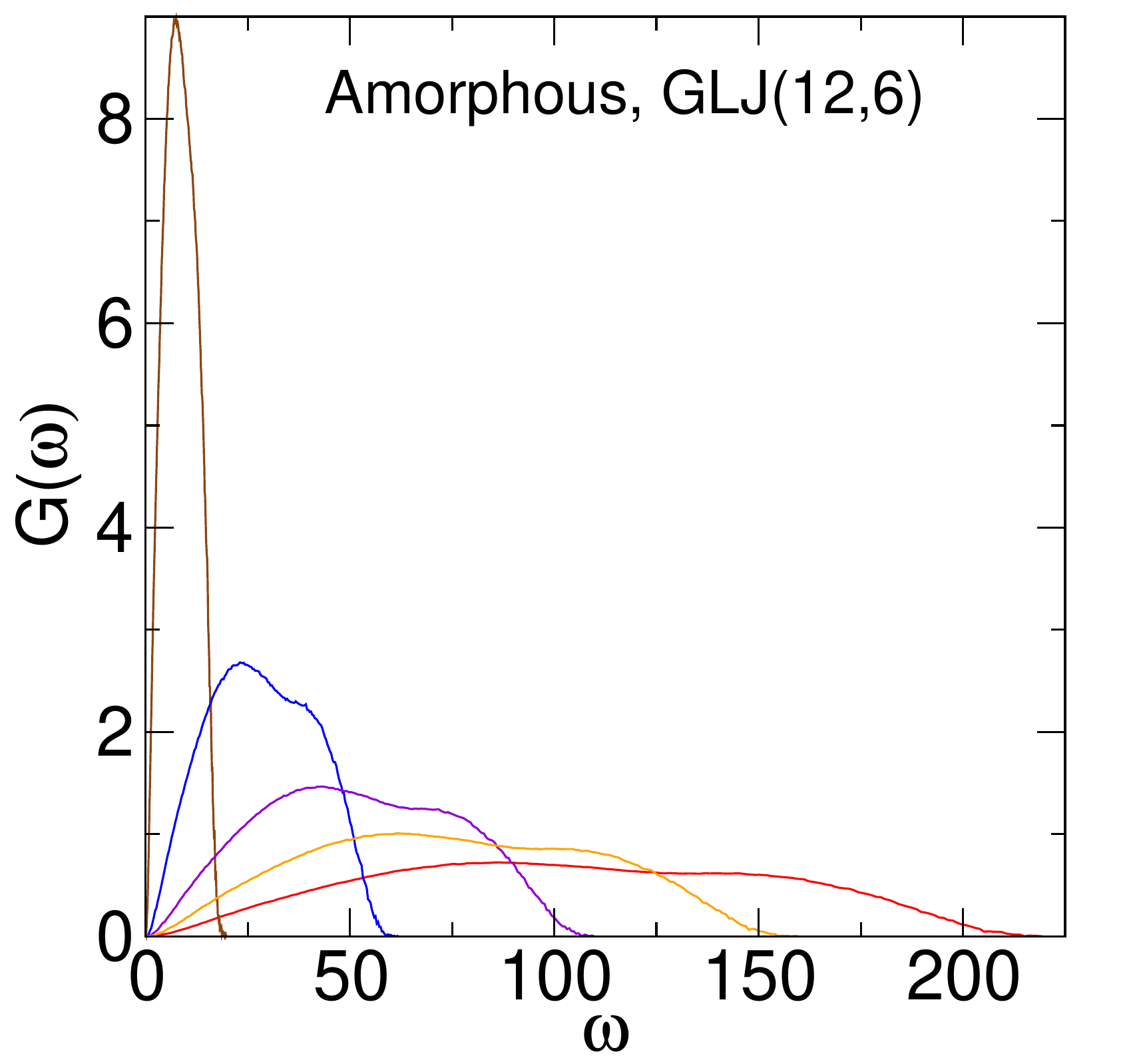}}
  \subfloat[]{\label{fig:(b)}\includegraphics[width=0.32\textwidth]{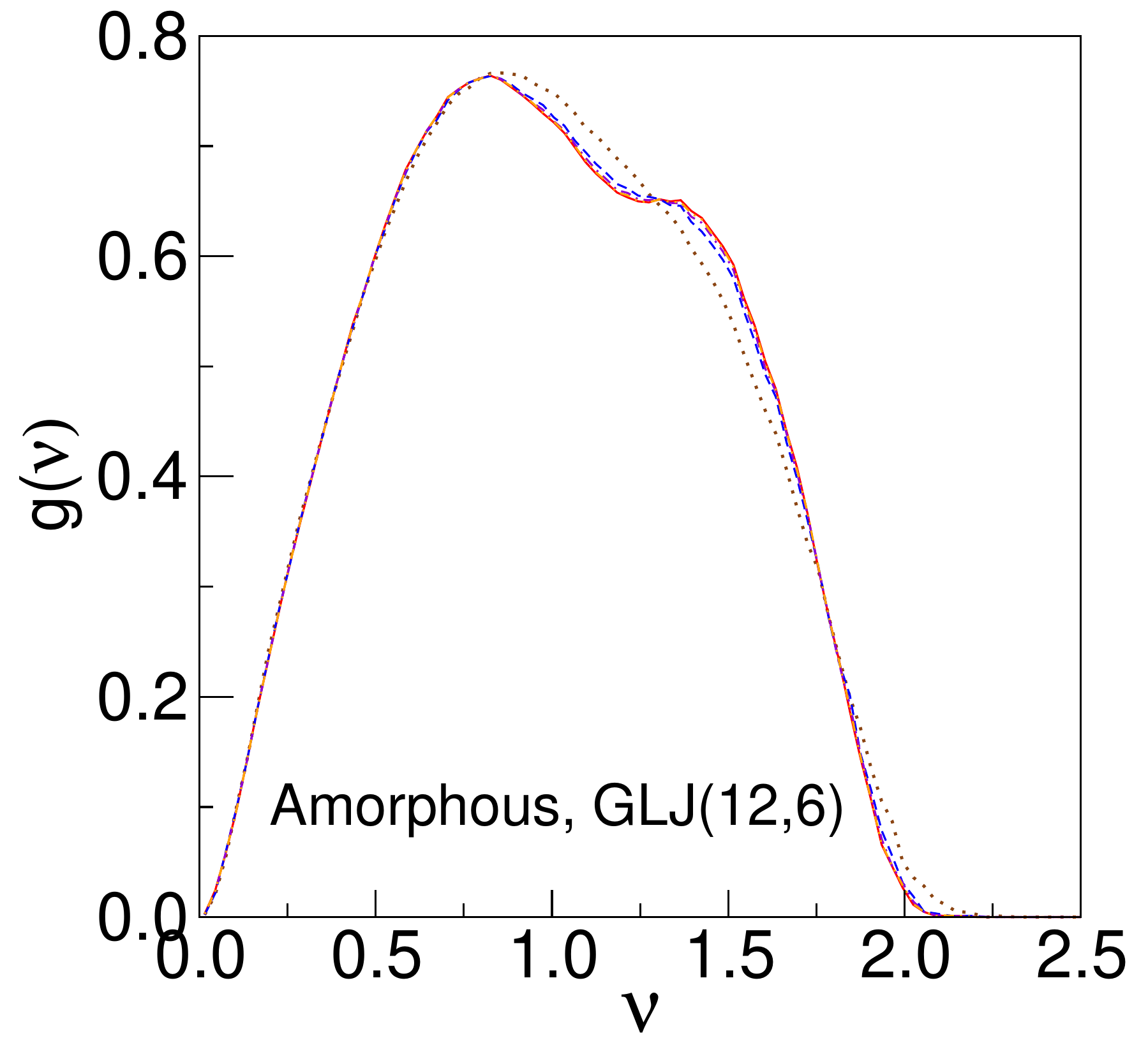}}
  \subfloat[]{\label{fig:(c)}\includegraphics[width=0.32\textwidth]{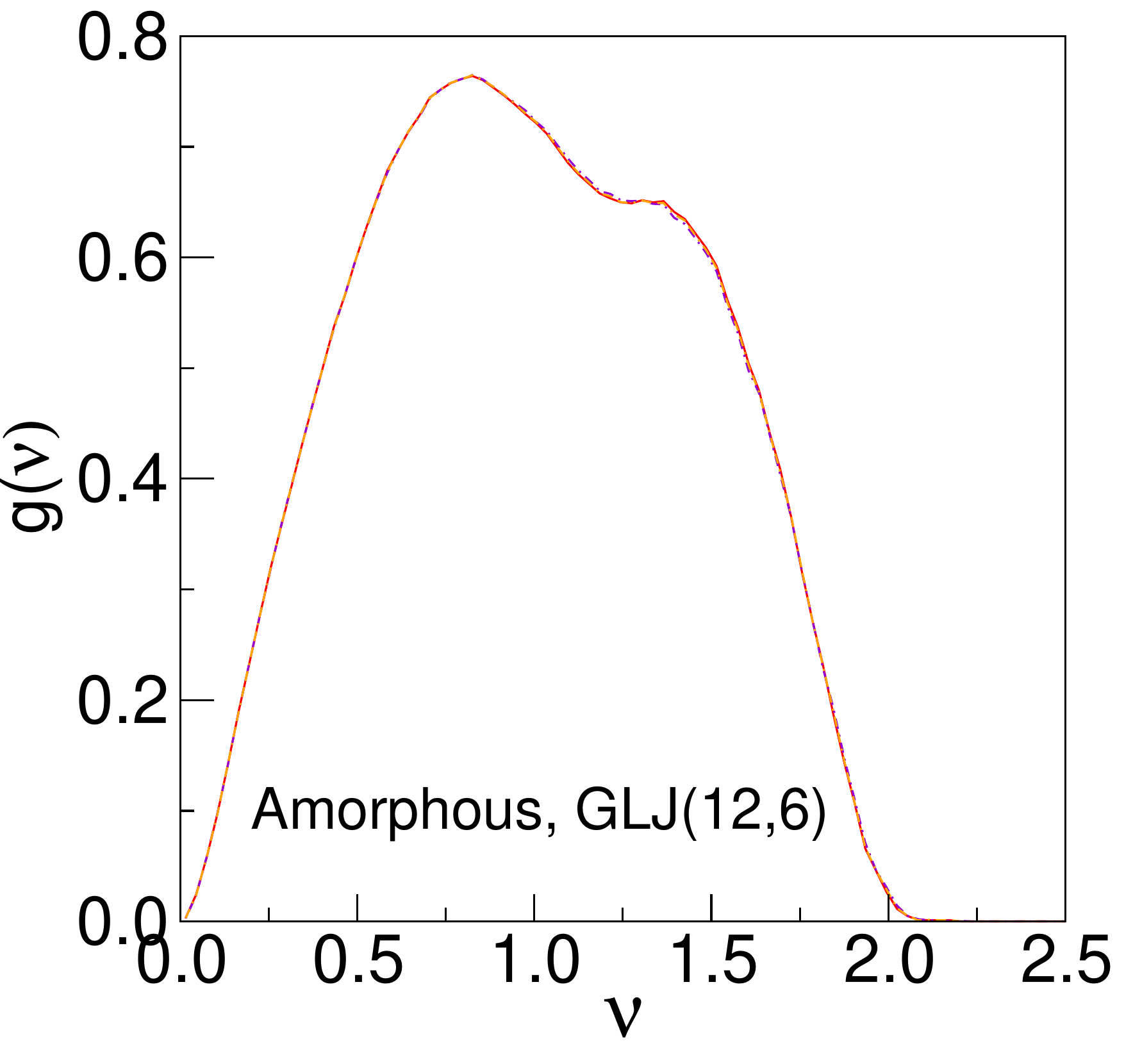}}
  \caption{\small{(a) Density of states (DOS) $G(\omega)$ for the amorphous
state of GLJ(12,6) is plotted against frequency ($\omega$) for (from left 
to right) pressure $P$ = 0, 44, 188, 444 and 932. Area under every curve is 
the same. (b) For the amorphous state normalized DOS $g(\nu)$ is
plotted against normalized frequency ($\nu$) for the same pressures as in
(a). Area under every curve is unity. (c) The three highest pressure cases
from (b) are reproduced to demonstrate apparent asymptotic convergence in
the normalized DOS function as pressure keeps increasing.}}
  \label{fig:fig6}
\end{figure}

Finally, figures 7, 8 and 9 show data analogous to figures
4,5 and 6 but for the Morse potential which belongs to type B. Inspection of
figures 7(c), 8(c) and 9(c) make it immediately clear that the inference of
asymptotic convergence of the NDOSNF is now less certain than in the case of the
GLJ(12,6) potential -- even though the evidence may be considered to be quite
strong by less rigorous standards.
%***************************************************
\begin{figure}[H]
  \centering
  \subfloat[]{\label{fig:(a)}\includegraphics[width=0.32\textwidth]{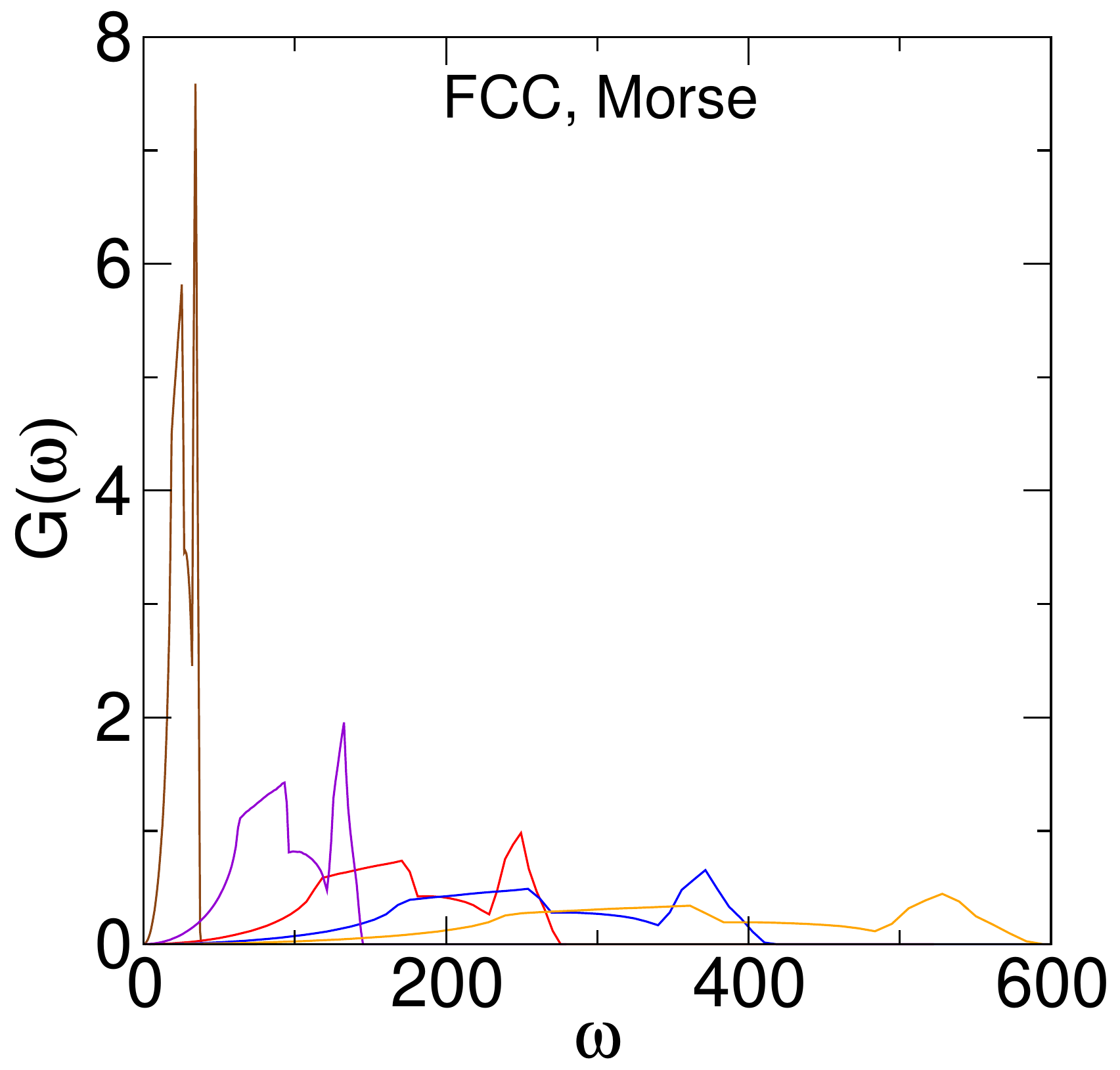}}
  \subfloat[]{\label{fig:(b)}\includegraphics[width=0.32\textwidth]{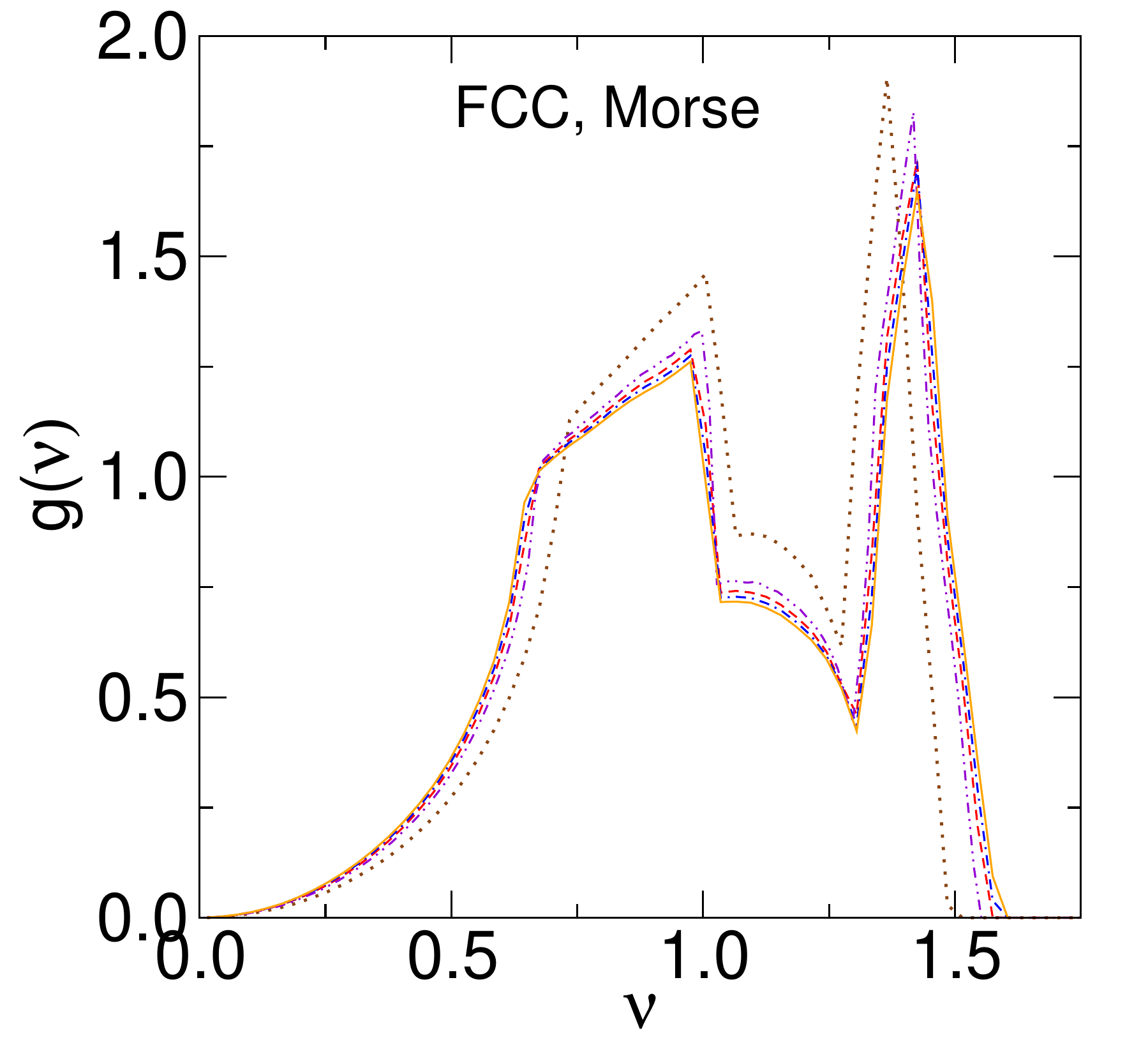}}
  \subfloat[]{\label{fig:(c)}\includegraphics[width=0.32\textwidth]{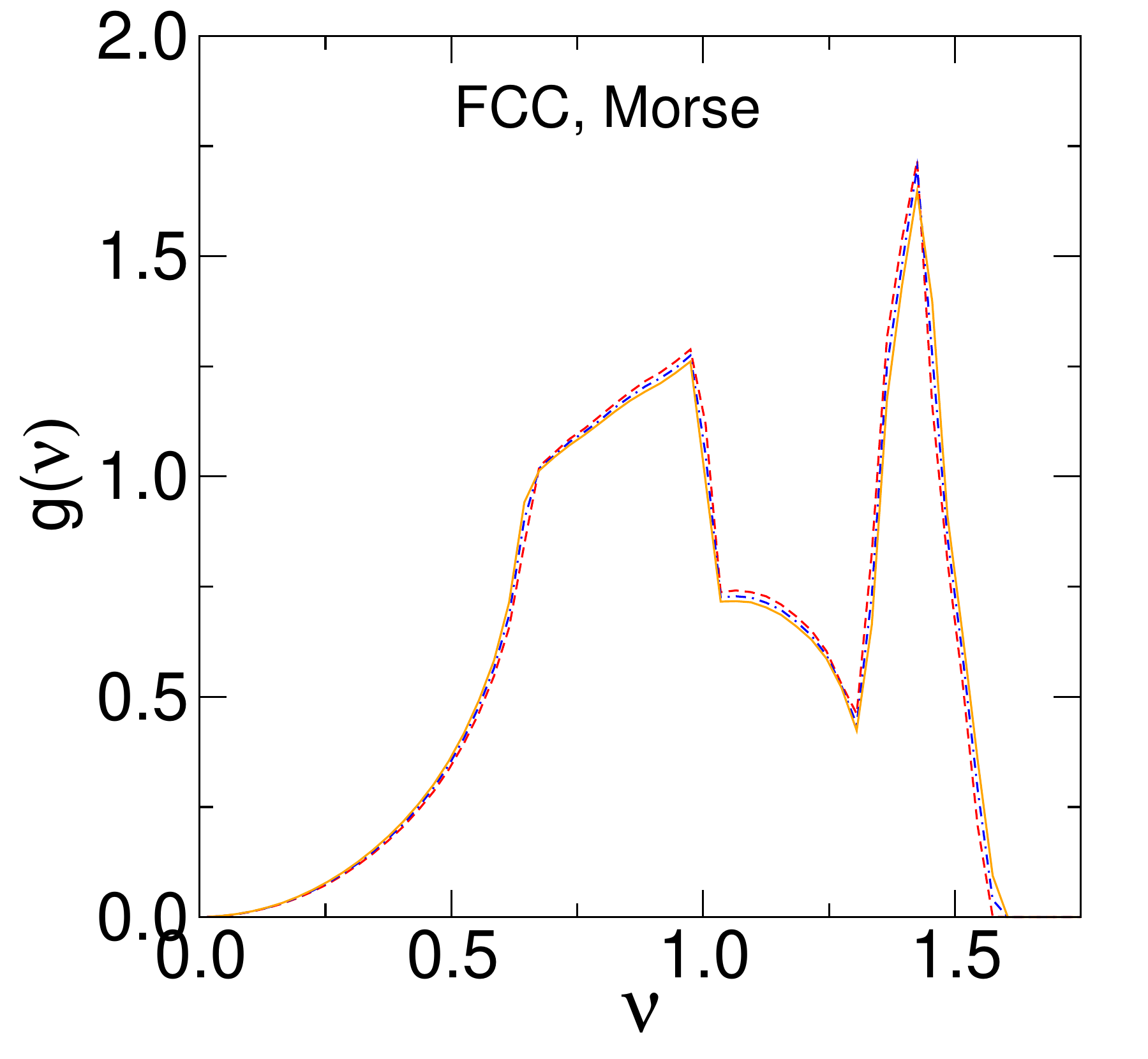}}
  \caption{\small{(a) Density of states (DOS) $G(\omega)$ for the FCC
state of the Morse potential is plotted against frequency ($\omega$) for 
(from left to right) pressure $P$ = 0, 439, 2087, 5453 and 12689. Area under 
every curve is 
the same. (b) For the FCC state normalized DOS $g(\nu)$ is
plotted against normalized frequency ($\nu$) for the same pressures as in
(a). Area under every curve is unity. (c) The three highest pressure cases
from (b) are reproduced to demonstrate the quality of evidence for asymptotic 
convergence in the normalized DOS function as pressure keeps increasing.
}}
  \label{fig:fig7}
\end{figure}
\begin{figure}[H]
  \centering
  \subfloat[]{\label{fig:(a)}\includegraphics[width=0.32\textwidth]{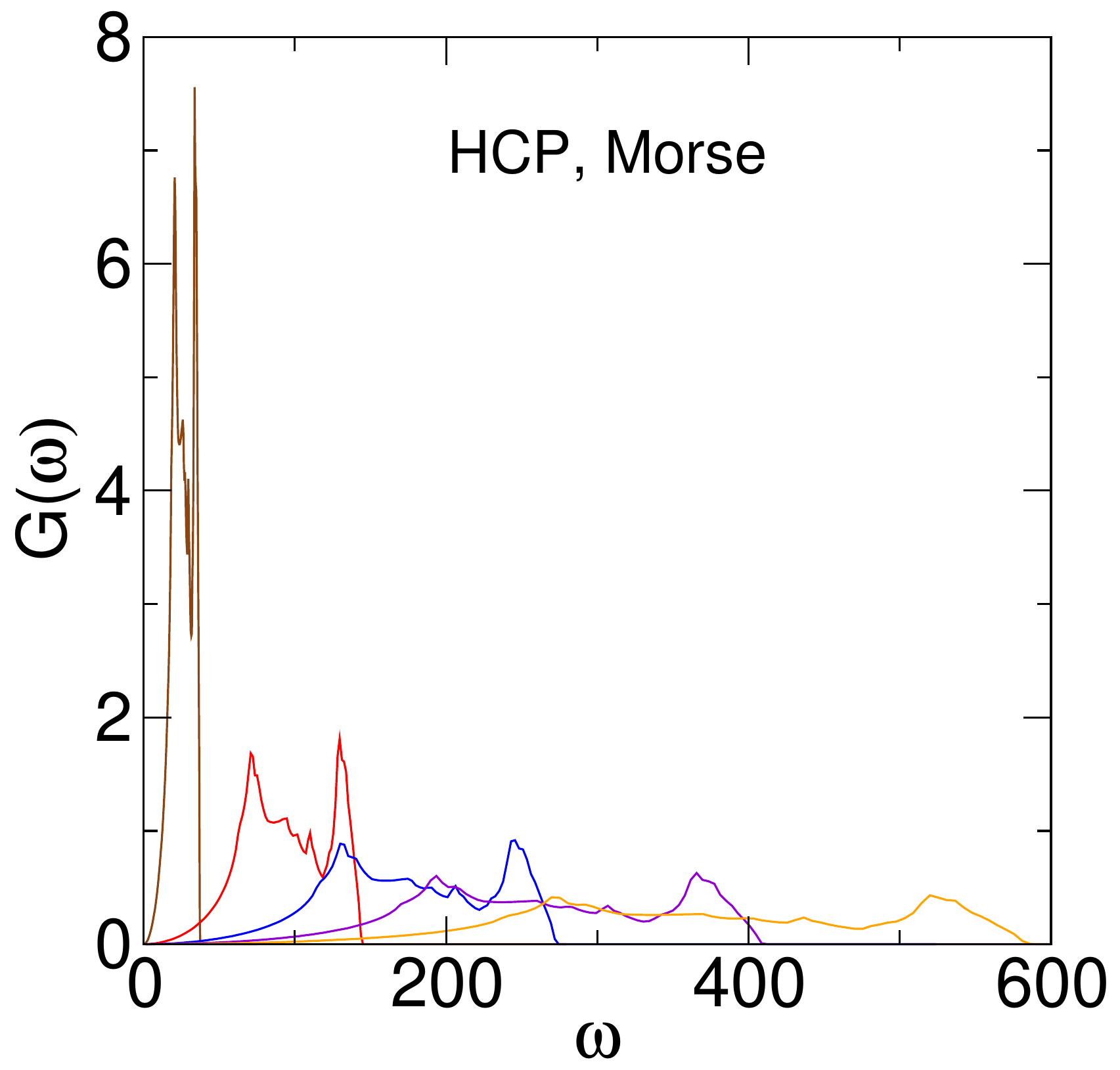}}
  \subfloat[]{\label{fig:(b)}\includegraphics[width=0.32\textwidth]{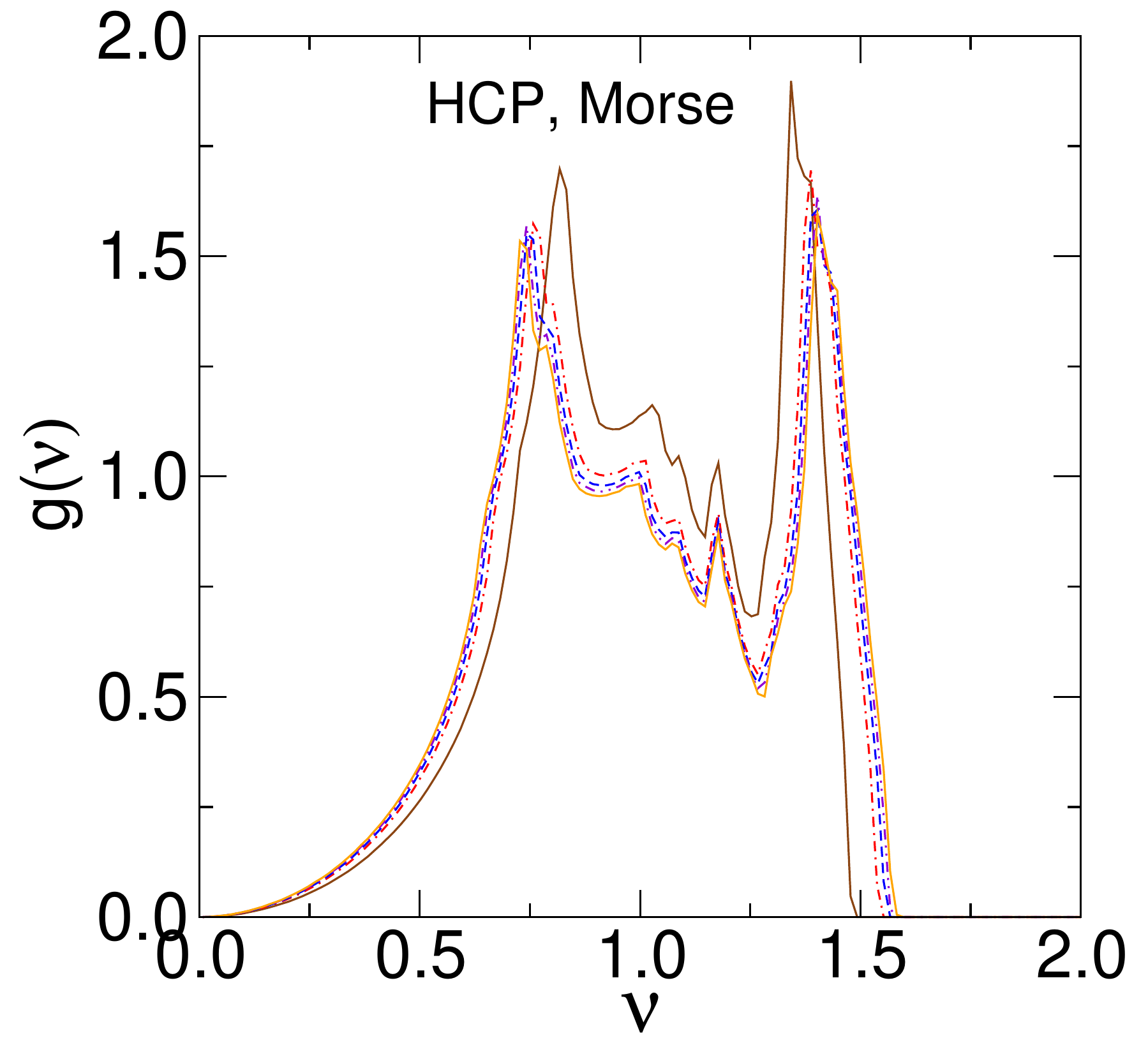}}
  \subfloat[]{\label{fig:(c)}\includegraphics[width=0.32\textwidth]{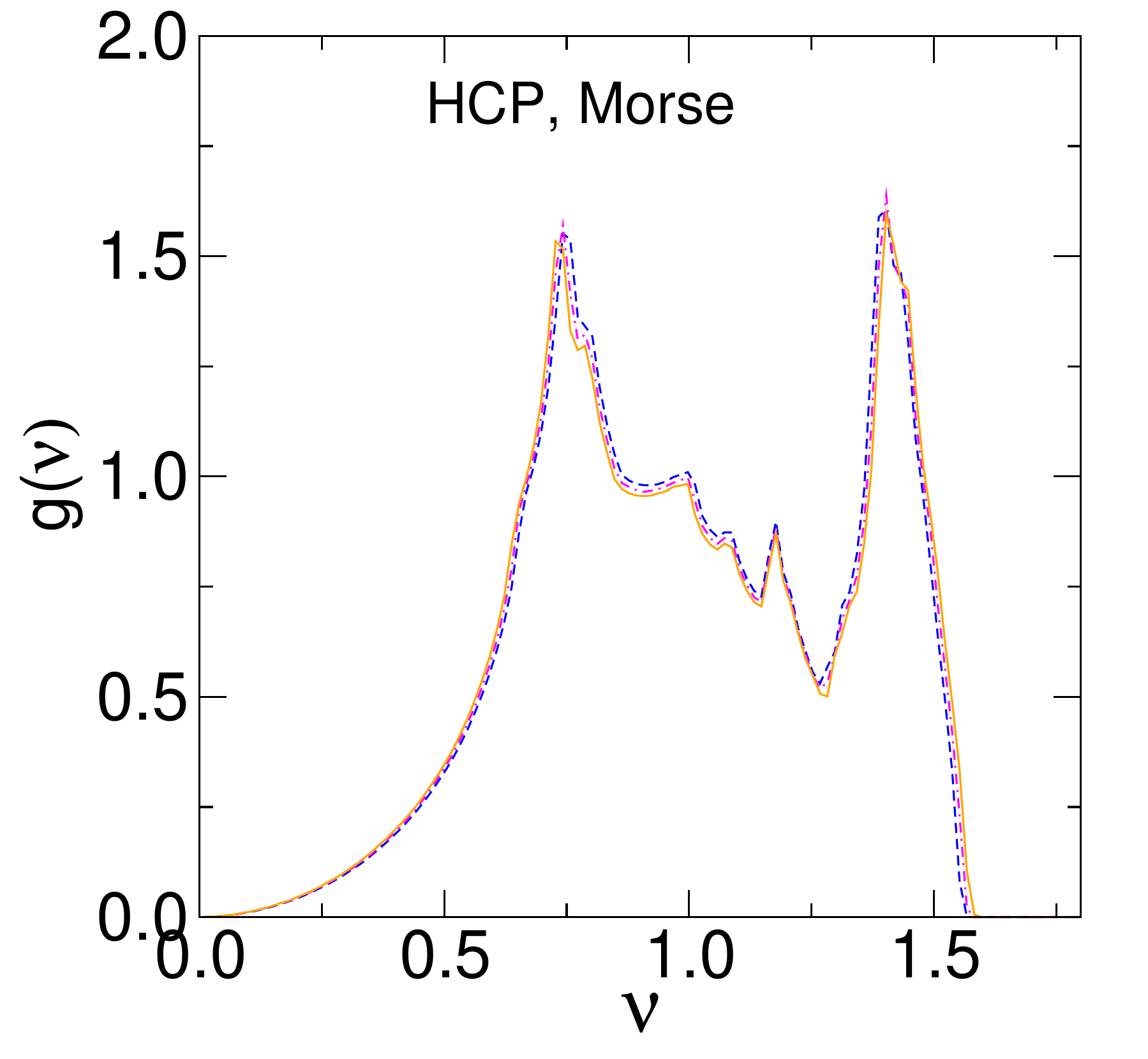}}
  \caption{\small{(a) Density of states (DOS) $G(\omega)$ for the HCP
state of the Morse potential is plotted against frequency ($\omega$) for 
(from left to right) pressure $P$ = 0, 439, 2087, 5453 and 12689. Area under 
every curve is 
the same. (b) For the HCP state normalized DOS $g(\nu)$ is
plotted against normalized frequency ($\nu$) for the same pressures as in
(a). Area under every curve is unity. (c) The three highest pressure cases
from (b) are reproduced to demonstrate the quality of evidence for asymptotic 
convergence in the normalized DOS function as pressure keeps increasing.}}
  \label{fig:fig8}
\end{figure}
\begin{figure}[H]
  \centering
  \subfloat[]{\label{fig:(a)}\includegraphics[width=0.32\textwidth]{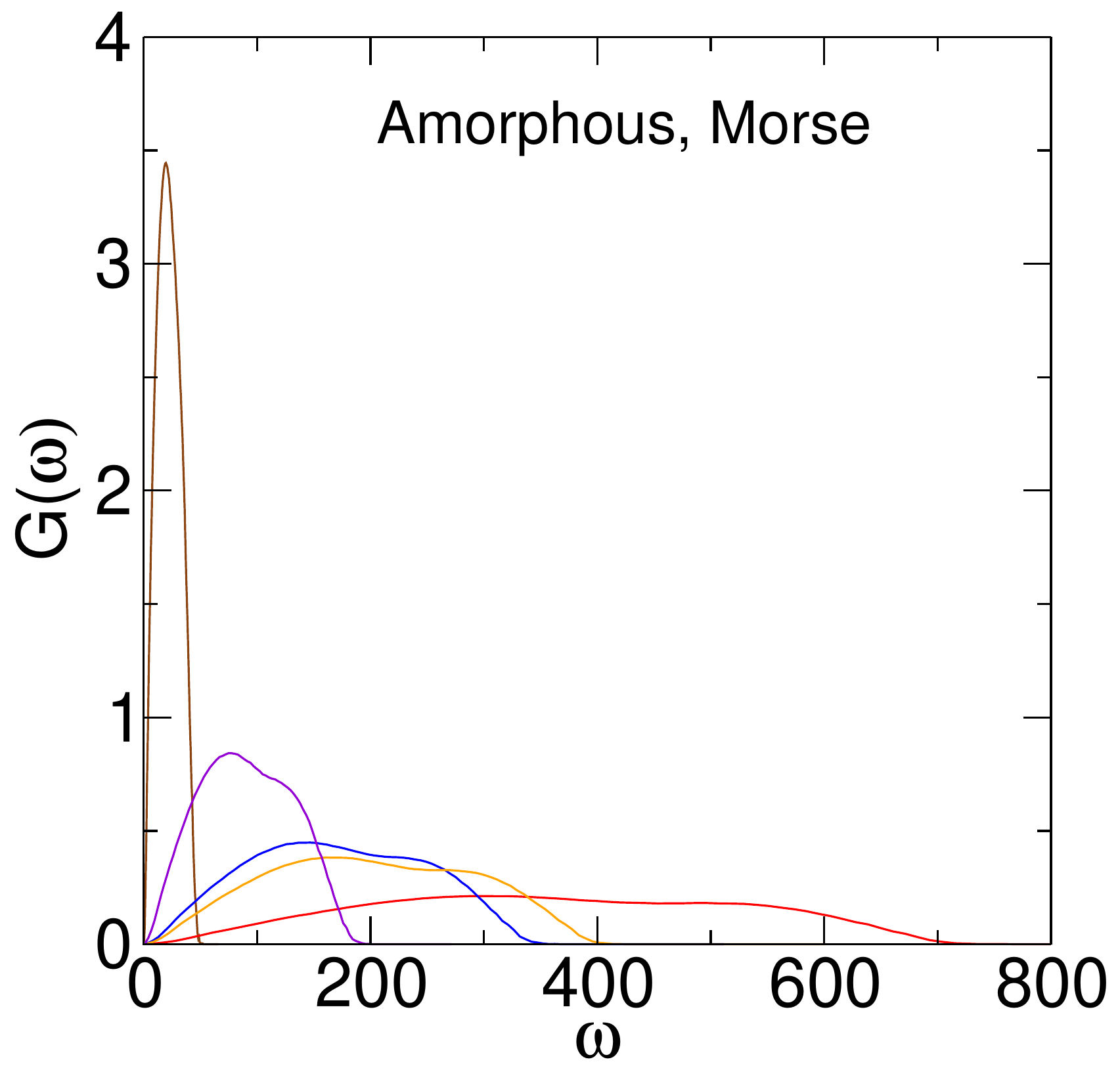}}
  \subfloat[]{\label{fig:(b)}\includegraphics[width=0.32\textwidth]{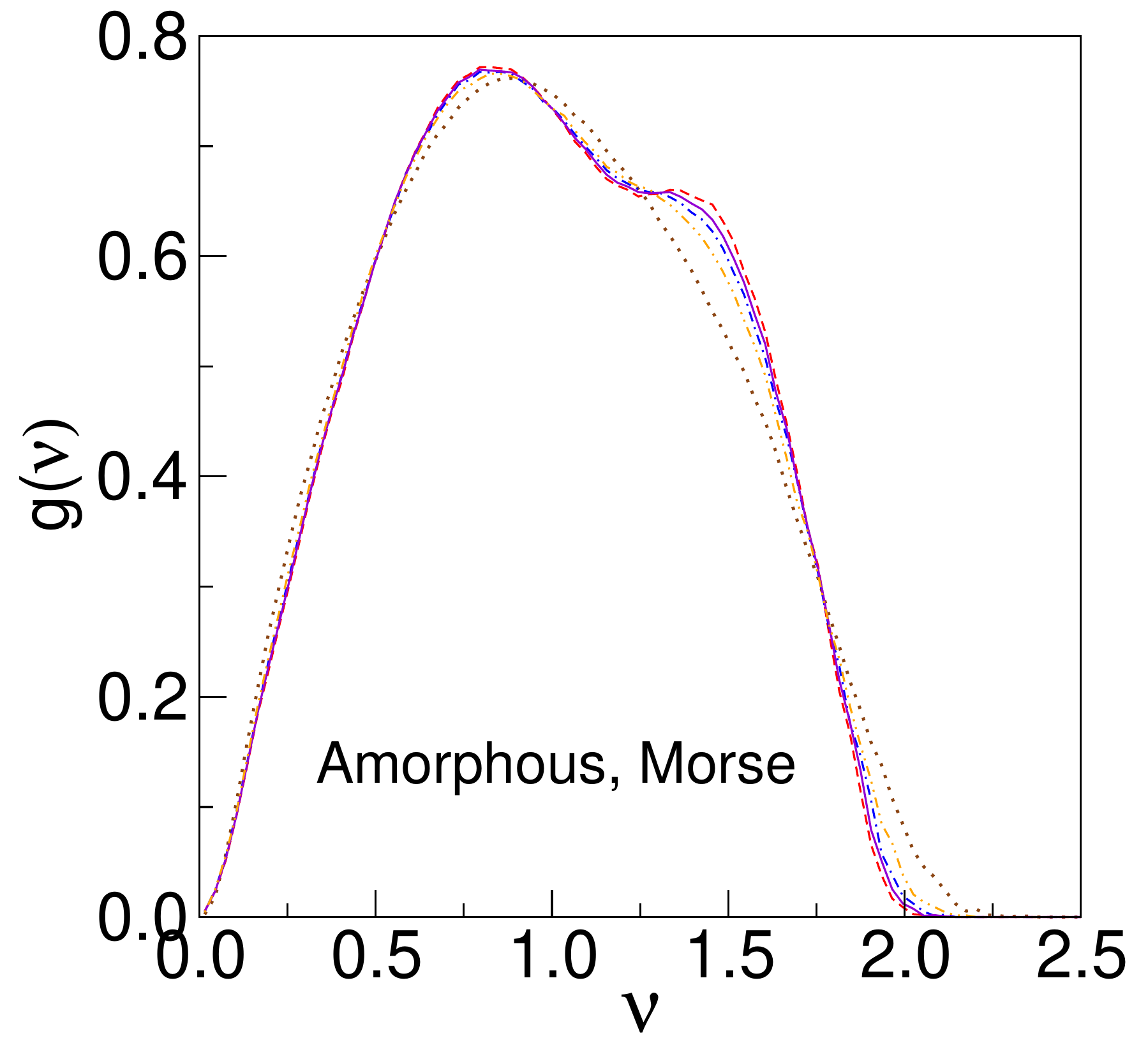}}
  \subfloat[]{\label{fig:(c)}\includegraphics[width=0.32\textwidth]{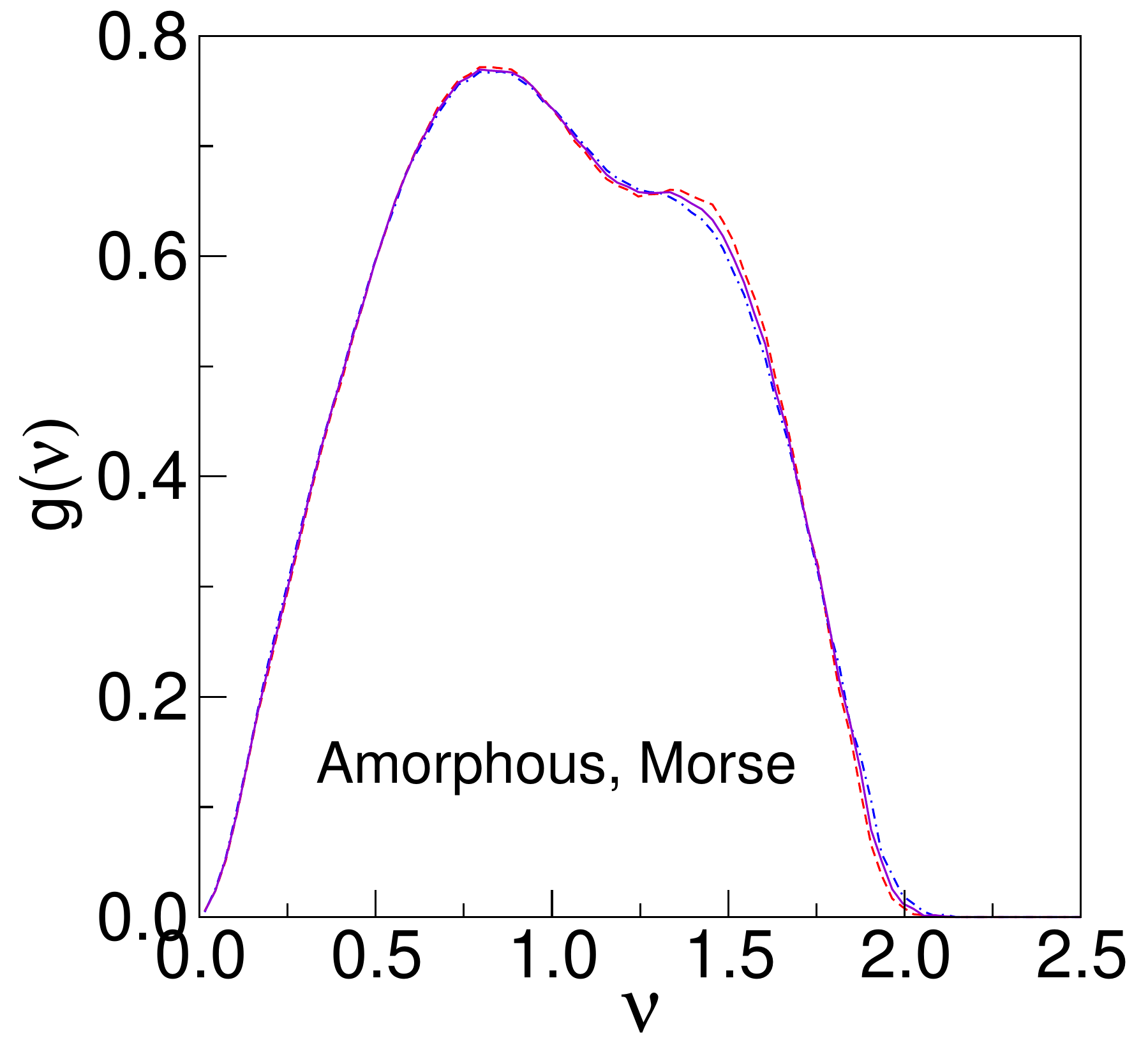}}
  \caption{\small{(a) Density of states (DOS) $G(\omega)$ for the amorphous
state of the Morse potential is plotted against frequency ($\omega$) for 
(from left 
to right) pressure $P$ = 0, 439, 2087,5453 and 12689. Area under every curve is 
the same. (b) For the amorphous state normalized DOS $g(\nu)$ is
plotted against normalized frequency ($\nu$) for the same pressures as in
(a). Area under every curve is unity. (c) The three highest pressure cases
from (b) are reproduced to demonstrate the quality of evidence for asymptotic 
convergence in the normalized DOS function as pressure keeps increasing.}}
  \label{fig:fig9}
\end{figure}
%****************************************************

%\newpage

\vskip0.3in
\begin{center}
{\bf C. Average frequency and Debye frequency}
\end{center}
\vskip0.2in

\noindent Inspection of the low frequency limit of the data in the part (c) of figures 4
through 9 suggests that the proposal of asymptotic shape
convergence made in section III(b) holds very accurately in that region.
However,
it needs to be remembered that the number of ${\bf k}$ points used in the
first Brillouin zone and , for the amorphous state, the size
of the unit cell are not too large. These issues compromise the accuracy
somewhat in the calculation of the low frequency limit of the density of
states. However,
using the method described in section II, we can calculate the sound velocities
very accurately since we do not use the more commonly used method of
numerically calculating the ratio ${\omega}/{k}$  in the limit
of $k$ going to zero. Hence an accurate alternative way of checking
whether asymptotic shape
convergence holds all the way down to the lowest frequencies is to calculate
the Debye frequency and check
whether, for higher pressures, it is proportional to the average frequency.
If it is so it will indeed provide a supplementary confirmation of the
primary implication of asymptotic shape
invariance i.e. that there is only one scale of frequency. We have calculated
the Debye frequency using the formula given in section II. Thus for each potential
there are three average frequencies (${<\omega>}_{FCC}$,${<\omega>}_{HCP}$
and ${<\omega>}_{Amor}$) and three Debye frequencies (${{\omega}^{D}}_{FCC}$
,${{\omega}^{D}}_{HCP}$ and ${{\omega}^{D}}_{Amor}$) corresponding
to the three states of aggregation that we have studied. Based on our data
we find that all these six frequencies are proportional to each other at
higher pressures. This can be seen from the figures 10 and 11 which present the data
for the GLJ(12,6) and the Morse potential, respectively (for Sutton-Chen and
Gupta potentials qualitatively indistinguishable results are obtained). Each of
these two figures show five plots (only five independent ratios can be formed
from six numbers) -- with each plot showing the two elements of one pair plotted against each
other.
The best fit straight line passing through the origin helps us judge
the validity
of the statement of proportionality for each pair. Following the earlier
observations regarding the power law scaling of the average frequency with
respect to pressure we see that the same scaling is applicable for the Debye
frequencies also.
%\newpage
%******************************************************************
\begin{figure}[H]
  \centering
  \subfloat[]{\label{fig:(a)}\includegraphics[width=0.4\textwidth]{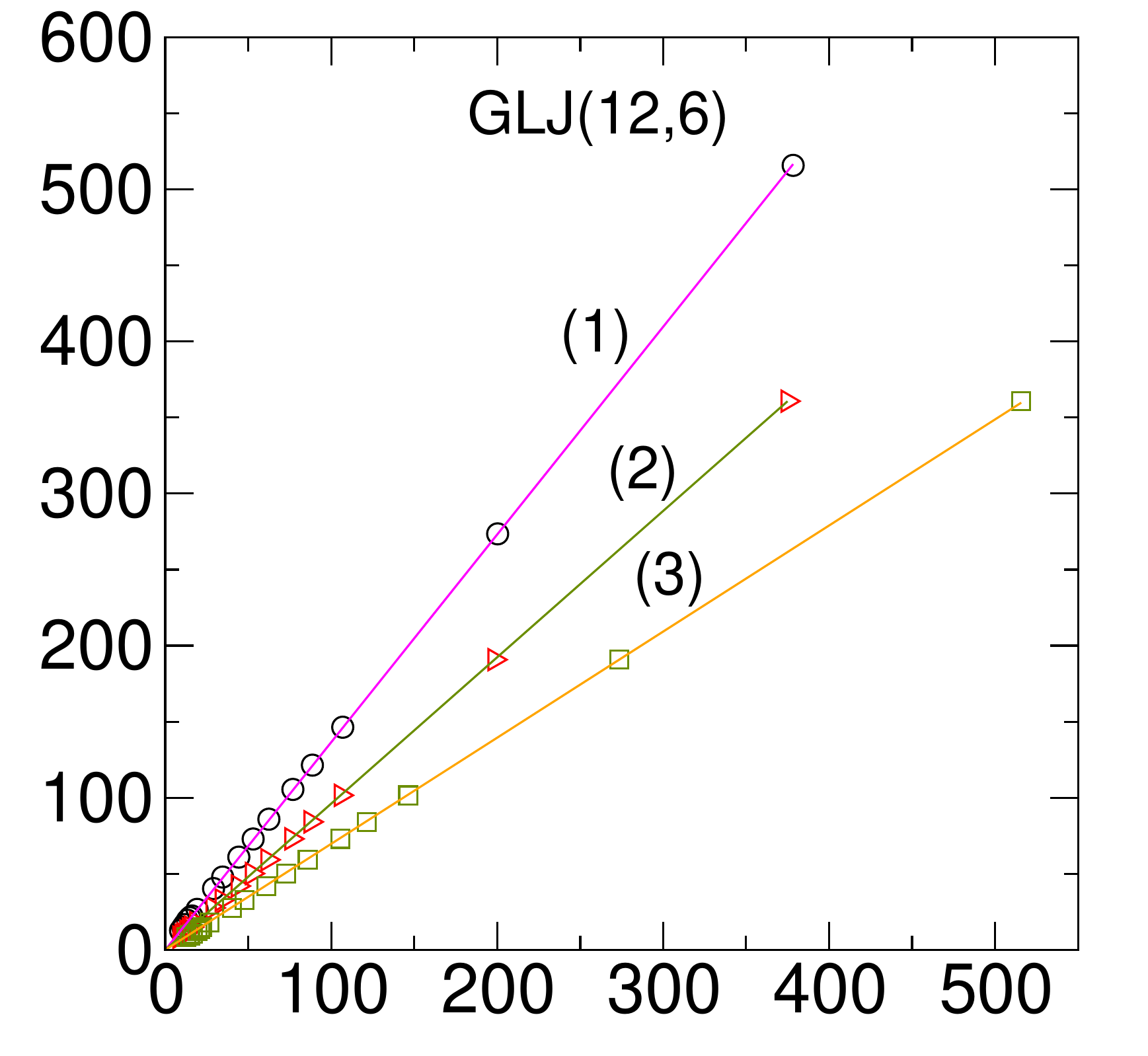}}
  \subfloat[]{\label{fig:(b)}\includegraphics[width=0.4\textwidth]{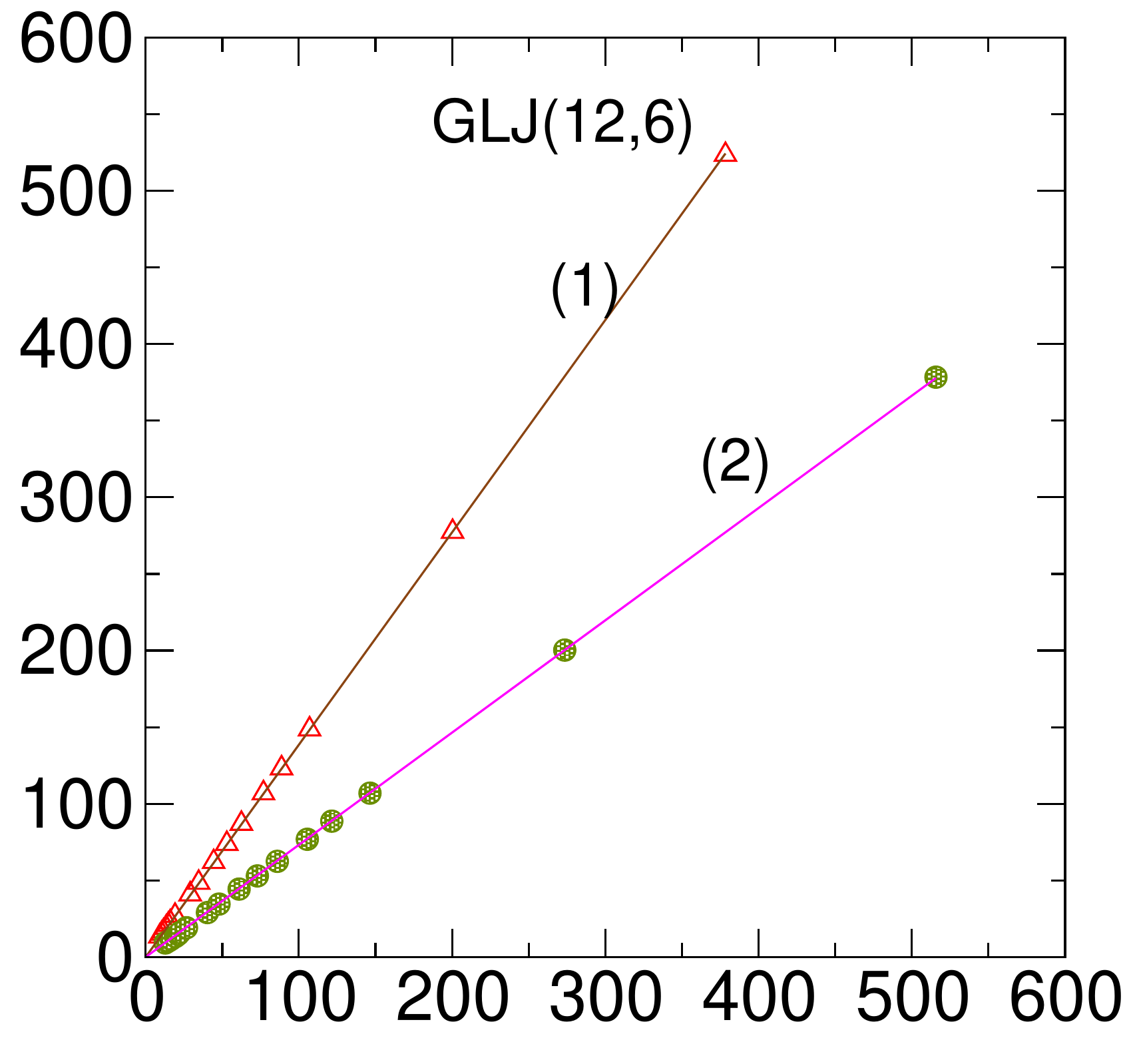}}
    \caption{\small{(a) Pairs of frequencies are plotted against each other at
various pressures -- as are the best fit straight lines passing through the
origin. The potential is GLJ(12,6). (1): ${<\omega>}_{FCC}$ (x-axis) vs. 
${\omega^{D}}_{FCC}$, (2): $<\omega>_{Amor}$
vs. ${\omega^{D}}_{Amor}$ and (3) ${\omega^{D}}_{FCC}$ vs. ${\omega^{D}}_{Amor}$. 
(b) Same as in (a) but the pairs plotted are: 
(1): $<\omega>_{HCP}$ (x-axis) vs. ${\omega^{D}}_{HCP}$ and  (2): ${\omega^{D}}_{FCC}$ 
vs. $<\omega>_{HCP}$.}}
  \label{fig:fig10}
\end{figure}
\begin{figure}[H]
  \centering
  \subfloat[]{\label{fig:(a)}\includegraphics[width=0.4\textwidth]{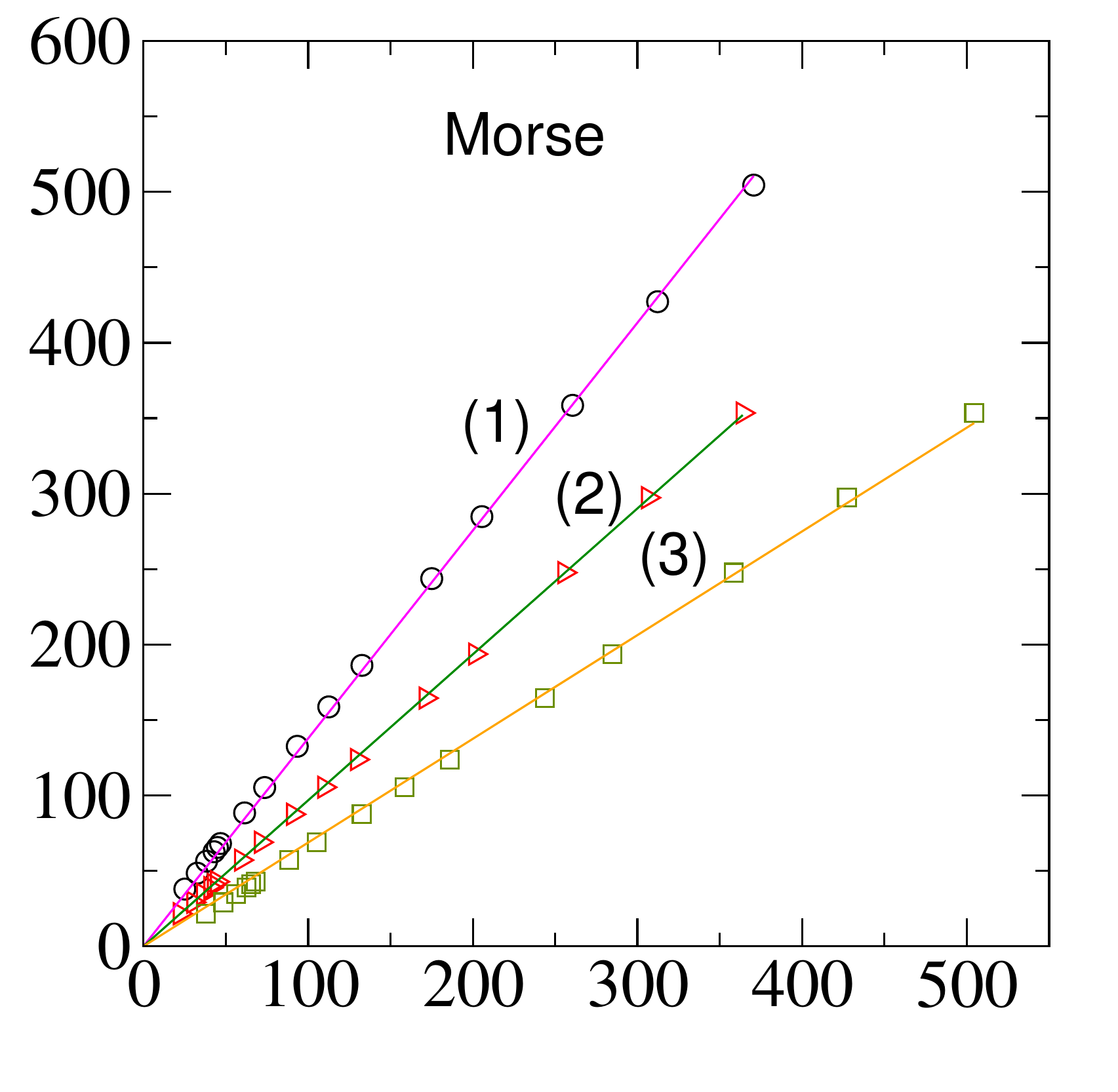}}
  \subfloat[]{\label{fig:(b)}\includegraphics[width=0.4\textwidth]{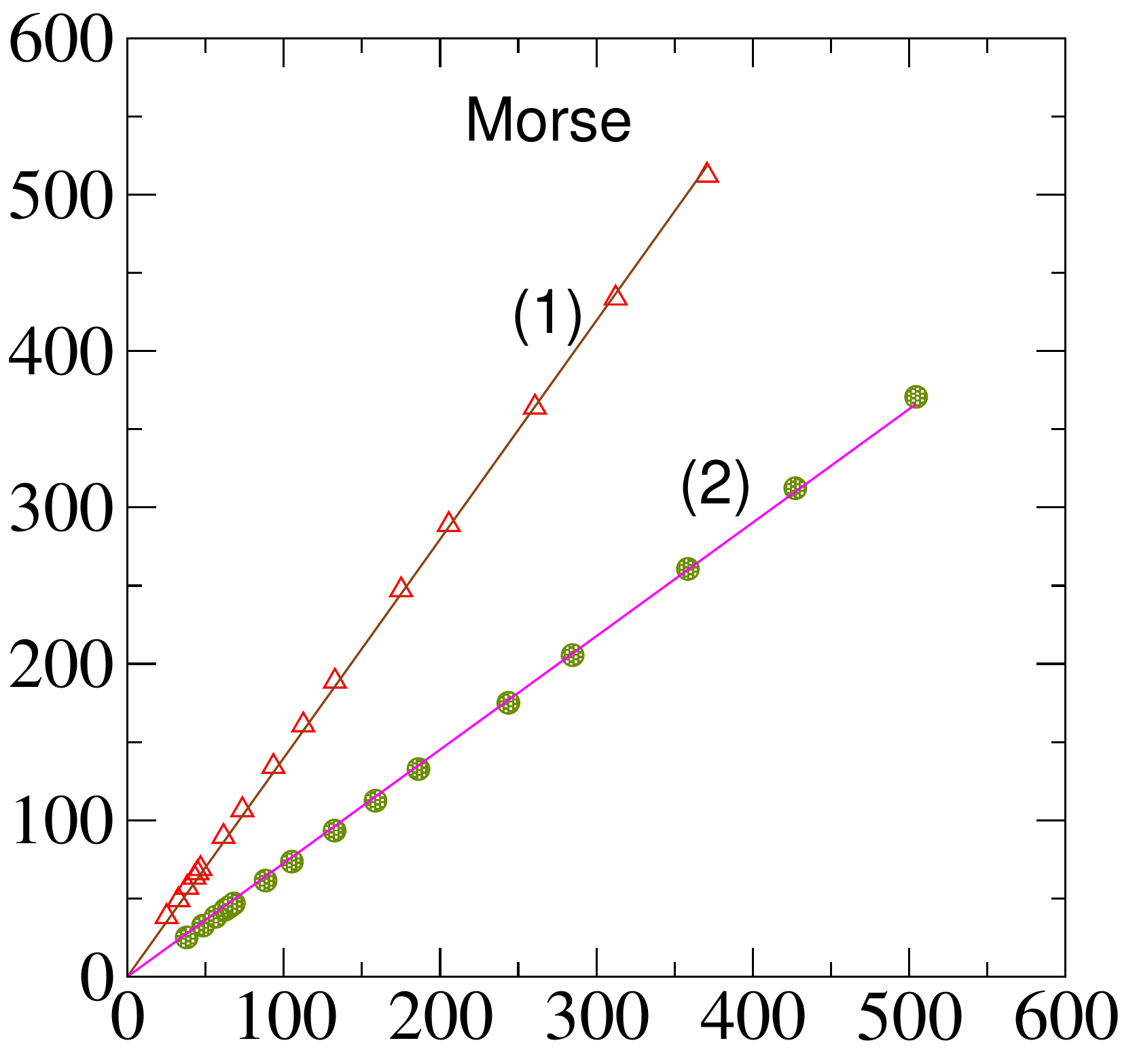}}
  \caption{\small{(a) For Morse potential pairs of frequencies are plotted against each other at
various pressures -- as are the best fit straight lines passing through the
origin. (1): ${<\omega>}_{FCC}$ (x-axis) vs. 
${\omega^{D}}_{FCC}$, (2): $<\omega>_{Amor}$
vs. ${\omega^{D}}_{Amor}$ and (3) ${\omega^{D}}_{FCC}$ vs. ${\omega^{D}}_{Amor}$. 
(b) Same as in (a) but the pairs plotted are: 
(1): $<\omega>_{HCP}$ (x-axis) vs. ${\omega^{D}}_{HCP}$ and  (2): ${\omega^{D}}_{FCC}$ 
vs. $<\omega>_{HCP}$.}}
  \label{fig:fig11}
\end{figure}

%******************************************************************

\vskip0.3in
\begin{center}
{\bf IV. Nearest neighbor model of vibrational spectrum for crystalline solids}
\end{center}
\vskip0.2in

\noindent In this section we demonstrate that a simple nearest neighbor model [see 
Chapter 22 of reference 46]
for calculating the vibrational spectra of the FCC and the (ideal) HCP solids
produces results in excellent agreement with the exact calculations at higher
pressures. We show that the model is exact for a special limit of the family 
of type A potentials. For type B
potentials it has a very wide range of validity but breaks down progressively
more and more when pressure keeps increasing beyond a limit i.e. the domain
of applicability is a wide range of {\it intermediate} pressures.

The first point to be noted is that the attractive part of the potential
energy becomes smaller and smaller as a fraction of the repulsive part when
the pressure keeps growing. This is a generic requirement that an expression
of potential energy has to satisfy if it has to qualify as a descriptor of a
stable solid. So, to describe a solid at higher pressures, we need to include
only the repulsive part. In the present work this repulsive part is always
a sum-over-pairs type expression. Let us denote this pair potential by
$\phi(r)$.

Now, for a FCC solid or an ideal HCP solid, the geometry of arrangement of the
particles does not depend on the applied pressure. We take the nearest neighbor
distance to be the only independent parameter that changes with pressure.
All other distances maintain a fixed ratio with respect to this distance. For
example, for FCC solid the next nearest neighbor distance is always $\sqrt{2}$
times the nearest neighbor distance. As stated earlier the repulsive part is
always taken to be a sum-over-pairs type expression. But this sum is over
{\it all} pairs to begin with. So it is obvious that  in a situation in which
the nearest neighbor contribution is far higher than the sum of all the
non-nearest neighbor contributions a quantitatively adequate description of
the vibrational spectrum can be given by including only the nearest neighbor
interactions. For type A potentials, where $\phi(r)$ is proportional to
$r^{-m}$, this is clearly true when the value of $m$ is reasonably
large (the lowest value we use is 9, for the Sutton-Chen potential). And the
approximation becomes exact in the limit of $m$ going to infinity.
For type B potentials the full range of higher pressures can be subdivided
into two ranges: intermediate and very high -- the exact ranges under these
two domains being controlled by the range of the exponential potential. To
see why the nearest neighbor description works in the intermediate range of
pressures (or the nearest neighbor distance) consider the ratio 
$T_{nn}$/$T_{nnn}$ where $T_{nn}$ and $T_{nnn}$ represent the contributions
of a nearest neighbor pair (separated by distance $d$) and a next nearest 
neighbor pair, respectively, to
the total energy. With  $\phi(r)$ being of the form $exp(-{\alpha}r)$ for the
type B potentials this ratio ($\theta$) is  $exp((p-1){\alpha}d)$ where
$p$, the ratio of the next-nearest-neighbor distance over the nearest neighbor
distance, is greater than unity by definition. For example, for FCC crystals,
$p$ is $\sqrt{2}$.  Clearly, higher the value of $\theta$, better is the nearest
neighbor approximation. However, with increasing pressure, the nearest
neighbor distance $d$ goes down and so does $\theta$ -- as can be seen from
the expression of $\theta$ above. Exactly how big is the range of pressures
for which $\theta$ is large enough is controlled by the value of $\alpha$.

Let us now assume that we are in a situation where it is permissible to ignore the 
interactions beyond the nearest neighbor. In such a situation
the vibrational spectrum can be calculated as follows:
\vskip0.3in
\begin{center}
{\bf A. FCC crystal}
\end{center}
\vskip0.3in
\noindent The frequencies of the three normal modes with wave vector $\bf k$
are proportional to the square roots of the three eigenvalues of the $3\times3$
dynamical matrix

\begin{equation} D({\bf k}) = \sum_{\bf{R}}\sin^2\left(\frac{1}{2}{\bf{k \cdot R}} \right)[A{\bf{I}}+B\, {\bf{\hat{R}\hat{R}}}]\end{equation}

\noindent where the sum is to be taken over the twelve nearest neighbors around
$\bf {R=0}$. $\bf{\hat{R}}$ is the unit vector in the direction of $\bf R$.
$\bf{I}$ is the $3\times3$ unit matrix and
$\bf{\hat{R}\hat{R}}$ is the diadic $3\times3$ matrix defined as
$(\bf{\hat{R}\hat{R}})_{\mu\nu}=\bf{\hat{R}}_\mu\bf{\hat{R}}_\nu$.
$A$ and $B$ are defined to be equal to $2\frac{\phi'(d)}{d}$ and
$2[\phi''(d)-\phi'(d)/d]$, respectively.
\vskip0.3in
\begin{center}
{\bf  B. Ideal HCP crystal}
\end{center}
\vskip0.3in
\noindent Here there are two particles ( located at $\bf p_1$ and $\bf p_2$ ) per unit cell and the algebra for
calculating the dynamical matrix is
somewhat more complicated than in the case of the
FCC crystal. But the result can be summarized as follows:

\noindent The dynamical matrix $D(\bf{k})$ has the following structure:

\[ D(\bf{k})=\Bigg[
 \begin{array}{c|c}
  (1,1) &(1,2) \\
  \hline
  (2,1) & (2,2)
 \end{array}\Bigg]
\]
where (1,1),(1,2),(2,1) and (2,2) are all $3\times3$  matrices. (1,1) and (2,2) are identical matrices of which the element at the $\mu^{th}$ row and $\nu^{th}$ column is given by
\[-\frac{\phi'(d)}{d} \Bigg[ \sum_{{\bf R}\in B}\left\{\left(1-d\frac{\phi''(d)}{\phi'(d)}\right)
\hat{n}_\mu({\bf R+p_2-p_1})\hat{n}_\nu({\bf{ R+p_2-p_1}})-\delta{\mu\nu}\right\} \]
\[ +2\sum_{{\bf R}\in C}\left\{\left(1-d\frac{\phi''(d)}{\phi'(d)}\right)\hat{n}_\nu({\bf R}) \hat{n}_\mu({\bf R})
-\delta_{\mu\nu}\right\}\sin^2\left(\frac{1}{2}\bf{R\cdot k}\right) \Bigg] \]
where $\hat{n}_{\alpha}(\bf{t})$ is the $\alpha^{th}$ component of the unit vector in direction of vector $\bf t$. $B$ and $C$ are sets of vectors of the form $n_1 {\bf a_1}+n_2{\bf a_2}+n_3{\bf a_3}$ $({\bf a_1,a_2 \mathrm{\,and\,} \bf \,a_3}\,$ are the edges of the unit cell). For the set $B$,
$(n_1,n_2,n_3)=(-1,0,-1),(-1,0,0),(0,-1,-1),(0,-1,0),(0,0,-1)\,\mathrm{and}\,(0,0,0)$ while for the set $C$, $(n_1,n_2,n_3)=(-1,0,0),(-1,1,0),$ $(0,-1,0),(0,1,0),(1,-1,0)\,\mathrm{and}\,(1,0,0)$. 
In block $(1,2)$ the element at the $\mu^{th}$ row and $\nu^{th}$ column is given by
\[-\frac{\phi'(d)}{d} \sum_{{\bf R}\in A}\Bigg[\delta_{\mu\nu}-\left(1-d\frac{\phi''(d)}{\phi'(d)}\right)\hat{n}_{\mu}({\bf R+p_1-p_2})\hat{n}_{\nu}({\bf R+p_1-p_2})
\Bigg] e^{-i{\bf k\cdot R}}\]
where $A$ contains the vectors $n_1 {\bf a_1}+n_2{\bf a_2}+n_3{\bf a_3}$ with $(n_1,n_2,n_3)=(0,0,0),(0,0,1),(0,1,0),$ $(0,1,1),(1,0,0)\,\mathrm{and}\,(1,0,1)$. In block (2,1), the element at the $\mu^{th}$ row and $\nu^{th}$ column is given by
\[-\frac{\phi'(d)}{d}\sum_{{\bf R}\in B} \Bigg[\delta_{\mu\nu}-\left(1-d\frac{\phi''(d)}{\phi'(d)}\right)\hat{n}_{\mu}({\bf R+p_2-p_1})\hat{n}_{\nu}({\bf R+p_2-p_1})
\Bigg] e^{-i{\bf k\cdot R}}\]

\vskip0.2in
\noindent While constructing the NDOSNF the overall scale of the
vibrational frequencies becomes irrelevant and hence it is clear that the
only parameter controlling the NDOSNF for both FCC and ideal HCP crystals
is the dimensionless number $\beta(d) \equiv 1- d{\phi}''(d)/\phi'(d)$ .
Thus, if
the nearest neighbor approximation is a valid one, two different potentials
will have identical FCC (or ideal HCP) NDOSNF if the pressures are adjusted such
that the value of the $\beta$ parameter is identical for the two situations.
This is illustrated in figure 12 where the FCC type NDOSNF are superposed
for GLJ(12,6)
at P = 1864 and for Morse potential at P = 12689, the (almost) common value
of $\beta$ being equal to 14.635 . 
%**************************************************************
\begin{figure}[H]
  \centering
  \includegraphics[width=0.5\textwidth]{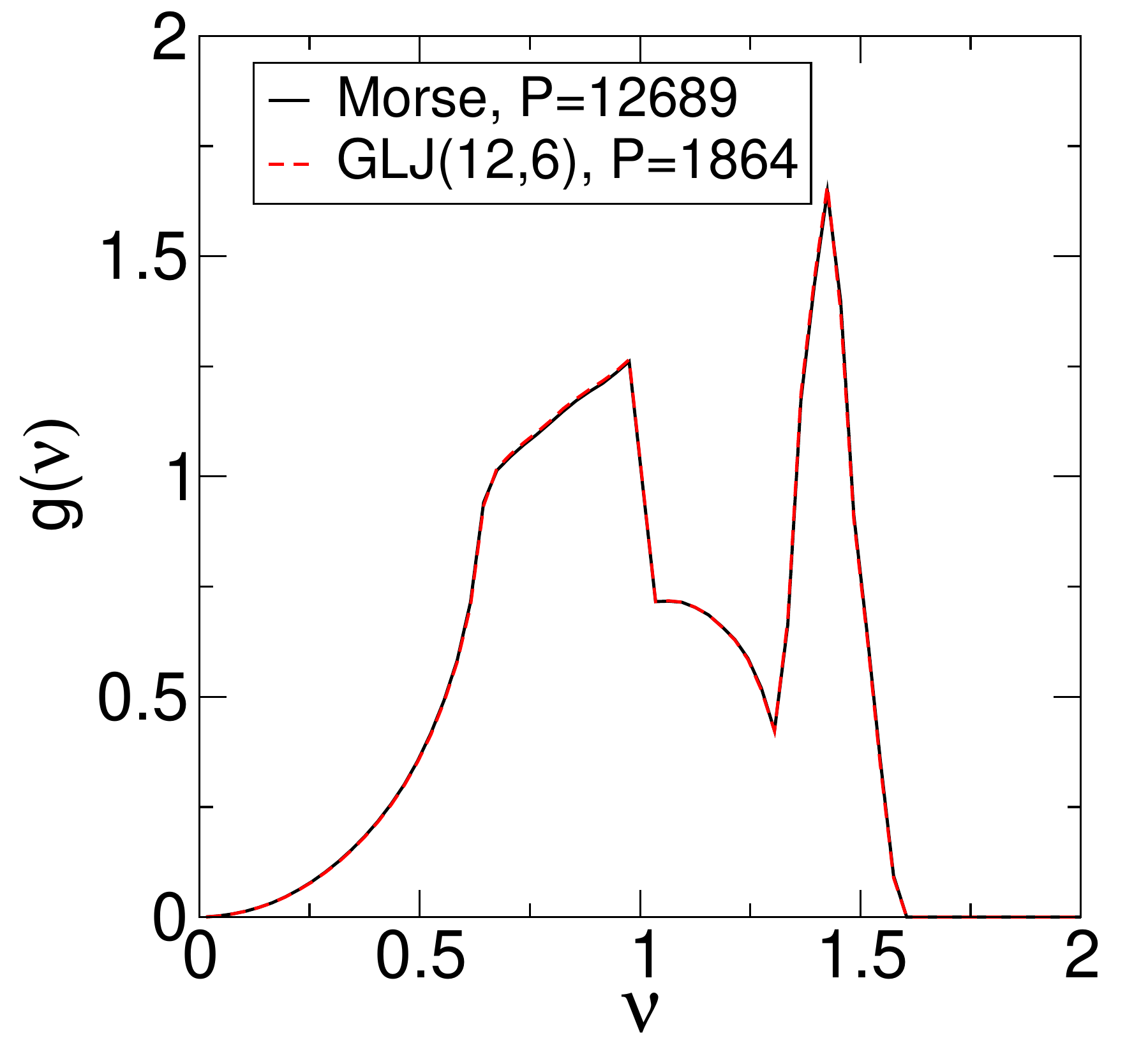}
  
    \caption{\small{Normalized density of states $g(\nu)$ for normalized frequency $(\nu)$ are
superposed for the GLJ(12,6) potential at $P = 1864$ and the Morse potential
at $P = 12689$. The two spectra are essentially indistinguishable.}}
  \label{fig:fig12}
\end{figure}
%**************************************************************

As we mentioned earlier the nearest neighbor approximation
breaks down at the highest pressures if the potential is of type B. This is
demonstrated in figure 13 for the Gupta potential. Fig.13(a)
corresponds to a pressure where the nearest neighbor approximation is still
a very good one. But for fig.13(b) the pressure is so high (the nearest
neighbor distance being correspondingly short) that the breakdown of the nearest 
neighbor approximation is quite visible. Needless to say the agreement worsens
progressively as the pressure increases even further.
%**************************************************************************
\begin{figure}[H]
  \centering
  \subfloat[]{\label{fig:(a)}\includegraphics[width=0.4\textwidth]{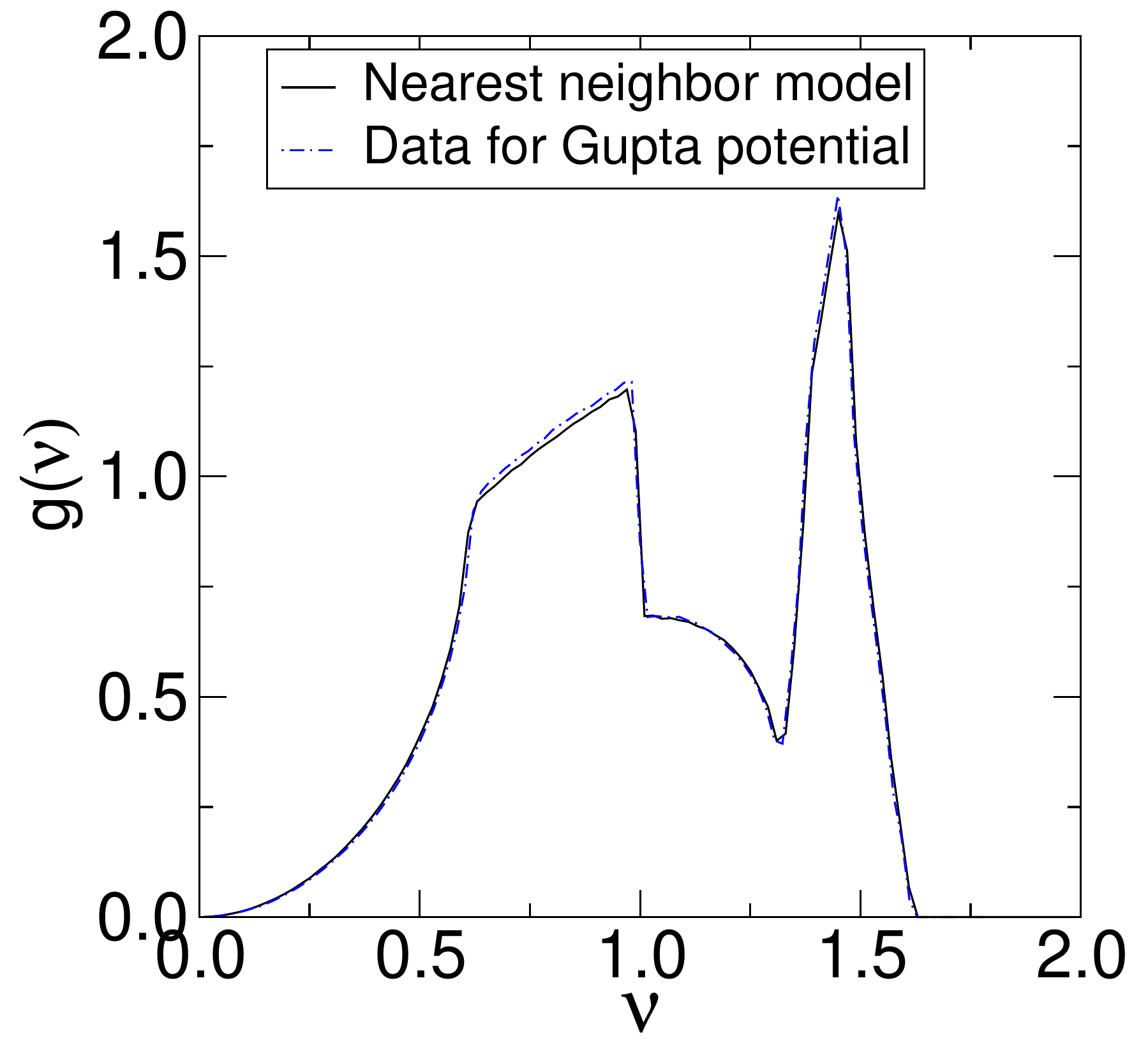}}
   \subfloat[]{\label{fig:(b)}\includegraphics[width=0.4\textwidth]{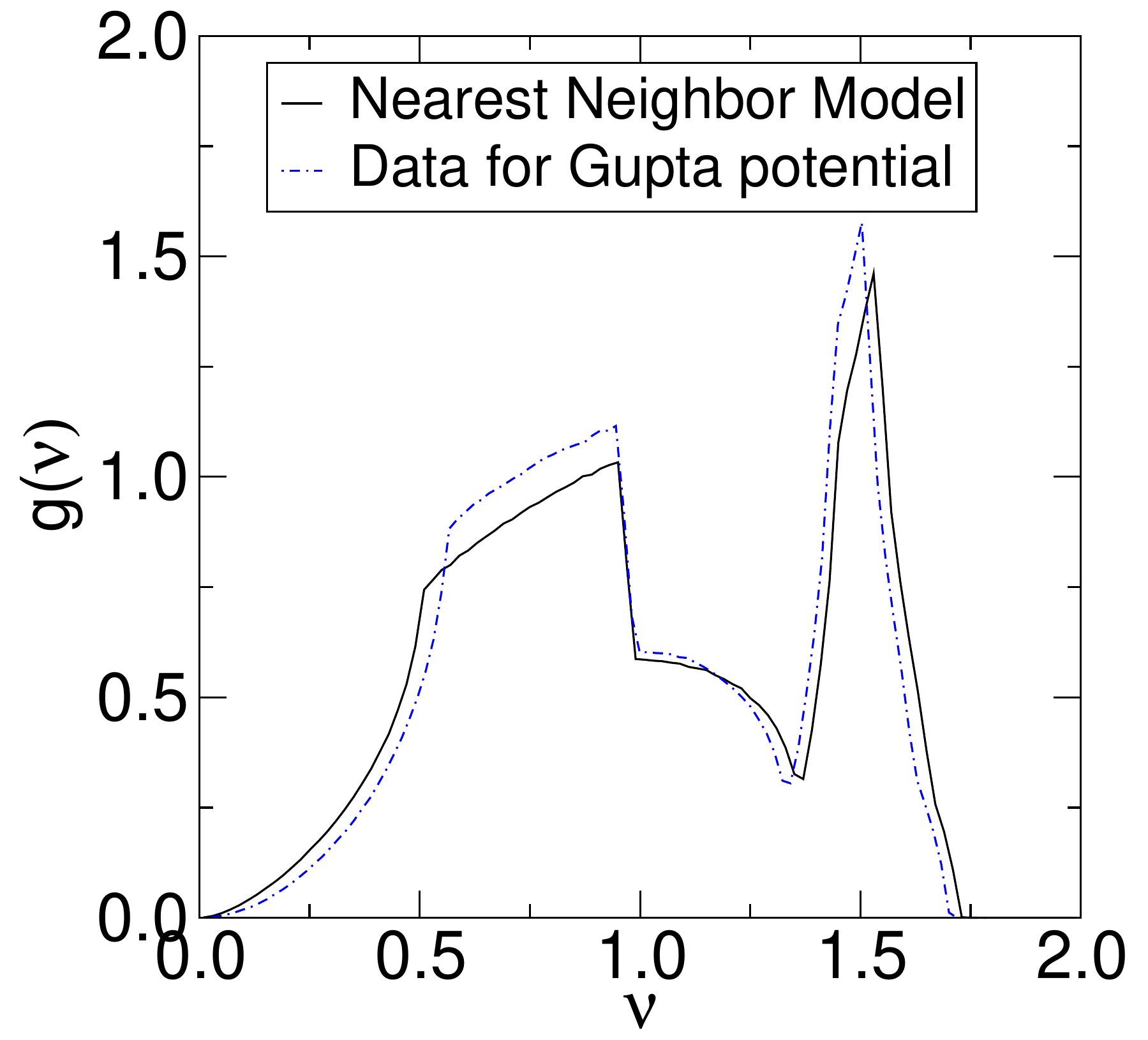}}
  
    \caption{\small{(a) Normalized density of states $g(\nu)$ for normalized frequency $(\nu)$ is
shown for Gupta potential at $P = 1650$ and the prediction for this case from
the nearest neighbor model. The agreement is very good. (b) Same as in (a) but for
$P = 160000$. Now the disagreement with the prediction from the nearest neighbor
model is quite prominent.}}
  \label{fig:fig13}
\end{figure}

%**************************************************************************

We have just described the procedure for calculating the NDOSNF for FCC
and (ideal) HCP crystal in the nearest neighbor approximation for a given
value of $\beta$. Two interesting aspects of this procedure are that: (i) In both the
cases vibrational spectrum exists only for
$\beta$ greater than 8 i.e. for a value of $\beta$ below this the dynamical matrix is not positive
definite for all $\bf k$ in the first Brillouin zone. There seems to be no
singularity
as $\beta$ approaches the value of 8 from above. (ii) In the limit
of $\beta$ going to infinity the NDOSNF is well defined. The first
observation is based on numerics and we do not have an analytical
understanding of this behaviour at $\beta$ = 8. We will see later that the
second observation is what enables us to establish connection with
experiments on real materials. Figures 14(a) and 14(b) show the evolution of
the shape of the vibrational spectrum with change in the value of $\beta$
within the framework of the nearest neighbor model for the FCC and the ideal
HCP geometries, respectively. In both the cases the shape function shrinks without
any change in the position of the center as $\beta$ goes to infinity. Due to 
normalization there is a corresponding increase in the height of the function.
One way to quantify this is to define a 'fatness coefficient' as the ratio of
the standard deviation of the frequency to the mean frequency. We find that for
both the FCC and the HCP geometries this coefficient changes from around 0.4
to 0.28 as $\beta$ goes from its lowest permissible value to infinity. 
As would be expected
from the existence of a well defined limit when $\beta$ goes to infinity the
dependence of the NDOSNF on $\beta$ becomes very weak for the higher values of $\beta$. 
%**************************************************************
\begin{figure}[H]
  \centering
  \subfloat[]{\label{fig:(a)}\includegraphics[width=0.4\textwidth]{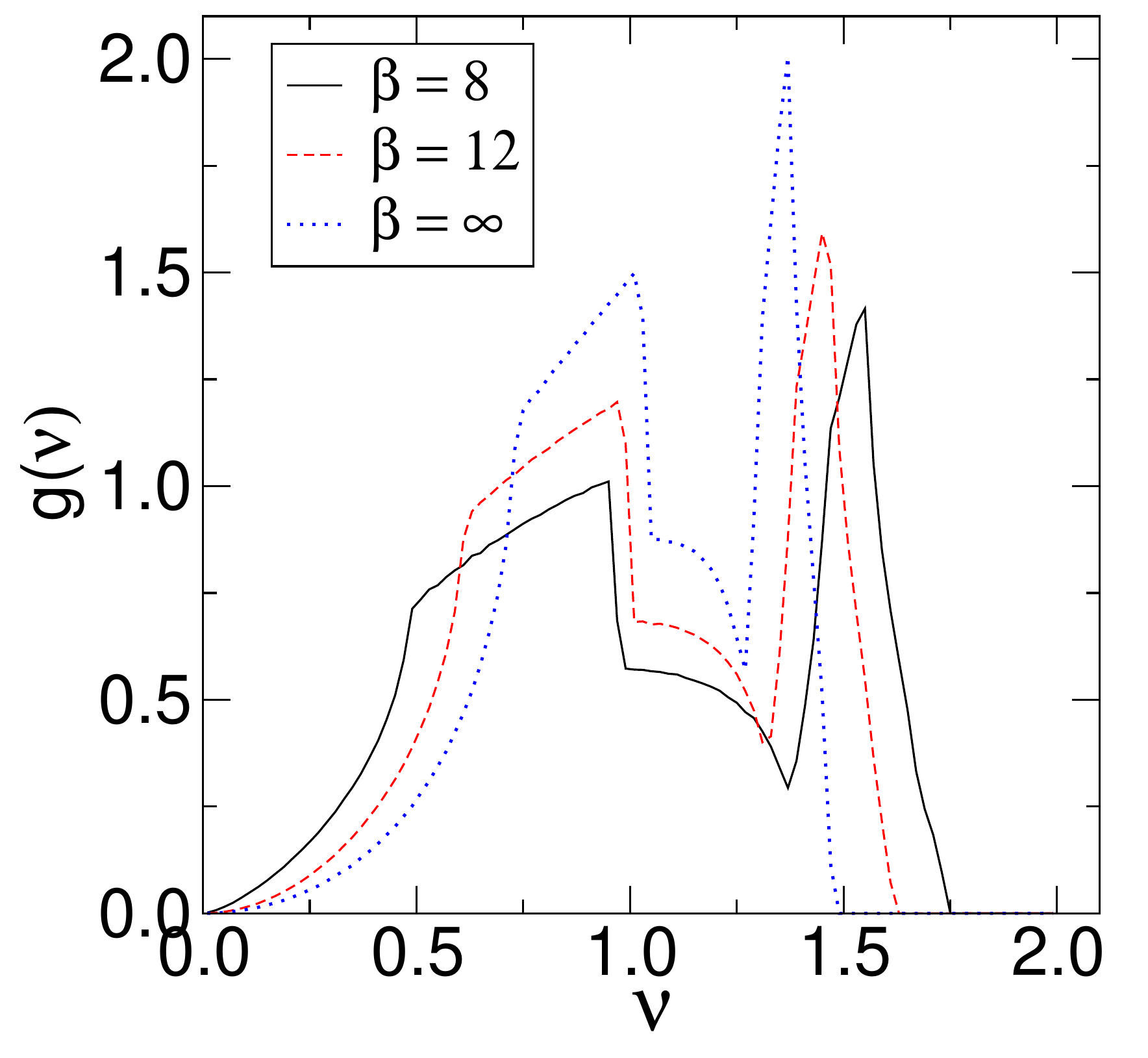}}
   \subfloat[]{\label{fig:(b)}\includegraphics[width=0.4\textwidth]{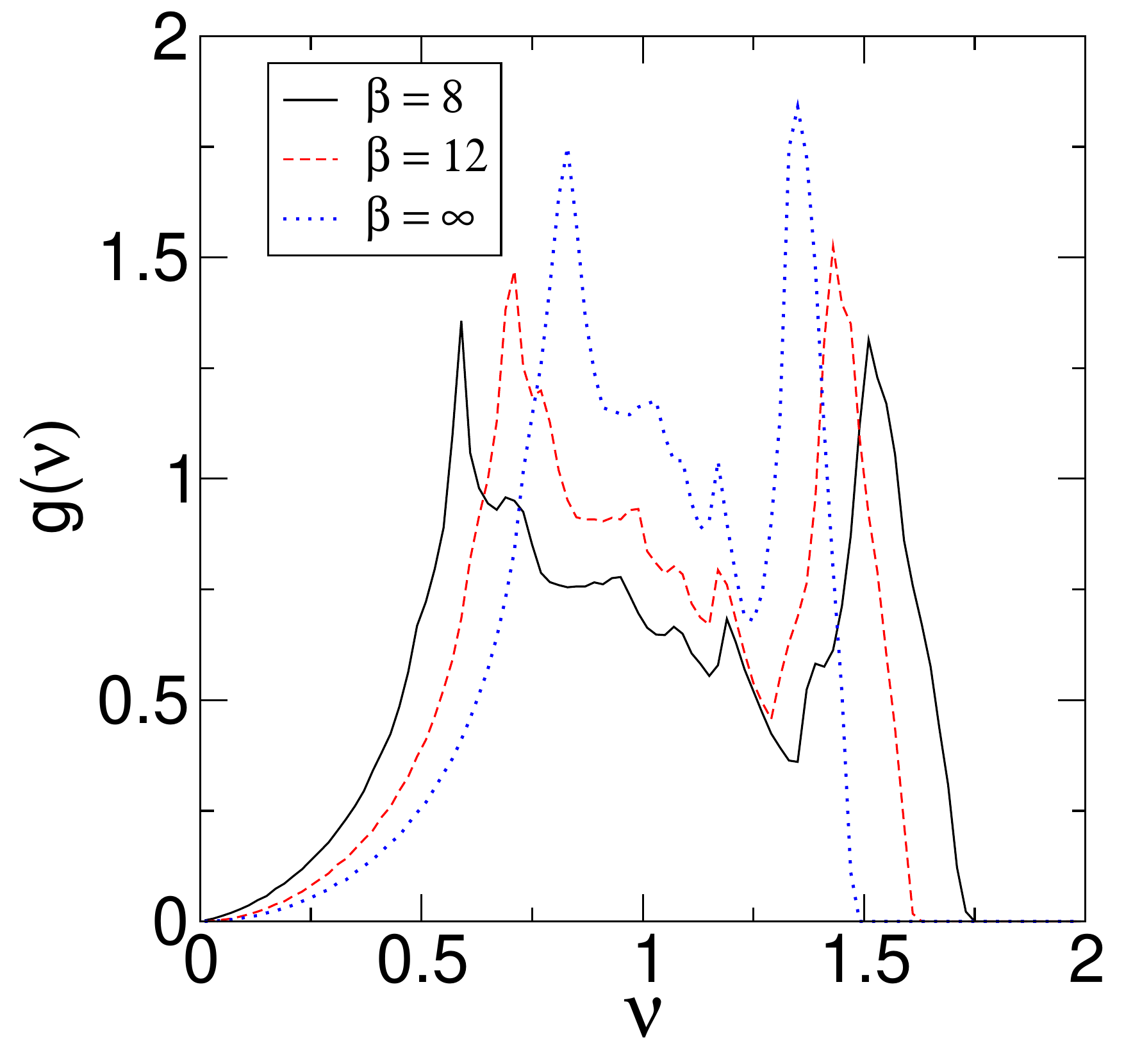}}
  
    \caption{\small{Evolution of the NDOSNF function for the nearest neighbor model with the 
control parameter $\beta$. For the sake of clarity plots are shown for only three values of 
$\beta$. (a) FCC (b) HCP.}}
  \label{fig:fig14}
\end{figure}

%**************************************************************

%\newpage
\vskip0.3in

\begin{center}
{\bf V. Scaling ansatz for the dispersion relation}
\end{center}
\vskip0.3in

\noindent The observation of shape convergence for the entire vibrational spectrum and
the proportionality of the average and Debye frequencies for the various
states of aggregation,
made in sections III.B and III.C, can be explained on the basis of a scaling
ansatz regarding the dispersion relations. The ansatz is that, at higher
pressures,

\begin{equation}{\omega}_{i}({\bf k}) = W(P){F}_{i}({\bf k} n^{-1/3})\end{equation}

\noindent where $n$ is the number density of particles and $W(P)$ depends on
the potential energy function but not on the state of aggregation. $i$ is
the index labeling a particular branch of the dispersion relations. For
every value of $i$ the function ${F}_{i}$, which depends on the state of
aggregation, is a positive definite function whose domain of definition is
finite and is independent of pressure. Here we assume that the shape of the first Brillouin zone 
does not change once the applied pressure becomes sufficiently high. In that case 
its linear dimension becomes
proportional to $n^{1/3}$. Also, for every $i$, $F_i$ is bounded
above. Hence, for the sake of definiteness, we normalize these functions such
that $\sum_{i} \int {F}_{i}({\bf k}')d^3{\bf k}' $ (the integrals are over the entire domain
of definition of the integrands) is
unity. This makes the definition of $W(P)$ unambiguous.
Let us now study the implications of the abovementioned scaling relationship
for the evolution of the shape of the density of states function as well as for the relationship
between the average frequency and the Debye frequency. If $h(\omega)$ is the 
density of states per degree of freedom (i. e. $\int h(\omega)d\omega = 1 $)
then

\begin{equation}h(\omega) = 1/(24n{\pi}^3) \sum_{i} \int 
\delta(\omega-{\omega}_{i}({\bf k})){d}^{3}{\bf k}\end{equation}
The expression on the right hand side can be reduced, by using the scaling ansatz given 
in equation (11), to  $(1/(24{\pi}^{3}W(P))$$H(\omega/W(P))$
where $H(x) =\sum_{i} T_i(x)$ -- with $T_i(x)$ being equal to
 $ \int \delta(x - F_i({\bf k}')){d^3}{\bf k}'$.
With these definitions
average frequency $<\omega>$ has the expression
\begin{equation}<\omega>  =  W(P) (\int xH(x) dx)/(\int H(x) dx)\end{equation}

\noindent It should already be obvious that the consequence of shape saturation is
immediate. Similarly it can be shown that the Debye frequency ${\omega}_{D}$
satisfies the following relationship:

\begin{equation}1/{{{\omega}_{D}}^3} = (1/18\pi^2W^3(P))\sum_{j}\overline{\left(\frac{1}{V_j^3(\hat s)}\right)}\end{equation}
\noindent where the summation index $j$ labels the three acoustic branches and 
$V_j(\hat s) =\lim_{x\rightarrow 0+}\frac{dF_j(x\hat s)}{dx}$
($\hat s$ is a unit vector labelling a particular direction).
The overhead bar denotes average over all directions.

To summarize, for a given potential, the scaling ansatz implies that both
$<\omega>$ and ${\omega}_{D}$ should be
proportional to $W(P)$ with the two constants of proportionality being
dependent on the state of aggregation. This would explain why the  values of
average frequency and Debye frequency computed for the three states of aggregation 
maintain the same mutual ratios at all
(higher) pressures. It should be mentioned here that the power law scaling of
average frequency (and hence Debye frequency, according to section III.C)
with respect to pressure is not implied by the scaling ansatz. At the empirical
level that is an additional feature.

We will now argue that for type A potentials the scaling ansatz should indeed
be valid exactly  as $P$ goes to infinity. In the same limit we will
show that the power law scaling of average or Debye frequency with respect to
pressure should be exact. The steps
involved in the argument are as follows: (i) With increasing pressure the
interparticle distances keep on shrinking. For any generic potential that
describes a stable solid the ratio of the repulsive part to the attractive
part of the potential will keep increasing and eventually it will become
permissible to ignore the attractive part altogether. For the potentials
we are studying the repulsive part is always a sum-over-pairs of the various
pair potentials. Thus, for type A potentials, the total potential energy
has the following {\it effective} form at high pressures:
$ \sum_{ i< j} (1/r_{ij}^m)$.
In our particular calculation m is always a positive integer. (ii) Consider a stable
equilibrium spatial
arrangement of the particles in the solid for any of the three relevant
states of aggregation at a pressure (let's say $P_1$) high enough for the
effective form of the potential energy given above to be valid. Suppose we
now consider the situation
at the somewhat higher pressure of ${P}_{2} = {\alpha}{P_1}$. Because of
the power law
scaling of the potential energy (and hence forces) it should be clear that
a configuration that has exactly the same geometry as the one at $P_1$
but has an overall contraction of the interparticle distances by a factor
of ${\alpha}^{1/(3+m)}$  will be a stable equilibrium geometry at this new
pressure $P_2$.
Now the elements of the dynamical matrix at wavevector $\bf k$ are
suitable Fourier transforms of the
second derivatives of the potential energy with respect to various
combinations of particle position coordinates. In this process of Fourier
transformation  the wavevector $\bf k$ is multiplied by the Bravais lattice vectors
which scale linearly with interparticle distances -- which
in turn is proportional to $n^{-1/3}$. Thus the $\bf k$ dependence of  the
dynamical matrix will be only through the combination ${\bf k}{n}^{-1/3}$. We
now relate the matrices of the second derivatives at $P_1$ and $P_2$ (before
Fourier transformation). As noted above the spatial arrangements at $P_2$ and
$P_1$ are different only by the scale of length. Since the pair potential
has a power law form ( proportional to $r^{-m}$ ) it is clear that the
dynamical matrix (before Fourier transformation) at $P_2$ is that at $P_1$
multiplied $\alpha^{(2+m)/(3+m)}$. Finally, since the frequencies are square
roots of the eigenvalues we get the desired result that $\omega(\bf k)$ for
a particular branch is indeed proportional to $P^{(2+m)/2(3+m)}$ multiplied
by a function of the product ${\bf k}{n}^{-1/3}$. This is the scaling form
that we have proposed earlier -- with the additional specification that
$W(P)$ itself has a power law form with the scaling exponent $\delta$ =
$(2+m)/(2(3+m))$ (which is obviously independent of the state of aggregation). 
So we see that the result III.A , which
in principle is independent of the results in III.B and III.C, actually
follows from
the same line of reasoning that leads to the scaling ansatz that explains the
results of III.B and III.C. In fact, after incorporating the power law 
dependence of density and the scale of vibrational frequency on pressure, the 
final scaling form of the dispersion relation is
\begin{equation}\omega_{i}({\bf k}) = ZP^{(2+m)/2(3+m)}G_{i}(P^{-1/(3+m)}{\bf k})\end{equation}
\noindent (The arguments of $G_{i}$ and $F_{i}$ are different only through an
overall constant of proportionality). 
This looks exactly like the standard scaling forms that are familiar in the theory of phase
transitions and critical phenomena [47].

Given the centrality of the power law form of the pair potential in the preceding 
arguments how do we explain the empirical observation that even for type B potentials
the properties of shape saturation of density of states and power law scaling
of average/Debye frequencies are observed to a very good approximation over
a very large ranges of pressure?
To understand this we note that the key aspect of the argument given above
is that at high pressures geometry of the arrangement of the particles should
not
change with pressure. For FCC and ideal HCP solids this is not an issue.
For this to be possible for amorphous geometries the following condition has
to be satisfied (we approximate the amorphous state by a periodic
lattice with a large unit cell containing $N$ particles) for $i = 1, 2, ...., N$:

\begin{equation}{\bf \nabla }V_i (b {\bf r}_1, b{\bf r}_2, ..., b {\bf r}_N) = q {\bf \nabla} V_i ({\bf r}_1, {\bf r}_2, ...., {\bf r}_N)\end{equation}

\noindent
where b is a positive rescaling factor for interparticle distances. q, which is a 
function of b, is the
rescaling factor for the forces. It is not difficult
to see that this can be satisfied exactly only for type A potentials.
Although type B potentials do not satisfy this exactly  they can still
satisfy, over a large but finite range of pressure, this to a very good
degree of approximation. To see this we construct an effective power law
exponent for pair potentials that are not actually of the power law form
(Morse potential, for example). Since an arbitrary constant can always be
added to a potential we construct the definition of the effective exponent in terms of the expression
for the force $f(r)$ between a pair. For a power law potential $ln(f(r))$ will
be linear in $ln(r)$ with a slope of $-(m+1)$. Hence a natural definition
of the effective exponent at an interparticle separation of d is $-
(dV''(d)/V'(d) + 1)$ ( Notice that this equals $\beta(d)-2$ where $\beta(d)$ is the
function that controlls the NDOSNF in the nearest neighbor approximation
(section IV)). As expected, for a potential that is not of the power
law type, this exponent is distance dependent. However, what really matters
is how large is the value of the exponent and how much does it vary over the
relevant range of pressures. For example, analysis of our data for the Morse potential shows
that this exponent varies in the range of approximately 5 to 7 (and $\beta$
varying from 7 to 9) in the regions
of  best linear fit in figure 3. It is this rather narrow range, caused by
the rapid distance dependence of an exponentially decaying force,
that is responsible for creating the impression of an almost saturating
NDOSNF. To see why the exponent $\delta$ seems to be so well defined in figure
3, recall that $\delta$ is predicted to be $(2+m)/(2(3+m))$ for a power law
potential with exponent $m$. This implies extremely weak dependence of $\delta$ on $m$. For
example, when $m$ varies from 5 to 10, $\delta$ varies only from 0.44 to 0.46.  
It is not thus surprising that the impression of an almost perfect linear
fit is created in figure 3. The weak dependence of $\delta$ on $m$ is due to
the following reason: For larger values of $m$ the nearest neighbor distance
varies  slowly with pressure but the spring constant that decides the
scale of vibrational frequency varies rapidly with the nearest neighbor
distance. For smaller values of $m$ exactly the opposite happens. Thus, of
the two factors that control the dependence of characteristic frequency on
pressure, rapid variation in one is largely compensated by the slow variation
of the other. It may be observed that the data in table I relating to the measured
values of the exponent $\delta$ are indeed consistent, within the uncertainties of
definition or measurement, with the prediction of
$(2+m)/(2(3+m))$ where $m$ is the true or effective power law exponent for the
potential. We also understand from the formula for $\delta$ why it is always below
1/2.

The considerations presented above lead us to conclude that the saturation
of the NDOSNF, mutual proportionality of the average and Debye frequencies and the 
power law dependence of the characteristic frequencies
on pressure are  rather robust results if one does not demand rigorous
exactness. Hence
these represent predictions that can be tested in experiments on real
materials or {\it ab initio} calculations provided the pressure is sufficiently
high.
%\newpage
\vskip0.3in
\begin{center}
{\bf VI. Double isosbestic points in the spectra of type A amorphous systems}
\end{center}
\vskip0.3in

\noindent The nearest neighbor model of vibrational spectrum for FCC and ideal HCP, as
discussed in section IV, has the property that the NDOSNF is decided for a
given type of crystal structure entirely by the value of the parameter
$\beta(d)$. However, this
parameter actually does not depend on the nearest neighbor distance $d$ for
a type A potential in the limit of high pressure (for $\phi(r)$ proportional to
$r^{-m}$ the asymptotic value of $\beta$ is $(2+m)$). This means that the NDOSNF
asymptotically converges to a well defined limit for FCC and ideal HCP for
type A potentials and this limit is decided by the value of $m$. At the
end of section IV we noted that the NDOSNF has a well defined limit as $\beta$
(or $m$) goes to infinity. Let us now consider the GLJ(m,n) family of
potentials
with a fixed value of (m-n) but m assuming all positive integer values
starting from, let's say, 12. If we consider the FCC or ideal HCP solids for
any of these potentials  the NDOSNF will
saturate (the limit being determined by $m$) as pressure keeps increasing.
Also this limiting curve itself will have a well defined limit when $m$ tends to infinity.

What happens if we consider the {\it amorphous} state for these potentials [43]? 
The arguments of section IV are no longer applicable. Here the NDOSNF does display
exact asymptotic saturation with respect to increasing pressure since we are dealing with 
type A potentials. However,
unlike in the case of FCC and ideal HCP solids, this asymptotic NDOSNF does not
have a well defined limit when $m$ tends to infinity. Instead what happens is that the asymptotic 
NDOSNF
displays two isosbestic points. To within the accuracy with which these two points can
be located, their coordinates do not depend on the  value of
$(m-n)$ (figure 15) . Isosbestic points in vibrational spectra have been 
reported earlier also in the
literature [9]. However, the interesting point about the present work is that the potentials
involved are essentially of the simplest conceivable kind (simple power law soft potential). 
Hence, although an analytical
understanding of this phenomenon of the existence of the isosbestic points is likely to be 
quite difficult there is atleast no needless theoretical
complication.
%\newpage
%**************************************************************
\begin{figure}[H]
  \centering
  \subfloat[]{\label{fig:(a)}\includegraphics[width=0.4\textwidth]{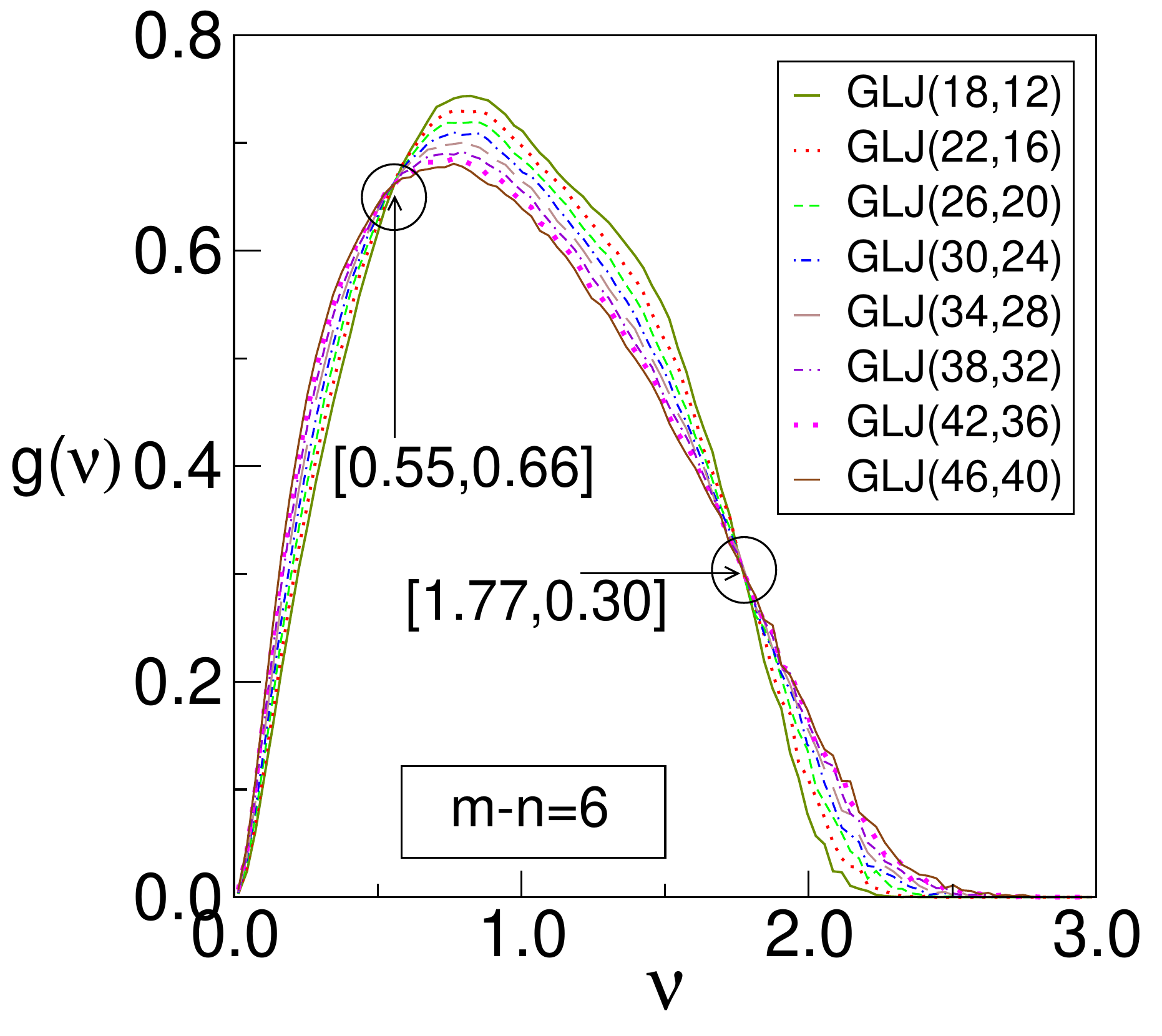}}
   \subfloat[]{\label{fig:(b)}\includegraphics[width=0.4\textwidth]{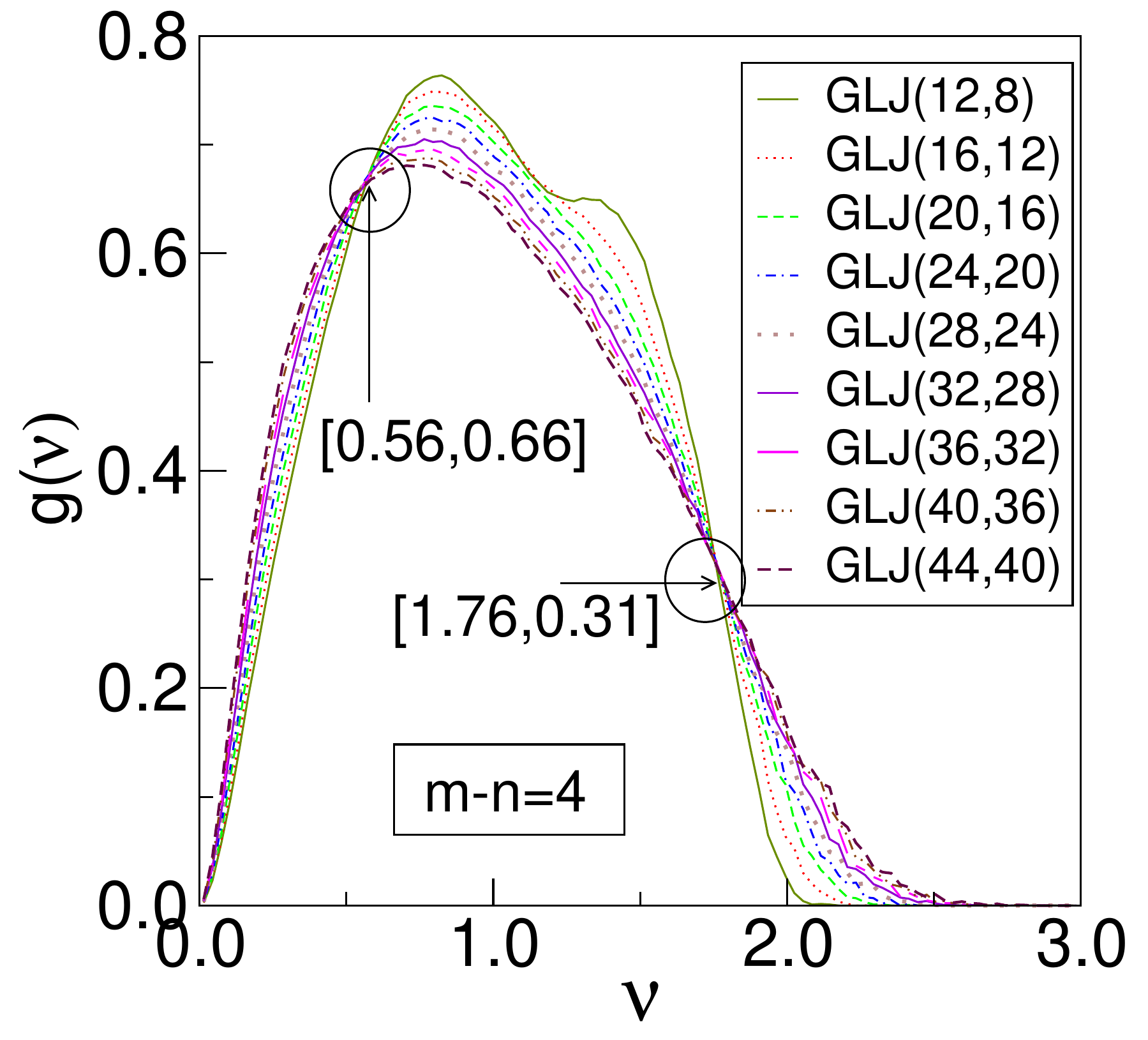}}
  
    \caption{\small{Existence of double isosbestic points in the superposition
of normalized density of states $g(\nu)$ for normalized frequencies $(\nu)$ for the family
of potentials GLJ(m,n). The two isosbestic points are indicated by arrows. Their
coordinates are also indicated. (a) For the family with $m-n = 6$. (b) For the
family with $m-n = 4$. The cooordinates of the two isosbestic points are the
same in (a) and (b).}}
  \label{fig:fig15}
\end{figure}

%********************************************************************
 
\vskip0.3in
\begin{center}
{\bf VII. Evolution of sound velocity with pressure}
\end{center}

\vskip0.3in
\noindent Study of sound propagation in materials under very high pressure is a subject of
key importance in planetary geophysics [48-54]. Our present investigations afford us a
way of studying this problem. Although our studies are at $T = 0$ this limitation is of much
less serious consequence at the extreme high pressures that we are interested in.
Inspired by the literature the two primary questions we have addressed in our studies 
are the following: (a) Does the ratio of transverse and longitudinal sound velocities 
saturate as pressure keeps growing? and (b) What are the limits of validity of the Birch's law [51] for
linear relationship between density and sound speed? 

Within the context of our studies based on model potentials answers to these questions can be 
found in the scaling form for the dispersion
relationship that we have introduced. In fact it follows quite simply from the scaling
law that  both longitudinal and transverse speed of sound, after averaging uniformly
over all angles,  are proportional
to $W(P){n}^{-1/3}$ -- immediately implying that ratio between the two averages approaches
a constant at high pressures (where the scaling law is valid). For a power law potential 
with exponent $m$ we have already found that: (1) the scaling law is exact, (2) $P$ is 
proportional to ${n}^{m/3+1}$, and (3) $W(P)$ is proportional to $P^{\delta}$ with
$\delta = (2+m)/(2(3+m))$. Thus, according to the scaling law, the two sound velocities should
be proportional to $n^{m/6}$ or $P^{m/(2(3+m))}$ for a type A potential with exponent $m$. Validity of this
can be seen from figure 16 where a  plot of ln(velocity) versus
ln(number density) is shown for the three states of aggregation with GLJ(12,6) potential.
In terms of the discussion above velocity should be proportional to the {\it square} of the 
density for 
this potential. This is indeed borne out by the data in fig.16. Also the ratio between 
the two computed average velocities 
approaches a constant as pressure keeps increasing. Thus we see that the Birch's law 
is certainly invalid for type A potentials at higher pressures except in the special 
case of $m = 6$. 
%***********************************************
\begin{figure}[H]
  \centering
  \subfloat[]{\label{fig:(a)}\includegraphics[width=0.32\textwidth]{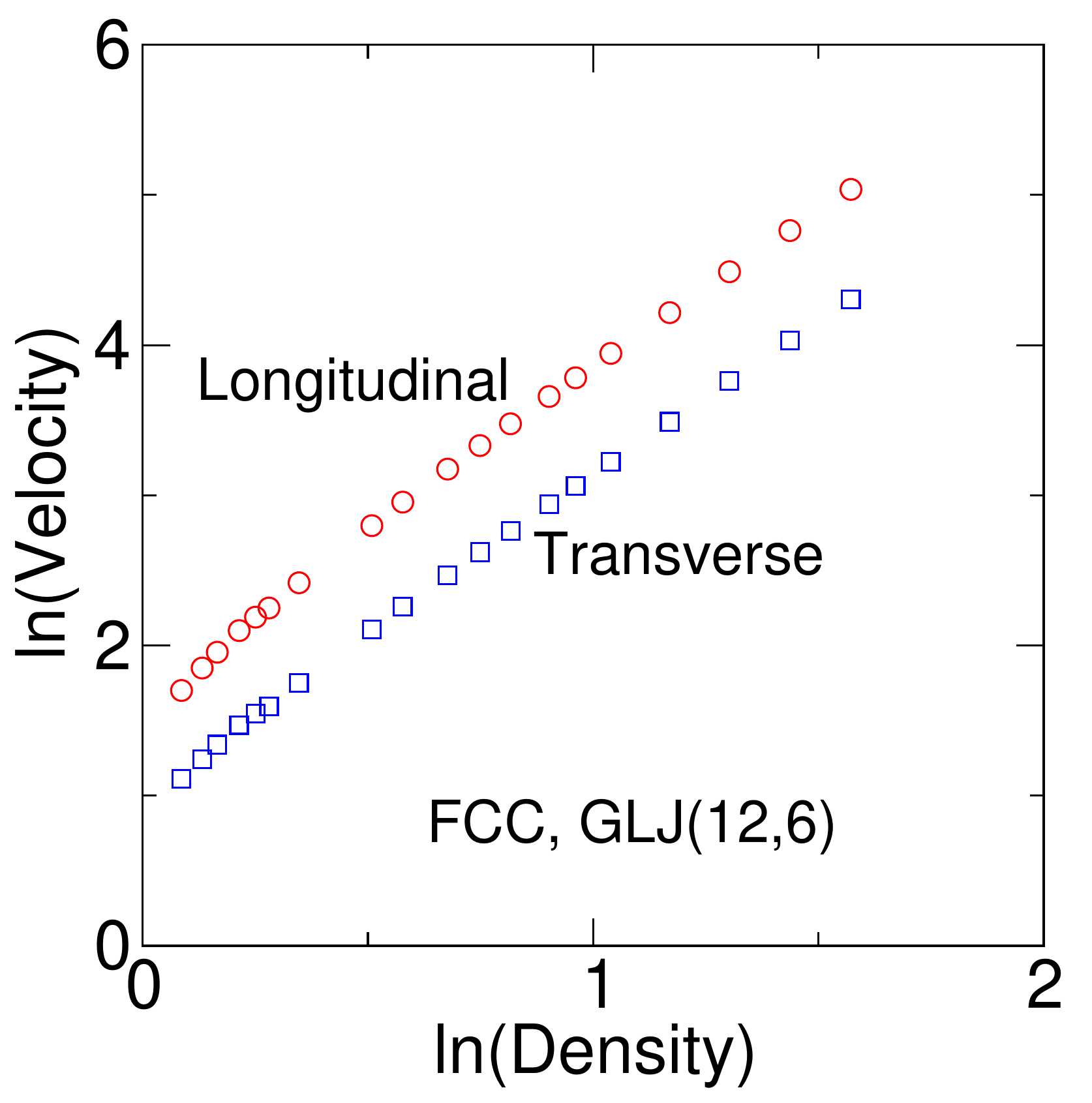}}
   \subfloat[]{\label{fig:(b)}\includegraphics[width=0.32\textwidth]{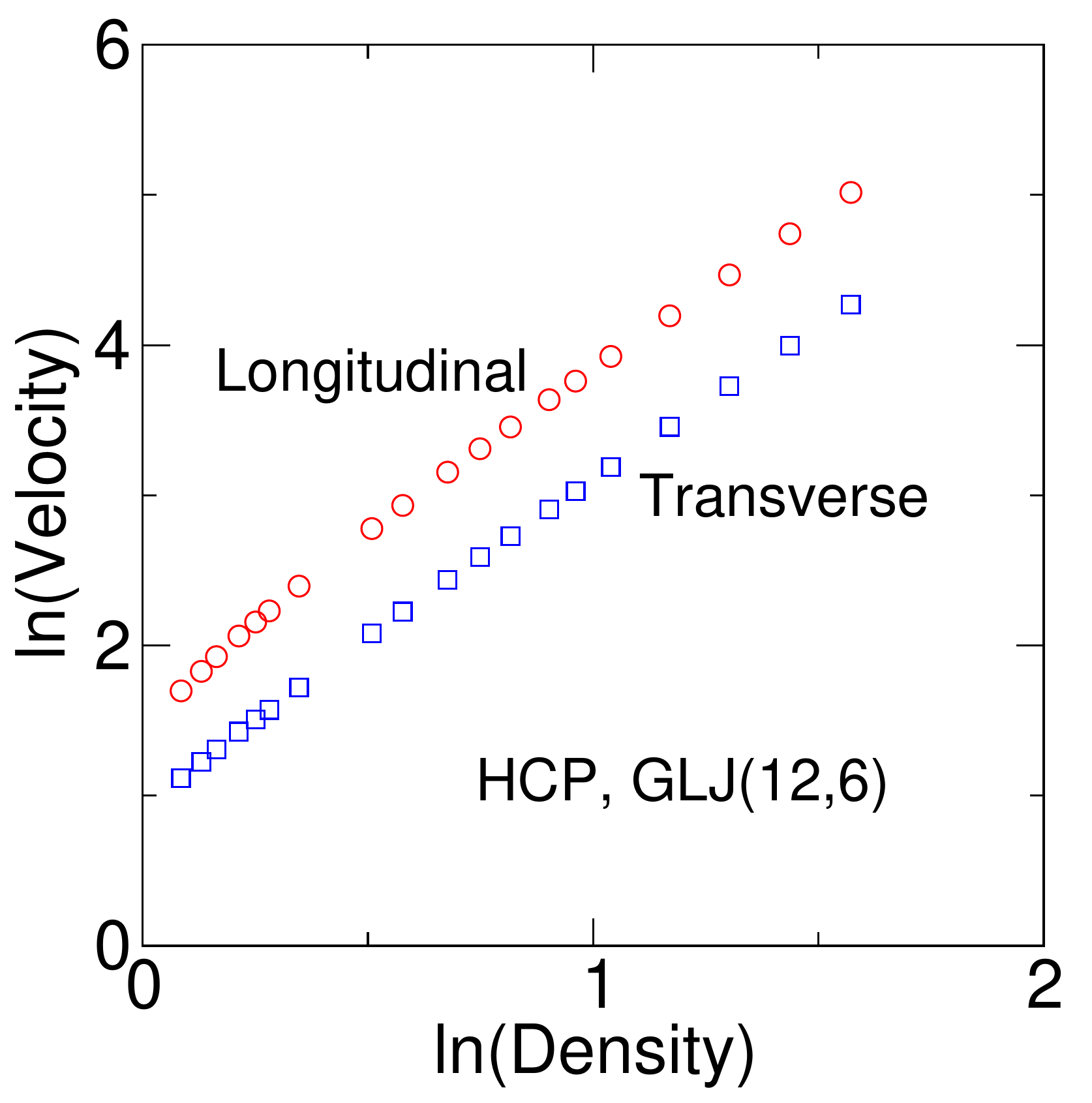}}
    \subfloat[]{\label{fig:(c)}\includegraphics[width=0.32\textwidth]{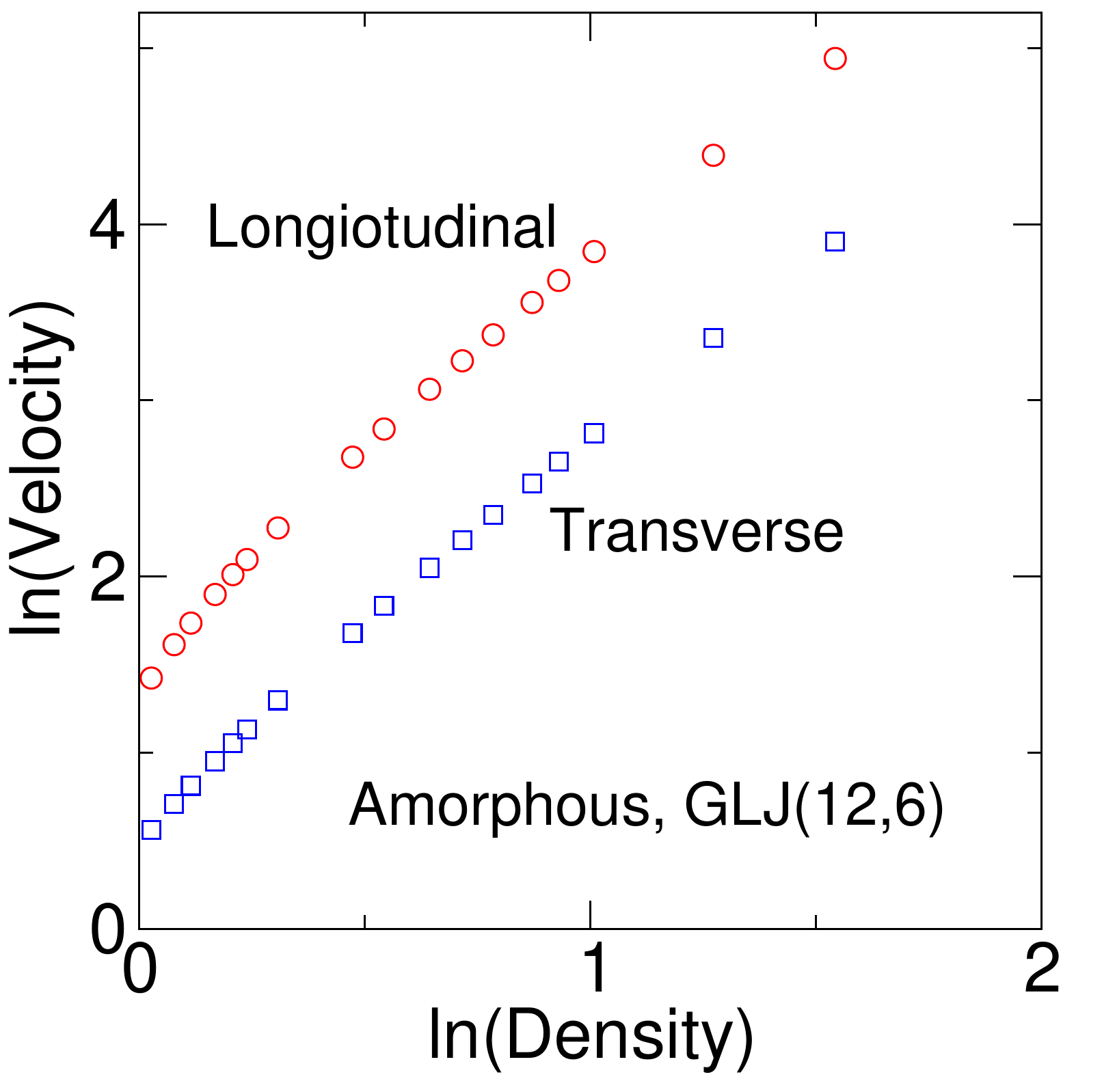}}
    \caption{\small{Log of sound velocity (longitudinal and transverse) versus log of
density for the GLJ(12,6) potential. (a) FCC (b) HCP and (c) Amorphous. In all
cases the slope at higher densities is 2 -- showing that sound velocity is proportional
to the square of density in that limit.}}
  \label{fig:fig16}
\end{figure}

%*************************************************
As a representative of type B potentials we examine the phenomenology of the Morse
potential. We find that the ratio of the two direction averaged velocities never quite 
saturates. However, the deviation
from saturation is very mild. For example, over a range of pressures (in the high
pressure region) where density increases by a factor of 2 the ratio increases only by
around 4 percent. Since this change is below the level of the typical error bar of an experimental
measurement of this quantity it may justifiably be described as {\it de facto} saturation.
At higher pressures average sound velocities show very close to linear variation with density for this
potential -- as can be seen from figure 17. 
%*********************************************
\begin{figure}[H]
  \centering
  \subfloat[]{\label{fig:(a)}\includegraphics[width=0.32\textwidth]{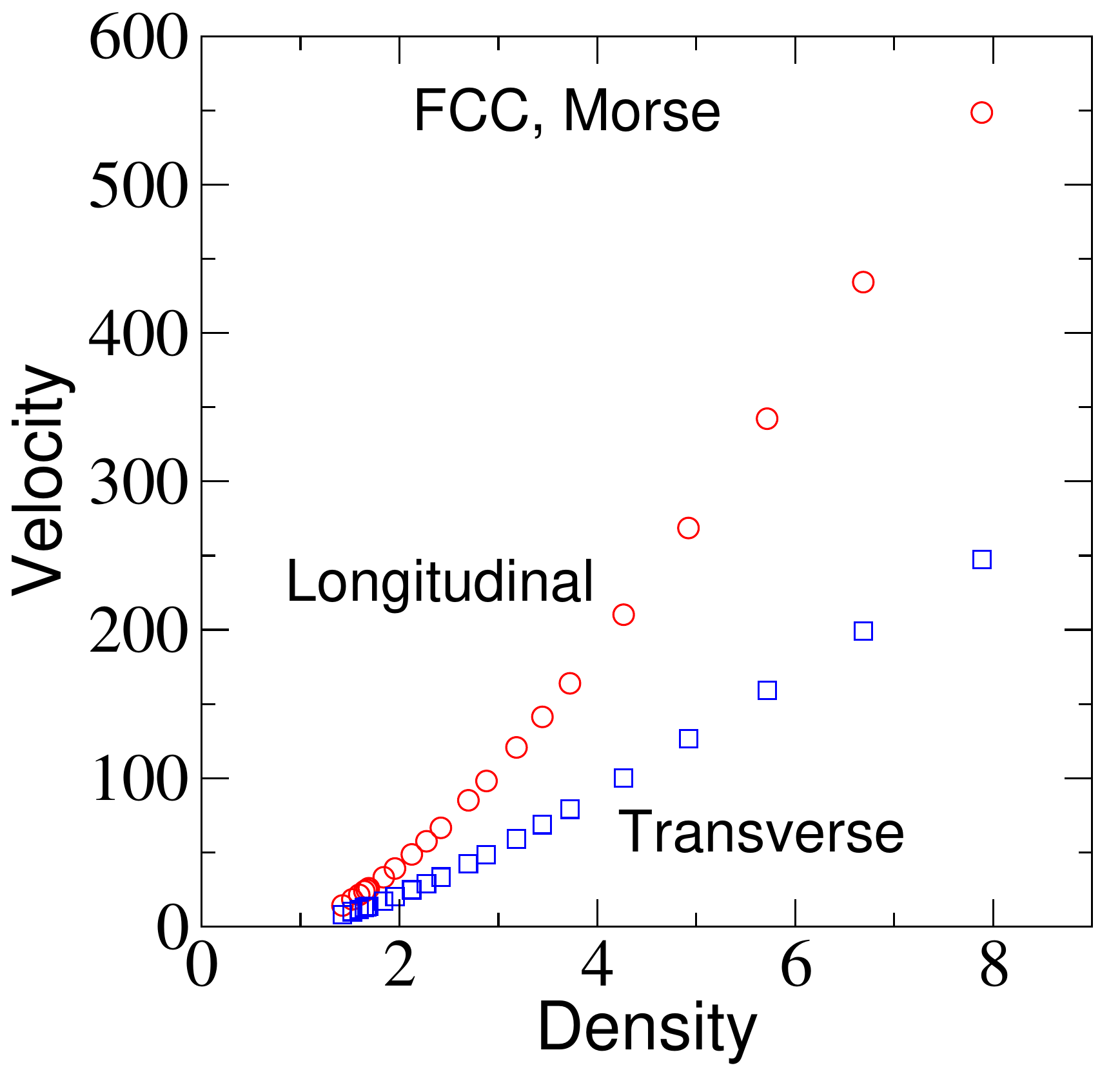}}
   \subfloat[]{\label{fig:(b)}\includegraphics[width=0.32\textwidth]{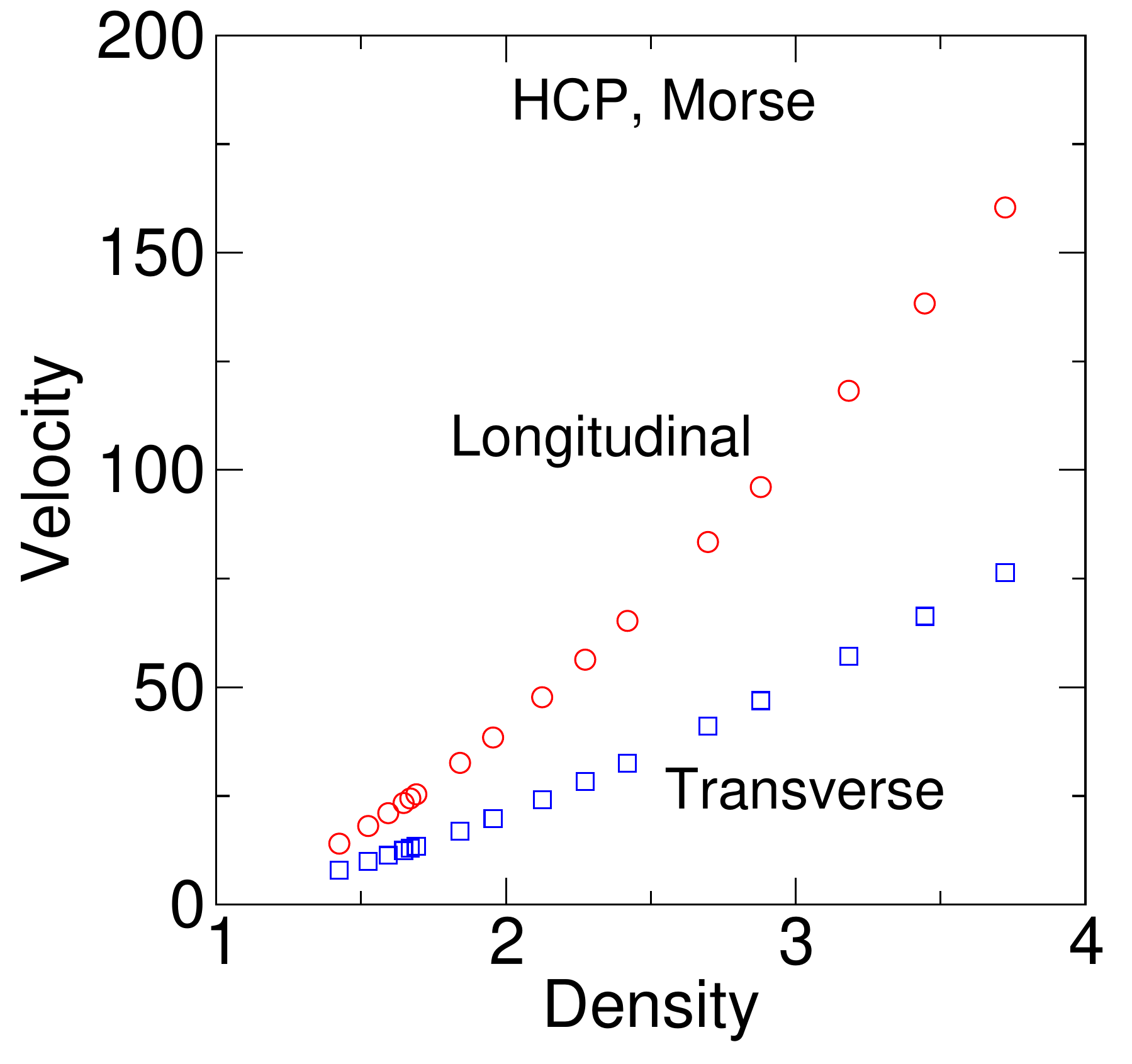}}
    \subfloat[]{\label{fig:(c)}\includegraphics[width=0.32\textwidth]{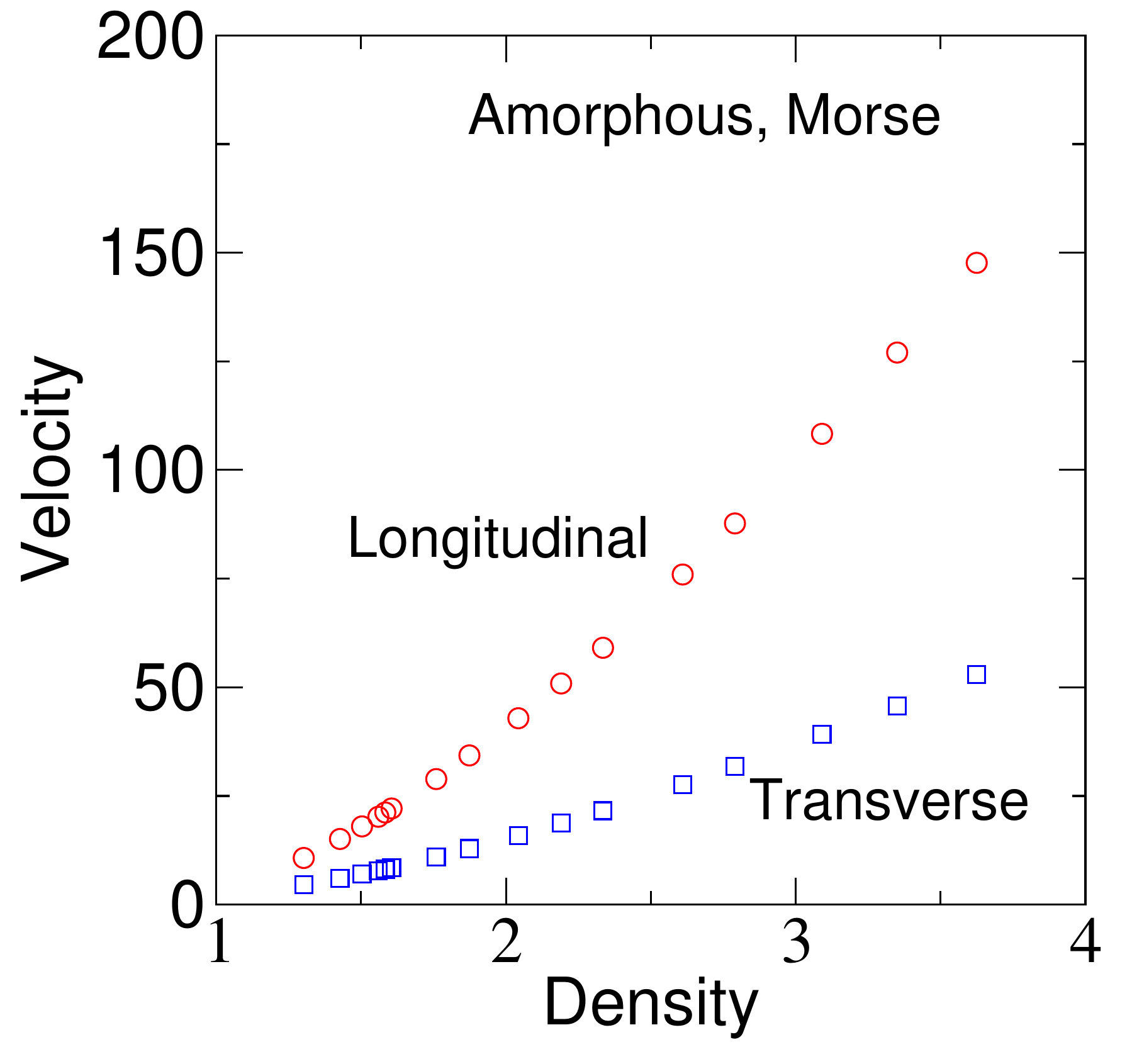}}
    \caption{\small{Plot of longitudinal and transverse sound velocities versus 
density for the Morse potential. (a) FCC (b) HCP and (c) Amorphous. In all
cases the variation of sound velocity with density is linear to a very good approximation
in a broad range of higher densities.}}
  \label{fig:fig17}
\end{figure}

%************************************************

This behaviour is actually a consequence of the fact that in the pressure range under reference the
effective exponent for the Morse potential varies in the range of 5 to 7. Combining this range for
$m$ with the 
prediction of $n^{m/6}$ variation of sound velocity, we can understand why an almost linear relationship 
(same as suggested by Birch's law) is found in this case over a rather large range of
densities. However, even in this case data over a sufficiently large range of densities will indeed bring out
the deviation from linear relationship quite clearly. The fact that the 
scaling law is not asymptotically exact in this case cannot be missed. 
%\newpage

\vskip0.3in
\begin{center}
{\bf VIII. Comparison with data from experiments and {\it ab initio} calculations}
\end{center}
\vskip0.3in

\noindent None of the potentials used in the present work are considered to be 
candidates for describing any elemental solid over the wide pressure ranges
in which we have performed our calculations. Moreover, the Morse and Gupta
potentials are bounded above. So obviously they cannot describe the
asymptotic pressure regions even in principle. Yet there are some qualitative
and quantitative aspects of our calculations that can  be tested
against laboratory experiments or {\it ab initio} calculations. For example,
we have argued at the end of section V that the mutual proportionality of average
and Debye frequencies or their power law dependence on pressure are expected to
be rather robust, even if not exact, results over rather wide range of high
pressures. Figure 18 shows the data on the aspect of proportionality that we
have found from reports of laboratory experiments. The three systems are
bcc-iron ,hcp-iron and bct-tin with data taken from references 16, 17 and 19
, respectively. In all the cases  the average frequency was calculated by us
from the published density of states data. The Debye frequency was taken directly 
from the
references quoted for bct-tin and bcc-iron whereas it was calculated by us from
the published vibrational density of states for hcp-iron.
%*********************************************************
\begin{figure}[H]
  \centering
  \includegraphics[width=0.52\textwidth]{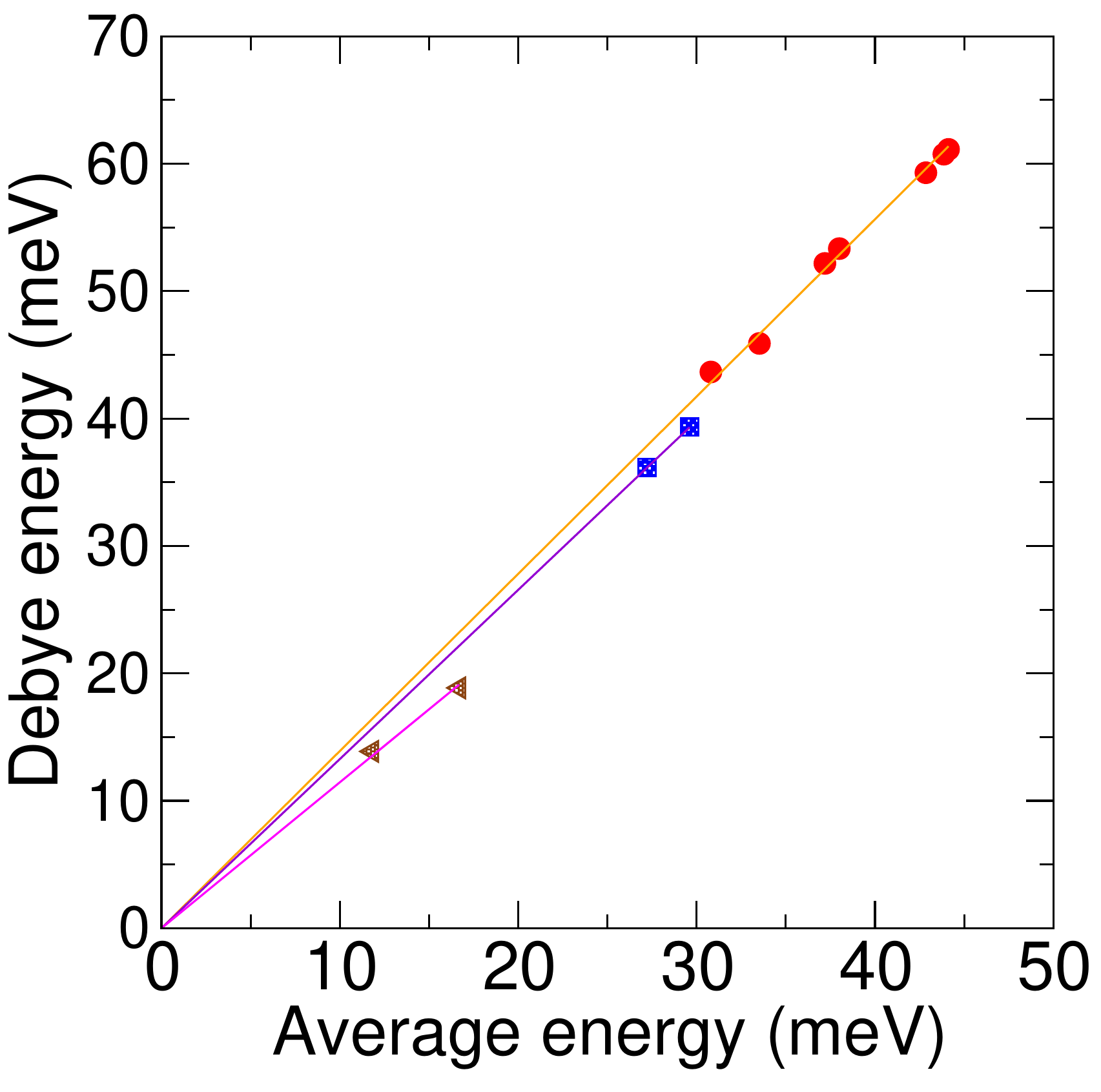}
    \caption{ \small{Average phonon energy is plotted against Debye energy for
the data on crystalline spectra collected from literature. Circles:
Experimental data for hcp-Fe [17]; Squares: Experimental data
for bcc-Fe [16]; Triangles: Experimental data for bct-Sn [19].
Best fit straight line passing through origin is shown in each case.}}
  \label{fig:fig18}
\end{figure}
%************************************************************
 
Although none of the three cases cover the range of pressures that would be desirable for a satisfactory
verification of the prediction of proportionality the
data in all the cases are consistent with the expectation. 

The verifiable quantitative predictions of our work follow from the following
line of reasoning: In any real material the repulsive potential at the shorter
distances is expected to grow faster than any power law i.e. the limit of pressure 
going to infinity should effectively be describable by a type A potential with the
effective $m$ becoming larger and larger at shorter distances -- at least when the 
sum-over-pairs type of potential for the
repulsive part is realistic. Several consequences follow from this. Below we
point out these consequences and compare them with data available from laboratory
experiments or {\it ab initio} calculations.

(i) When the applied pressure keeps increasing the NDOSNF for elemental solids should
approach the $\beta$ going to infinity limit of the NDOSNF in the nearest neighbor
model. Earlier we mentioned that this limit of the nearest neighbor model is well
defined. In fact the NDOSNF in this limit can be seen in figures 14(a) and 14(b) for
FCC and ideal HCP, respectively. Figure 19 compares the NDOSNF of HCP iron (the
only HCP data we could find at such high pressures) from laboratory experiment
as well as {\it ab initio} calculation at the
highest available pressure (153 GPa) in [17] with the $\beta$ going to 
infinity limit of the ideal HCP NDOSNF as shown in fig. 14(b). The agreement is quite 
satisfactory over the entire spectrum. We have not found analogous high pressure data 
for the FCC case that can be compared with the prediction shown in fig. 14(a).
%**********************************************
\begin{figure}[H]
  \centering
  \includegraphics[width=0.52\textwidth]{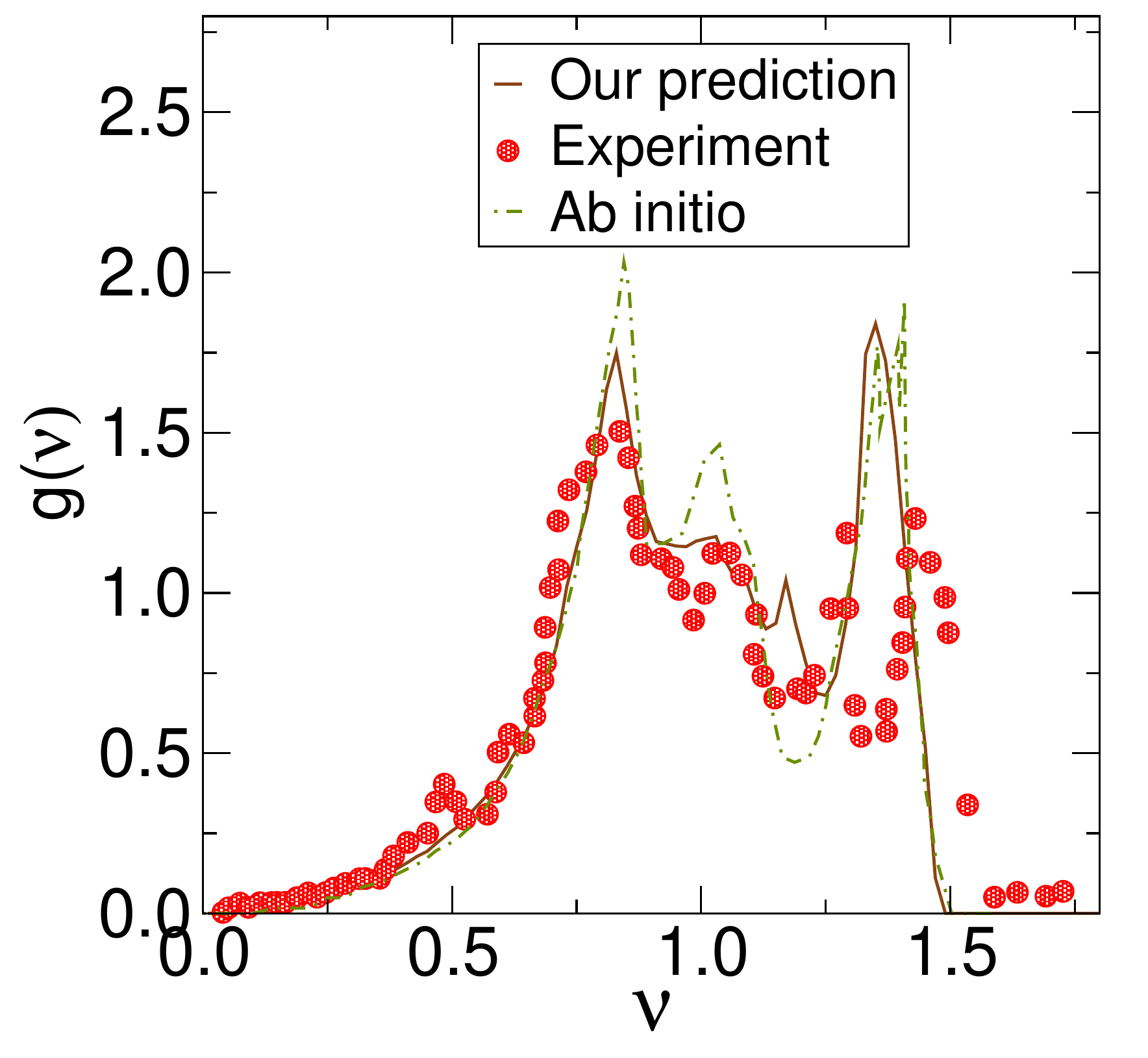}
    \caption{\small{Normalized density of states $g(\nu)$ for normalized frequency $(\nu)$: comparison
between our prediction and results from experiments/{\it ab initio} calculations for hcp-iron [17].}} \label{fig:fig19}
\end{figure}
%*************************************************
 
(ii) From the nearest neighbor model discussed in section IV we can calculate the
ratio of the direction averaged longitudinal speed of sound ($<c_{L}>$ and the
corresponding value for transverse speed ($<c_{T}>$). This ratio depends on
the crystal type and the value of $\beta$. As we have discussed above the extreme high pressure
limit should correspond to $\beta$ going to infinity. For the ideal HCP case
$<c_{L}>$/$<c_{T}>$ is predicted by this procedure to be 1.77 (in comparison, this ratio
is 3.65 when $\beta$ approaches 8 from above) and this is {\it independent of
the material}. We have found experimental/{\it ab initio} calculation based values for
this ratio at high pressure for hcp-rhenium and hcp-iron from [17,18,48]. The highest 
pressure experimental values of the ratio are 
around 1.85 and 2.00 for rhenium and iron, respectively. The corresponding values found
from {\it ab initio} calculations of the phonon density of states in the GGA approximation
are  1.96 (rhenium) and 1.93 (iron). These values, from experiments and calculations, are to
be compared with the prediction of 1.77 from our approach.

(iii) The considerations for the HCP structure given in the immediately preceding 
paragraph can be extended to the FCC type also. Here the reported high pressure experimental 
data are available in the form of the three independent elastic constants $C_{11}$, $C_{12}$
and $C_{44}$. We have found data only for the inert element solids helium, neon, xenon, argon 
and krypton [20-23]. From these three independent elastic constants we can calculate two 
independent ratios involving the direction dependent speeds of transverse acoustic (TA) and 
longitudinal acoustic (LA) waves [21]: $(c_{LA,max}/c_{TA,max})^{2}$
= $(C_{11} + 2C_{12} + 4C_{44})/3C_{44}$ and $(c_{LA,max}/c_{LA,min})^{2}$ = 
$(C_{11} + 2C_{12} + 4C_{44})/3C_{11}$ .\par

In table II we show how the experimental values for these two ratios from
the laboratory experiments compare with our predictions. Please note that {\it all}  
FCC elemental solids are predicted to have the same values for these two ratios. While the
agreement between the predicted and measured values of $(c_{LA,max}/c_{LA,min})^{2}$ is
remarkably close in all cases, the situation is somewhat mixed for the other ratio.  It is 
interesting to speculate on
the possible role of many-body forces in causing these deviations. Presence of such forces,
which is reflected in the deviation from Cauchy relationship between $C_{12}$ and $C_{44}$,
has been demonstrated in the case of argon [21]. In our models the potential is always effectively 
of the sum-over-pairs type at the highest pressures.
\begin{table}[H]
\begin{center}
\caption{\small{Comparison for ratios of sound velocities in FCC systems: experiment versus our prediction}}
\begin{tabular}{|c|c|c|c|c|c|}
 \hline
\multicolumn{1}{|c|}{\multirow{3}{*}{System}} &\multicolumn{1}{|c|}{ Highest Pressure}& \multicolumn{2}{|c|}{$\frac{C_{11}+2C_{12}+4C_{44}}{3C_{44}}$}
& \multicolumn{2}{|c|}{$\frac{C_{11}+2C_{12}+4C_{44}}{3C_{11}}$}  \\ \cline{3-6}
& \multirow{2}{*}{\small{(GPa)}} & {\small{Experimental}} &\small{Prediction of} & {\small{Experimental}}& \small{Prediction of}\\ & &\small{value} & \small{present work}&\small{value} &\small{present work}\\\cline{1-6}
  He (Ref. 20)& 0.493 & 2.65 & & 1.34 &\\ \cline{1-3} \cline{5-5}
  Ne (Ref. 23)& 7 &3.65& &1.36&\\ \cline{1-3} \cline{5-5}
  Ar (Ref. 21)& 70 & 3.04& 2.67&1.30&1.34\\ \cline{1-3} \cline{5-5}
  Kr (Ref. 23)& 8&3.47&&1.31&\\ \cline{1-3} \cline{5-5}
  Xe (Ref. 23)&10.6&3.19&&1.34&\\ \hline
\end{tabular}   
\end{center}
\end{table}
%

%\newpage
\vskip0.3in
\begin{center}
{\bf IX. Concluding remarks}
\end{center}
\vskip0.3in

In this paper we have demonstrated that certain aspects of the vibrational spectra and related
properties in isotropically interacting solids  at high pressures can be captured
quite well by extrapolating the behaviour of simple model solids. Towards this end we have studied 
numerically various model solids in three different states of aggregation. For the crystalline states 
we have also studied a nearest neighbor interaction model. Although we have an analytical expression
for the dynamical matrix for this model calculation of the density of states has been done only on computers.
It would be desirable to 
solve this problem analytically for a finite value of the sole parameter ($\beta$) of the model for
the various crystal structures involved. A somewhat simpler problem to solve would be the version
where $\beta$ is set to infinity right at the beginning.

We have also formulated a scaling law for the dispersion relations alongwith a justification for
its applicability to certain types of potentials. An aspect of this work that perhaps needs 
improvement is the formulation of this scaling law for the amorphous case. Here the presence of a 
unit cell is artificial and the use of a Bloch vector is problematic. The challenge is to formulate
the scaling properties in a manner that does not depend in an essential way on these artificialities.

For crystalline cases we should note that the asymptotic predictions made for various properties here are
characteristic only of the crystal type and hence are applicable to {\it all} elemental solids 
belonging to
that crystal type provided the interaction is isotropic and the repulsive part can be adequately
described in terms of sum-over-pairs type potential in the high pressure limit. Thus there are 
significant elements of
universality in the predictions. Data available from experiments are presently rather limited. 
In particular we have not found any data on the density of states at high enough pressures
for an FCC system.
But whatever data we have access to show reasonable to very good agreement with our
predictions. Similar experiments and {\it ab initio} calculations on other elemental 
solids would provide further testing ground for the validity and utility of this work. Finally, we would
like to draw attention to the extraordinary agreement between experimental data for five different
FCC type inert element solids and our prediction for the ratio $(c_{LA,max}/c_{LA,min})^{2}$. Does it
imply something more fundamental than we are able to see? This question becomes even more relevant when
we notice that the agreement between experimental data for these same systems and our prediction is
not nearly as close for the ratio $(c_{LA,max}/c_{TA,max})^{2}$.

\begin{center}
{\bf Acknowledgements}
\end{center}
D.S. thanks  CSIR, India  for financial support. Computational facility for
this work has been funded by the FIST and UPOE programs of the Government of
India.

\newpage

{\bf References}
\begin{enumerate}
\item R. J. Bell, Reports on Progress in Physics {\bf 35}, 1315 (1972).

\item T. S. Grigera, V. Mart\'{i}n-Mayor, G. Parisi, and P. Verrocchio, Phys. Rev.
 Lett. {\bf 87}, 085502 (2001); Nature (London) {\bf 422}, 289 (2003).
 
\item T. S. Grigera, A. Cavagna, I. Giardina, and G. Parisi, Phys. Rev. Lett.
{\bf 88}, 055502 (2002).

\item W.Schirmacher, G. Diezemann, and C. Ganter, Physica B {\bf 284-288},
1147 (2000); Phys. Rev. Lett. {\bf 81}, 136 (1998).

\item F. H. Stillinger and T. A. Weber, Phys. Rev. A {\bf 25}, 978 (1982).

\item A. Rahman, M. J. Mandell, and J. P. McTague, J. Chem. Phys. {\bf 64}, 1564
(1976).

\item M. Parrinello and A. Rahman, Phys. Rev. Lett. {\bf 45}, 1196 (1980).

\item M. Sampoli, P. Benassi, R. Eramo, L. Angelani, and G. Ruocco, J. Phys. :
Cond. Matt. {\bf 15}, S1227 (2003);.

\item C. A. Angell {\it et al}, J. Phys.: Cond. Matt. {\bf 15},S1051 (2003).

\item  H. R. Schober, J. Phys.: Condens. Matter {\bf 16}, S2659-S2670(2004).

\item L. V. Heimendahl and M. F. Thorpe, J. Phys. F: Metal Physics {\bf 5},
L87 (1975). 

\item J. J. Rehr and R. Alben, Phys. Rev. B {\bf 16}, 2400 (1977).

\item {\it Amorphous Solids: Low Temperature Properties}, edited by
W. A. Philips (Springer-Verlag, Berlin, 1981).

\item S. N. Taraskin and S. R. Elliott, {\it Vibrations in Disordered Systems}
(Oxford University Press Inc., USA, 2004).

\item R. L\"{u}bbers, H. F. Gr\"{u}nsteudel, A. I. Chumakov, and G. Wortmann,
Science {\bf 287}, 1250 (2000); G. Shen {\it et al}, Phys. Chem. Minerals
{\bf 34}, 353 (2004).

\item S. Klotz and M. Braden, Phys. Rev. Lett. {\bf 85}, 3209 (2000).

\item H. K. Mao {\it et al}, Science {\bf 292}, 914 (2001).

\item G. Steinle-Neumann, L. Stixrude, and R. E. Cohen, Phys. Rev. B {\bf 60}, 791 (1999).

\item H. Giefers {\it et al}, Phys. Rev. Lett. {\bf 98}, 245502 (2007).

\item J. Eckert, W. Thomlinson, and G. Shirane, Phys. Rev. B {\bf 16}, 1057 (1977).

\item H. Shimizu, H. Tashiro, T. Kume, and S. Sasaki, Phys. Rev. Lett. {\bf 86}, 4568 (2001).

\item H. Shimizu, N. Saitoh, and S. Sasaki, Phys. Rev. B {\bf 57}, 230 (1998).

\item S. Sasaki, N. Wada, T. Kume, and H. Shimizu, J. Raman Spectroscopy  {\bf 40}, 121 (2009).

\item J. K. Dewhurst, R. Ahuja, S. Li, and B. Johansson, Phys. Rev. Lett.
{\bf 88}, 075504 (2002).

\item  H. Kobayashi {\it et al}, Phys. Rev. Lett. {\bf 93}, 195503 (2004). 

\item V. V. Struzhkin {\it et al}, Phys. Rev. Lett. {\bf 87}, 255501 (2001). 

\item S. Ghose {\it et al}, Phys. Rev. Lett. {\bf 96}, 035507 (2006).

\item A. P. Cantor {\it et al}, Phys. of the Earth and Planetary Interiors {\bf 164}, 83 (2007).

\item M. Yamaguchi, T. Nakayama, and T. Yagi, Physica (Amsterdam) {\bf 263B}
258 (1999); S. Sugai and A. Onodera, Phys. Rev. Lett. {\bf 77}, 4210 (1996).

\item B. Frick and C. Alba-Simionesco, Physica (Amsterdam) {\bf 266B}, 13
(1999); Appl. Phys. A {\bf 74} [Suppl.], S549-S551 (2002).

\item Y. Inamura, M. Arai, T. Otomo, N. Kitamura, and U. Buchenau, Physica B
{\bf 284-288}, 1157 (2000); Y. Inamura, M. Arai, M. Nakamura, T. Otomo,
N. Kitamura, S. M. Bennington, A. C. Hannon, and U. Buchenau, J. of
Non-Crystalline Solids {\bf 293-295}, 389 (2001).

\item P. Jund and R. Jullien, J. Chem. Phys. {\bf 113}, 2768 (2000); O. Pilla,
L. Angelani, A. Fontana, J. R. Goncalves, and G. Ruocco, J. Phys.: Condens.
Matter {\bf 15}, S995 (2003).

\item R. J. Hemley, C. Meade, and H. K. Mao, Phys. Rev. Lett. {\bf 79}, 1420
(1997). 

\item J. Schroeder, W. Wu, J.L. Apkarian, M. Lee, L. -G. Hwa, and C. T. Moynihan, 
J. Non-Cryst. Solids {\bf 349}, 88 (2004).

\item K. S. Andrikopoulos, D. Christofilos, G. A. Kourouklis, and
S. N. Yannopoulos, J. of Non-Crystalline Solids {\bf 352}, 4594 (2006).

\item B. Begen, A. Kisliuk, V. N. Novikov, A. P. Sokolov, K. Niss,
A. Chauty-Cailliaux, C. Alba-Simionesco, and B. Frick, J. Non-Cryst. Solids
{\bf 352}, 4583 (2006).

\item K. Niss, B. Begen, B. Frick, J. Ollivier, A. Beraud, A. Sokolov,
V. N. Novikov, and C. Alba-Simionesco, Phys. Rev. Lett. {\bf 99}, 055502
(2007).

\item A. Monaco, A. I. Chumakov, Y. -Z. Yue, G. Monaco, L. Comez, D. Fioretto,
W. A. Crichton, and R. R\"{u}ffer, Phys. Rev. Lett. {\bf 96}, 205502
(2006); A. Monaco, A. I. Chumakov, G. Monaco, W. A. Crichton, A. Meyer,
L. Comez, D. Fioretto, J. Korecki, and R. R\"{u}ffer, Phys. Rev. Lett.
{\bf 97}, 135501 (2006).

\item  A. I. Chumakov, I. Sergueev, U. van B\"{u}rck, W. Schirmacher,
T. Asthalter, R. R\"{u}ffer, O. Leupold, and W. Petry, Phys. Rev. Lett.
{\bf 92}, 245508 (2004).

\item N. Xu, M. Wyart, A. J. Liu, and S. R. Nagel, Phys. Rev. Lett. {\bf 98},
175502 (2007); M. Wyart, L. E. Silbert, S. R. Nagel, and T. A. Witten, Phys.
Rev. E {\bf 72}, 051306 (2005).

\item S. K. Sarkar, G. S. Matharoo and A. Pandey, Phys. Rev. Lett. {\bf 92},
215503 (2004).

\item  G. S. Matharoo, S. K. Sarkar, and A. Pandey, Phys. Rev. B {\bf 72}, 075401 (2005). 

\item G. S. Matharoo and S. K. Sarkar, Phys. Rev. B {\bf 74}, 144203 (2006).

\item P. B. Allen, W. Garber, and L. Angelani, arXiv:cond-mat/0307435v2.

\item M. Born and K. Huang, {\it Dynamical Theory of Crystal Lattices}
(Oxford University Press,1954).

\item N. W. Ashcroft and N. D. Mermin, {\it Solid State Physics} (Saunders College, Philadelphia, 1976).

\item Shang-Keng Ma, {\it Modern Theory of Critical Phenomena} (Benjamin/Cummings, Reading, 1976).

\item H. K. Mao {\it et al}, Nature {\bf 396}, 741 (1998); Nature {\bf 399}, 280 (1999).

\item G. Fiquet, J. Badro, F. Guyot, and H. Requardt, Science {\bf 291}, 468
(2001).

\item L. S. Dubrovinsky, N. A. Dubrovinskaia, and T. Le Bihan, Proc. Natl. Acad. Sciences
{\bf 98}, 9484 (2001).

\item F. Birch, Geophys. J. R. Astron. Soc. {\bf 4}, 295 (1961); J. Geophysical Research
{\bf 66}, 2199 (1961).

\item A. J. Campbell and D. L. Heinz, Science {\bf 257}, 66 (1992).

\item Jung-Fu Lin {\it et al}, Science {\bf 308}, 1892 (2005).

\item D. H. Chung, Science {\bf 177}, 261 (1972).

\end{enumerate}

\end{document}